\documentclass[11pt]{article}

\usepackage[utf8]{inputenc}
\usepackage{setspace,graphicx,amsmath,amsfonts,amssymb,amsthm}
\usepackage{enumitem}

\usepackage{xcolor}
\definecolor{LovelyLink}{RGB}{52,92,150}
\definecolor{LovelyCite}{RGB}{72,130,100}
\definecolor{LovelyURL}{RGB}{70,115,180}
\usepackage[colorlinks=true,linkcolor=LovelyLink,citecolor=LovelyCite,urlcolor=LovelyURL,filecolor=LovelyURL,breaklinks=true,hypertexnames=false]{hyperref}

\usepackage{float}

\usepackage[longnamesfirst]{natbib}

\usepackage{booktabs}
\usepackage{pdflscape}
\usepackage{chngcntr}
\def\cpar{\egroup\centerline\bgroup}
\usepackage{longtable}
\usepackage{subfig}
\usepackage{graphicx}
\usepackage{adjustbox}
\usepackage{array}
\usepackage{multirow}
\usepackage{titlesec}

\titleformat{\subsubsection}[runin]
{\normalfont\bfseries}{\thesubsubsection}{1em}{}

\usepackage[margin=1in]{geometry}

\newtheorem{corollary}{Corollary}
\newtheorem{proposition}{Proposition}

\newtheorem{lemma}{Lemma}
\newtheorem{remark}{Remark}
\newtheorem{hypothesis}{Hypothesis}

\newcommand{\E}{\mathbb{E}}
\newcommand{\insertfloat}[1]{\begin{center}\textit{[Insert #1 about here]}\end{center}}
\DeclareTextCommandDefault{\textbullet}{\ensuremath{\bullet}}
\DeclareTextCommandDefault{\textasteriskcentered}{\ensuremath{\ast}}
\begin{document}

	\setlist{noitemsep}  %
	\onehalfspacing      %

	\thispagestyle{empty}
	\begin{center}
		{\Large\bf Too Late to Prune: Generative AI, the Screen, and \\ Team Creative Decisions}

		\vspace{0.6cm}
		{\large Difang Huang$^{1,\dagger,*}$, Shumin Yu$^{2,\dagger,*}$, Fadong Chen$^{2,*}$, Weihua Zhou$^{2,*}$}

		\vspace{0.4cm}
		\begin{singlespace}
		$^{1}$Academy of Mathematics and Systems Science, Chinese Academy of Sciences, Beijing, China\\[0.15cm]
		$^{2}$School of Management, Zhejiang University, Hangzhou, China\\[0.3cm]
		$^{\dagger}$These authors contributed equally to this work.\\[0.3cm]
		This version: October 10, 2026
		\end{singlespace}
	\end{center}

	\vspace*{0.35in}

	\centerline{\bf Abstract}
	\medskip
	Generative AI makes options cheap and shifts the bottleneck of creative work from generating options to screening them. We argue that AI's value depends on whether its options arrive before or after the team's screen. In our model, early options compete for a place in the portfolio; late options arrive after that competition and are kept unscreened. In a preregistered experiment, we randomly assigned 148 three-person teams to five schedules of access to the same language model during a product-design task with timed search and selection stages. Moving the same 15 minutes of access from search to selection lowered expert-rated creativity by 0.62 standard deviations, although teams used the model at similar rates. Platform logs locate the difference at the screen. Early access enlarged the pool of ideas fivefold, and teams pruned it. Late-access teams reached the screen with the same small pool as teams without AI and then searched again, devoting 29\% of their selection-stage queries to generating new options, compared with 11\% for teams with access throughout. Late access added breadth, not quality: these teams submitted 73\% more proposals than did teams without AI. Language-model judges detected the benefit of access but not the cost of late timing.
	 


	\clearpage
	\singlespacing
	
\section{Introduction}\label{sec:intro}

Generative artificial intelligence (AI) has made options cheap. A language model can produce a dozen fully written
product concepts, feature sets, or strategic alternatives in seconds, and their average quality is comparable to that
of options generated by people \citep{wang2026large,Boussioux2024,Doshi2024}. When options are cheap, the binding
constraint in creative work moves from generating options to screening them: comparing alternatives, discarding most
of them, and committing to the few that survive \citep{TerwieschUlrich2009,Girotra2010,KornishHK2017}.
\citet{DeFreitas2026} argue that generative AI moves the bottleneck of innovation in exactly this direction. Yet
organizations typically deploy generative AI as standing infrastructure. They license a model, connect it to their
teams' workspaces, and leave it on. They then treat the managerial question as one of adoption: should a team have the
tool, and how much should it use it?

This framing overlooks a choice that every team using generative AI makes, deliberately or not: \emph{when} the AI
assistant's options enter the work. The leading evidence on this choice looks inconsistent. Generative AI improves design work
at the ideation stage, but at the implementation stage it helps novices and slows experts \citep{Hou2025}. Writers who
consult a language model before their own ideation converge on its suggestions, whereas writers who consult it
afterward do not \citep{Qin2025}. Assistance can also change later, unassisted performance \citep{Kumar2025}. If
timing is measured by the clock, the first two findings point in opposite directions: one favors AI at the start of
the work, the other favors consulting it later. We argue that the clock is the wrong reference point. What matters is
whether an option arrives before or after the team's \emph{screen}, the point at which the team compares its
alternatives and commits to a few. Measured against the screen, the findings no longer conflict. Both timings that
\citet{Qin2025} compare precede any selection, so their contrast concerns fixation rather than screening; the
assistance that \citet{Hou2025} find helpful precedes selection, whereas the assistance that slows experts follows it.

The argument rests on a simple asymmetry. Creative teams \emph{search}, building a set of alternatives, and then
\emph{select}, narrowing the set and developing the survivors into a deliverable \citep{March1991,Girotra2010};
product development formalizes the transition as a gate \citep{Cooper1990}, where the screen is applied. An option
that arrives before the gate must compete for a place in the team's portfolio of proposals. An option that arrives
after the gate reopens a search the team has already closed, and it comes too late to prune. Little time remains to
compare it with the options already retained, and that time is better spent developing the chosen proposals. A team
that expects some credit for each proposal it submits therefore keeps the new option without comparing it. Every
option it keeps then draws on a development budget fixed by the deadline. Moreover, the assistant delivers its output
to the member who asked for it, so a late option can enter the deliverable without the rest of the team ever seeing
it. The same option that enlarges the pool before the gate competes for scarce development time after it.

The asymmetry also predicts which teams reopen the search. An unscreened option is worth adding only if its expected
quality is not too far below that of the proposals already in hand. A team that enters selection with a strong set,
drawn from a large pool, gains little from another draw and uses the assistant to develop what it has. A team that
enters with a thin set gains more from each new option and uses the same assistant to search again. Early access thus
changes how a team uses later access, so continuous access need not inherit the cost of late access.

We formalize this argument in a model of team creative decisions and test it in a preregistered laboratory
experiment. We randomly assigned 148 three-person teams (444 participants) to five schedules of access to the same
language model during a 40-minute product-design task with a 15-minute search stage, a 15-minute selection stage, and a
10-minute review stage. In the two focal conditions, teams had access to the assistant for the same 15 minutes, either
during search, before the gate (\emph{early access}), or during selection, after it (\emph{late access}). They used it
at similar rates in the two conditions (14.9 and 13.8 queries per team). The other conditions provided continuous
access, no access, or early access with a standing warning against overreliance. Four expert raters scored every
proposal. Platform logs let us reconstruct each team's idea funnel: the pool of ideas it held at the gate, the share of
that pool that reached the submission, and the origin of the submitted text.

Four results bear out this account. First, timing decided whether access helped. Teams with early access produced
submissions rated 0.36 points higher than those of teams with late access (Hedges' $g=0.62$, $p=0.018$) and 0.49
points higher than those of teams without AI ($g=0.83$). Late access, with the same model for the same duration,
produced no detectable gain over working without AI, and continuous access did no better than early access alone.

Second, the funnel shows where the difference arises. Early access enlarged the pool at the gate fivefold, from about 4
to about 20 distinct ideas, and teams pruned it to fewer than three proposals; a calibration of the model indicates
that a pool of this size can account for the early-access gain. Late-access teams reached the gate with the same small
pool as teams without AI and then largely abandoned it: only 5\% of their submitted text existed before the gate,
compared with 22\% for teams without AI and 71\% for teams with early access. The cost of late access lies in the
larger pool it forgoes rather than in a measurable penalty for discarding the team's own ideas.

Third, teams that first received the assistant after the gate used it to search again. In the same selection window
and under the same system prompt, which directed the assistant toward refinement, teams with late access devoted 29\%
of their queries to generating new options, compared with 11\% for teams with continuous access. The two groups differ
in whether they had used the assistant during search, and because this access history was randomly assigned, the
difference is a causal effect of the schedule on how teams used an identical tool. Late output was also more likely to
enter the submission nearly verbatim and without ever appearing in the team chat. Late access added breadth, not
quality: these teams submitted 73\% more proposals than did teams without AI, with no detectable gain in rated
creativity.

Fourth, whether the timing effect is visible depends on the evaluator. Because late access adds breadth, crediting
breadth changes the verdict: adding as little as 0.1 points per proposal to each team's score is enough for late
access to outscore working without AI. A panel of ten frontier language models detected the benefit of access but not
the cost of late timing.

Our paper contributes to three literatures. The first studies how generative AI changes knowledge and creative
work. AI raises productivity and output quality \citep{Noy2023,DellAcquaJagged2026,Brynjolfsson2025} and creative
performance \citep{Doshi2024,Boussioux2024,Lee2024}; the gains depend on how people collaborate with the
model \citep{ChenChan2024,Jia2024} and on the task \citep{Vaccaro2024}, and they extend to teams
\citep{DellAcqua2026,Li2024}. Exposure to AI ideas can also fixate people on the model's suggestions
\citep{Wadinambiarachchi2024} and reduce the diversity of what they produce
\citep{Doshi2024,Anderson2024,Meincke2025}, although passive exposure to AI examples can instead raise
collective diversity \citep{Ashkinaze2025}. The studies that vary when AI enters define timing relative
to the clock, to a stage label, or to an individual's own ideation
\citep{Hou2025,Qin2025,Kumar2025,HuangCreativePhases2026,Hutter2026}. We instead define timing relative to the team's
screen and hold the model and the duration of access fixed; teams in the two focal conditions also used the assistant
at similar rates. We argue that this reference point reconciles findings that conflict when timing is measured by the
clock. The design also clarifies what studies with continuous access estimate: a mixture of early and late effects
whose weights depend on how participants schedule their own use. In our data, teams with continuous access used the
assistant most during search and, in the selection stage, mainly to develop their proposals, so continuous access
performed like early access.

The second literature studies idea generation and selection. The value of an ideation process depends on the ideas that
survive selection rather than on the average idea
\citep{DahanMendelson2001,TerwieschUlrich2009,Girotra2010,Kornish2011,Boudreau2011,KornishHK2017}, and additional
alternatives can impair choice \citep{HauserWernerfelt1990,Iyengar2000,Chernev2015}. With generative AI, screening
rather than supply becomes the constraint \citep{Terwiesch2026,DeFreitas2026}, and recent work asks whether AI can do the
screening \citep{Lane2025,Kireyev2025}. We study the complementary question of when AI-generated options should reach a
human screen. The model links this literature to search theory \citep{McCall1970,Weitzman1979}: it derives when a late
option is kept without comparison and why a team without an early supply of options keeps searching after the gate.
The data, in turn, measure the funnel that this literature models. The team setting matters: because the assistant
answers only the member who asked, its output resembles the unshared information that groups often fail to pool
\citep{Stasser1985}. In our data, some late output entered submissions without ever being discussed.

The third literature studies advice taking and algorithmic evaluation. The timing of advice relative to a decision
maker's own judgment shapes how much the advice is used \citep{Sniezek1995,Bonaccio2006}; people sometimes avoid and
sometimes prefer algorithmic input \citep{Dietvorst2015,Logg2019}; and unrestricted access to generative AI can harm
learning, whereas guardrails largely mitigate the harm \citep{Bastani2025}. We add a dynamic dimension: earlier access
changed not only how much teams later relied on the assistant but also what they asked it to do. Language models are
also increasingly used to evaluate creative and managerial output
\citep{ChenJudge2024,Doshi2025SMJ,Lane2025}, and they can favor verbose text and their own generations
\citep{ZhengJudge2023}. In our data, a large, internally consistent panel of language-model judges reproduced the
gross effect of AI access but not the effect of its timing, so organizations that use AI both to generate and to
evaluate ideas may miss the consequences of when AI enters the work.

For managers, the timing of AI access is a design parameter. The assistant should be available while the team builds
its option set; when that is not possible, late output should pass a screen before it enters the deliverable, or late
use should be limited to developing proposals the team has already chosen. Because a small credit for breadth reverses
the verdict on late access, organizations should also decide whether they value depth or breadth before judging
whether a deployment works.

The rest of the paper proceeds as follows. Section~\ref{sec:model} presents the model, and Section~\ref{sec:design} describes the experiment.
Section~\ref{sec:results} reports the timing effect, the funnel, the effect of access history, and the fate of late
output; Section~\ref{sec:robust} examines robustness and alternative evaluators; and Section~\ref{sec:discussion}
concludes. Proofs and additional analyses are in the Online Appendix.

\section{Model}\label{sec:model}

We model a team that must deliver a portfolio of proposals within a fixed time. The model has four ingredients: a
screen that the team applies when search ends, a development budget that the retained proposals share, a decision
about how much to keep searching once selection has begun, and an evaluator who scores the portfolio. It is
deliberately stylized; its purpose is to show how one primitive, the arrival of options relative to the screen, generates predictions about quality, breadth, use of the assistant, and measurement.

The building blocks are standard. We represent screening through order statistics of independent quality draws, as in
extreme-value models of concept selection and innovation tournaments
\citep{DahanMendelson2001,TerwieschUlrich2009,Girotra2010,Kornish2011}. Whether a late option is compared with the
retained ones or kept without comparison follows from a trade-off between comparison and development time
(Lemma~\ref{lem:retention}). The same valuation governs the decision to keep searching
(Proposition~\ref{prop:latesearch}), a sequential search problem in the tradition of \citet{McCall1970} and
\citet{Weitzman1979}. Online Appendix~\ref{ec:model} contains the proofs, a result on how the timing effect depends on
the way a portfolio is valued, a remark on warnings, and a numerical analysis.

\subsection{Setup}

\paragraph{Options.} Each candidate idea has an intrinsic quality $\theta$, drawn independently from a continuous
distribution $F$ with mean $\mu$ and finite variance. Ideas generated by the team and ideas supplied by the assistant
come from the same distribution; we do not assume that AI ideas are better or worse.\footnote{If AI ideas had a
higher mean, early access would gain more relative to no access, and unscreened late ideas would dilute portfolio
quality less, because they would pull it toward a higher mean. Equal means isolate the role of timing. This assumption is supported by the large scale comparison of divergent creativity in humans and large language models \citep{wang2026large}.} For $N\geq k$
draws, let $M_k(N)$ denote the expected mean quality of the $k$ best draws. $M_k(N)$ increases in $N$, and $M_k(N)>\mu$
whenever $N>k$.

\paragraph{Search and screening.} In the search stage, the team generates $n>k$ ideas of its own. If the assistant is
available during search, it supplies $a$ additional ideas. At the end of search, the team applies a screen: it
compares the ideas in its pool and retains the $k$ best. In our experiment, the instructions for the second stage
asked teams to select the best ideas from the first stage and develop them.

\paragraph{Selection, late options, and development.} In the selection stage, the team develops its retained ideas into
proposals. A proposal's delivered quality is $q=\theta+h(e)$, where $e$ is the development effort it receives and $h$
is strictly increasing and strictly concave with $h(0)=0$. Total development effort is limited by a budget $B$ (the
fixed session time), which the $K$ proposals in the portfolio share equally, so that $e=B/K$. An assistant available
during selection can help develop the retained ideas, which raises the effective budget to $B+\beta$ with
$\beta\geq0$, and it can supply new ideas on request. A new idea obtained during selection arrives already written
out, after the screen has been applied. In the main analysis, late ideas are kept without comparison (Lemma
\ref{lem:retention} derives when), and the team retains each with probability $\rho\in(0,1]$ for reasons unrelated to
its quality. Let $r$ denote the number of late ideas retained.

\paragraph{Evaluation.} An evaluator scores the portfolio by the mean delivered quality of its proposals,
$\bar Q=\frac{1}{K}\sum_{j}q_j$, possibly plus a credit $\gamma\geq 0$ per proposal: $S_\gamma=\bar Q+\gamma K$. The
case $\gamma=0$ corresponds to an evaluator who judges each proposal on its merits, as our expert raters were
instructed to do.

\paragraph{Schedules.} Under \emph{no access} ($N$), the pool contains $n$ ideas, $K=k$, and the budget is $B$. Under
\emph{early access} ($E$), the pool contains $n+a$ ideas, $K=k$, and the budget is $B$. Under \emph{late access}
($L$), the pool contains $n$ ideas, the team retains $r$ late ideas, $K=k+r$, and the budget is $B+\beta$. Under
\emph{full access} ($F$), the pool contains $n+a$ ideas, and the team also chooses how many late ideas to request.

\subsection{When Are Late Options Compared?}

Consider a team that holds $K$ proposals with mean quality $\bar Q$ and has $R$ units of development time left, shared
equally among its proposals. An option of unknown quality $\theta\sim F$ arrives, already written out. The team can
discard it, keep it without comparison, or spend $\tau>0$ units of time comparing it with the retained proposals, which
reveals $\theta$, and then decide. The team values its portfolio by $W=\bar Q+h(R/K)+\gamma_T K$, where $\gamma_T\geq0$
is the credit it expects for each proposal; in our experiment, teams were told that scoring would also take quantity
into account. Keeping an option of quality $\theta$ changes $W$ by
\[
\Delta(\theta,R)=\frac{\theta-\bar Q}{K+1}+h\!\left(\frac{R}{K+1}\right)-h\!\left(\frac{R}{K}\right)+\gamma_T ,
\]
which increases in $\theta$; let $\hat\theta(R)$ solve $\Delta(\hat\theta,R)=0$. Comparing lets the team discard
options below $\hat\theta(R)$. Relative to keeping without comparison, its expected benefit is
$b(R)=\E[(\hat\theta(R)-\theta)^+]/(K+1)$, and its cost is the development it takes away from the retained proposals,
$c(R)=h(R/K)-h((R-\tau)/K)$.

\begin{lemma}[Optimal retention]\label{lem:retention}
Suppose that $e\,h'(e)$ is nondecreasing, as for $h(e)=A\,e^{\alpha}$ with $A>0$ and $0<\alpha<1$, and that an average option is
worth keeping, $\Delta(\mu,R)\geq0$. Then (i) the team never discards an option without comparing it; (ii) there is a
threshold $R^*\geq\tau$ such that the team compares the option if and only if $R\geq R^*$; and (iii) the probability that
the option is retained is $\rho(R)=1$ for $R<R^*$ and $\rho(R)=1-F(\hat\theta(R))<1$ for $R\geq R^*$.
\end{lemma}

Lemma~\ref{lem:retention} makes retention a consequence of timing. Comparing an option costs development time, and
because the returns to development are concave, that time is most valuable when little of it is left. An option that
arrives during search is compared when the team screens its pool, with the whole selection stage still ahead. An option
that arrives late is kept without comparison, because the time a comparison would take is worth more as development of
the proposals the team already has; the team keeps rather than discards it because it expects some credit for the
number of proposals. In the numerical example of Online Appendix Table~\ref{tab:ec-retention}, a team compares a new
option only when at least half of the selection stage remains.

In a team, comparison has a further requirement: the option must reach the other members. Because the assistant
delivers its output to the member who asked, the comparison time $\tau$ includes bringing the option to the team, and
an option kept without comparison can enter the deliverable without the team having seen it. The lemma therefore has
an observable counterpart (Section~\ref{sec:fate}): options obtained after the screen should more often enter the
deliverable without passing through the team's common channel, and in the form in which they arrived. The analysis
below treats late options as kept without comparison ($R<R^*$); options that arrive early enough in the selection stage
to be compared face the cutoff $\hat\theta(R)$, which weakens dilution without removing it.

\subsection{Timing, Quality, and Composition}

Expected mean quality under the first three schedules is
\begin{align*}
\bar Q_N &= M_k(n) + h(B/k),\qquad \bar Q_E = M_k(n+a) + h(B/k),\\
\bar Q_L &= \frac{k\,M_k(n)+r\mu}{k+r} + h\!\left(\frac{B+\beta}{k+r}\right).
\end{align*}

\begin{proposition}[Timing]\label{prop:timing}
(i) Early access raises mean quality relative to no access: $\bar Q_E>\bar Q_N$.
(ii) The advantage of early over late access decomposes as
\[
\bar Q_E-\bar Q_L=\underbrace{M_k(n+a)-M_k(n)}_{\text{screening premium}}
+\underbrace{\frac{r}{k+r}\bigl(M_k(n)-\mu\bigr)}_{\text{dilution of quality}}
+\underbrace{h\!\left(\tfrac{B}{k}\right)-h\!\left(\tfrac{B+\beta}{k+r}\right)}_{\text{dilution of development}} .
\]
For $a\geq1$ and $r\geq1$, the first two terms are strictly positive. Hence $\bar Q_E>\bar Q_L$ whenever development help
in the selection stage does not more than offset the spreading of the budget, $\beta\leq rB/k$.
(iii) Late access lowers mean quality relative to no access when $\beta\leq rB/k$; it can raise mean quality only if
development help exceeds the spreading of the budget, $\beta>rB/k$, by enough to outweigh the dilution of quality.
\end{proposition}

Proposition~\ref{prop:timing} isolates two forces that favor early supply. An option that arrives during search competes
for a place in the portfolio, whereas an option that arrives during selection takes a place with little competition and
pulls the expected quality of the portfolio toward the unconditional mean $\mu$. In addition, every unscreened option the
team retains spreads the development budget more thinly. The two selection terms predict a gap in the quality of the
ideas themselves (their originality and fit); the development term predicts a gap in how far proposals are
elaborated. The rated dimensions let us distinguish these channels (Section~\ref{sec:dims}).

\begin{proposition}[Breadth and depth]\label{prop:composition}
Late access increases the number of proposals in the portfolio, $K_L=k+r>K_E=K_N=k$, and, if $\beta<rB/k$, lowers
the development each proposal receives, $e_L=(B+\beta)/(k+r)<e_E=e_N=B/k$.
\end{proposition}

\begin{remark}[Displacement]\label{rem:displacement}
The schedules above add late options to the screened set. If late options instead replace the screened ideas, the
late portfolio consists of $r$ unscreened options with expected quality $\bar Q_L=\mu+h\bigl((B+\beta)/r\bigr)$. The
early-versus-late gap is then $M_k(n+a)-\mu+h(B/k)-h\bigl((B+\beta)/r\bigr)$: late access forfeits not only the
increment from a larger pool but also the team's own screening premium $M_k(n)-\mu$. Partial displacement lies between
this case and Proposition~\ref{prop:timing}(ii).
\end{remark}

Because late access trades depth for breadth, it also matters how a portfolio is valued. An organization that runs an
innovation tournament may care about its best idea \citep{Girotra2010,TerwieschUlrich2009}, and one that pays per idea
may care about the total. Online Appendix Proposition~\ref{prop:aggregation} shows that the early-access advantage is
largest in the mean, smaller but positive in the best proposal, and negative in the total.

\subsection{Why Late Access Becomes Late Search}

So far, the number of late options has been taken as given. In practice, the team decides query by query whether to
use an available assistant to generate a new option or to develop the options it already has. We model this decision
with the same portfolio value $W$ that governs retention. Suppose the team enters selection holding $K$ screened
proposals with mean quality $\bar Q_0$ and has $R<R^*$ units of development time, so that, by
Lemma~\ref{lem:retention}, any option it obtains is kept without comparison. A request costs $\eta>0$ in attention and
$\delta\geq0$ in development time, and the number of selection-stage queries a team can issue is fixed, so queries not
spent on generation go to development.

\begin{proposition}[Endogenous late search]\label{prop:latesearch}
(i) The expected gain from requesting and keeping one more option,
\[
G(\bar Q,K,R)=\frac{\mu-\bar Q}{K+1}+h\!\left(\frac{R-\delta}{K+1}\right)-h\!\left(\frac{R}{K}\right)+\gamma_T-\eta,
\]
is strictly decreasing in the mean quality $\bar Q$ of the proposals in hand.
(ii) If the team keeps requesting new options while this expected gain is positive, then for any sequence of draws the
number of late requests is nonincreasing in the mean quality $\bar Q_0$ with which it enters selection.
(iii) Early access weakly raises $\bar Q_0$ for every realization of the team's own ideas. Hence teams with early access
make weakly fewer generation requests in the selection stage than teams without it, retain fewer late options, and
devote a larger share of their selection-stage queries to development.
\end{proposition}

Proposition~\ref{prop:latesearch} explains why continuous access need not inherit the cost of late access. An option
kept without comparison is worth adding only if its expected quality, $\mu$, is not too far below that of the proposals
already in hand. A team that enters selection with a strong, AI-enriched set has little to gain from further draws and
uses the assistant to develop what it has. A team that enters selection without such a set gains more from each new
option and uses the same assistant to reopen its search after the screen. This is precisely the behavior the model penalizes.
In the numerical example of Online Appendix~\ref{ec:model}, teams without early access make 4.8 late requests on average
and teams with early access make 2.1, and the ordering holds in every simulated draw.\footnote{The rule in part (ii)
requests while the expected gain from one more option is positive. If the team instead valued only its single best
idea, as a team competing solely for a prize would, a standard reservation-value argument
\citep{McCall1970,Weitzman1979} yields the same ordering, because the best of $n+a$ draws first-order stochastically
dominates the best of $n$.}

\subsection{Evaluators and Gates}

\begin{corollary}[Evaluator weight]\label{cor:evaluator}
For an evaluator with quantity credit $\gamma$, the early-versus-late gap is $S_{\gamma,E}-S_{\gamma,L}=(\bar Q_E-\bar
Q_L)-\gamma r$. The gap is strictly decreasing in $\gamma$ and vanishes at $\gamma^*=(\bar Q_E-\bar Q_L)/r$.
\end{corollary}

An evaluator who penalizes additional proposals less, or who credits breadth more, therefore ranks the two schedules
closer together, and whether the timing effect is detected depends on how the evaluator weighs depth against breadth.

\begin{corollary}[Reopening the gate]\label{cor:gate}
Suppose that a late option may enter the portfolio only after it has been compared with the retained proposals. By
Lemma~\ref{lem:retention}(iii), it is then kept only if $\theta\geq\hat\theta(R)$, so a retained late option has expected
quality $\E[\theta\mid\theta\geq\hat\theta(R)]>\mu$, at a cost in development time $c(R)$ per comparison. The dilution of
quality in Proposition~\ref{prop:timing}(ii) falls from $\frac{r}{k+r}\bigl(M_k(n)-\mu\bigr)$ to
$\frac{r'}{k+r'}\bigl(M_k(n)-\E[\theta\mid\theta\geq\hat\theta(R)]\bigr)$ with $r'\leq r$, and it becomes zero or
negative once the cutoff exceeds the mean of the retained ideas.
\end{corollary}

An enforced screen therefore recovers part of the value of late access, and it removes the dilution of quality entirely
when comparisons are cheap relative to the development they displace. Corollary~\ref{cor:gate} is the model's
prescription for organizations that cannot open the assistant before the gate, and it is directly testable.

\subsection{Hypotheses}

We preregistered the following hypotheses; the model supplies their common logic.

\begin{hypothesis}\label{hyp:access}
Assistant access during search raises rated creativity relative to no access (Proposition~\ref{prop:timing}(i)).
\end{hypothesis}
\begin{hypothesis}\label{hyp:timing}
Assistant access during search raises rated creativity relative to access of equal duration during selection
(Proposition~\ref{prop:timing}(ii)).
\end{hypothesis}
\begin{hypothesis}\label{hyp:composition}
Assistant access during selection increases the number of proposals that teams carry into their submissions and
lowers the development of each proposal (Proposition~\ref{prop:composition}).
\end{hypothesis}
\begin{hypothesis}\label{hyp:warning}
A standing warning against overreliance reduces teams' reliance on assistant output without raising rated creativity
(Online Appendix Remark~\ref{rem:warning}).
\end{hypothesis}

The model has further implications, which we also examine: early access enlarges the pool at the screen and is
followed by pruning, whereas late options enter after the screen and can displace the screened set
(Remark~\ref{rem:displacement}); early access crowds out late generation (Proposition~\ref{prop:latesearch}); late
options pass into the deliverable without comparison and in the form delivered (Lemma~\ref{lem:retention}); and the
measured effect depends on the evaluator's weight on breadth (Corollary~\ref{cor:evaluator}).

\section{Experimental Design}\label{sec:design}

\subsection{Task and Platform}

Three-person teams completed a product-design task on a custom web platform. Each team received one of three
object-combination prompts asking it to combine two everyday objects into one or more products or services that were
both creative and feasible: a backpack and a pet accessory, a refrigerator and a calendar, or shoes and a map. The
platform had three modules: a team chat, a personal draft editor for each member, and, in conditions with access, a
private assistant panel through which each member could query the assistant. Assistant responses were visible only to
the member who asked; teammates saw them only if they were shared in the chat. The final submission could contain one
or more proposals, each consisting of a product name and a description of its main features and functions in no more
than 250 Chinese characters.

The session lasted 40 minutes, divided into three fixed stages that the platform enforced and displayed. In the
15-minute divergent stage (S1), teams were instructed to brainstorm as many product concepts as possible and not to
make final decisions. In the 15-minute focusing stage (S2), they were instructed to select the best ideas from the
first stage (in any number) and develop them, naming each and describing its functions. In the 10-minute final-review
stage (S3), they checked feasibility and wording, and one member submitted the team's document. The fixed stages make
timing manipulable: they fix when the screen is applied and let us open the assistant before or after it, much as a
stage-gate process fixes when a project is reviewed. In eight teams, the logged search or selection stage deviated
from 15 minutes by more than three minutes, in most cases because a member joined late; excluding these teams does not
change the results (see Section~\ref{sec:robust}).

The assistant was DeepSeek V3.2, accessed through the provider's API at temperature 0.1. Each team had its own
conversation instance, and at each query the assistant received the team's 100 most recent chat messages and a
stage-specific system prompt. The search-stage prompt instructed the assistant to provide as many ideas as possible;
the selection-stage prompt, to help refine features, evaluate potential, and screen for the best solutions; and the
review-stage prompt, to check and polish. The prompts were set by stage and were identical across conditions. Queries
were unlimited, and outside a team's assigned window the assistant panel was visibly disabled. All materials were in
Chinese; Online Appendix~\ref{ec:materials} reproduces the prompts and instructions.

\subsection{Conditions}

\begin{figure}[!htbp]
		\centerline{\includegraphics[width=0.95\textwidth]{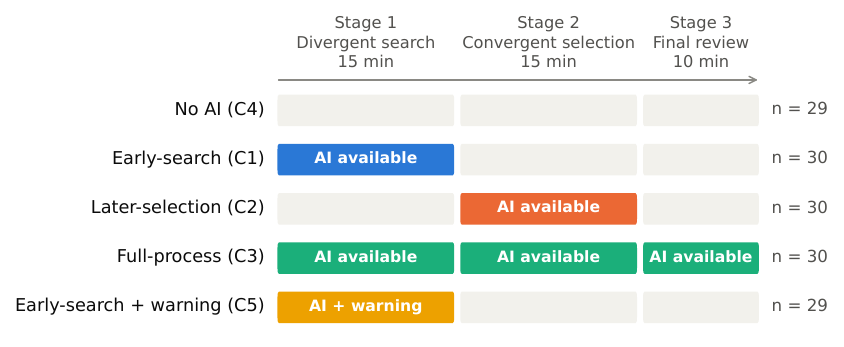}}
		\noindent\caption{{\bf Experimental design.} Teams worked for 40 minutes in three platform-enforced stages. Colored cells mark when the assistant was available. All conditions used the same model (DeepSeek V3.2, temperature 0.1) with the same stage-specific system prompts.}
		\label{fig:design}
	\end{figure}

Figure~\ref{fig:design} and Table~\ref{tab:design} summarize the five conditions. Under \emph{early-search access}
(C1), the assistant was available during S1 only; under \emph{later-selection access} (C2), during S2 only; under
\emph{full-process access} (C3), throughout the session; and under the \emph{no-AI control} (C4), never. Under
\emph{early-search access with warning} (C5), the assistant was available during S1 only, and every response was
followed by a statement in bold italics: ``\textbf{\textit{AI serves only as an initial reference point and its outputs may not be
accurate or comprehensive. We highly encourage you to discuss these ideas further with your team members.}}'' Teams
learned their access window from a tool notice worded identically across conditions except for the window it named;
they were not told that other conditions existed.

\insertfloat{Figure~\ref{fig:design}}
\insertfloat{Table~\ref{tab:design}}

The treatment is the window in which the assistant could be used. The platform enforced it, and every one of the 2,124
queries fell inside the assigned window. The stage tasks were the same in every condition, and how teams actually used
each stage is an outcome. The stage tasks did shape behavior: teams with full-process access devoted 58\% of their
search-stage queries but only 11\% of their selection-stage queries to generating new options. Continued search by
teams that first received the assistant in the selection stage is therefore a response to the timing of access, not a
failure of the stage manipulation.

The comparison of C1 and C2 is the core of the design. The two conditions use the same model for a window of the same
length and differ in whether the window precedes or follows the screen. Because the system prompt was set by stage,
they also differ in what the assistant was instructed to do: C1 teams consulted an assistant instructed to generate
ideas, and C2 teams one instructed to refine them. This difference is part of what late access means in practice, since an assistant opened during selection is configured for selection work. The comparison
of C2 and C3 within the selection window holds the window, the prompt, and the model fixed and varies only whether the
team had used the assistant before. C3 also shows whether adding selection-stage access to search-stage access helps or
hurts, and C5 tests a reliance intervention while holding the window fixed.

\subsection{Participants, Randomization, and Incentives}

We recruited 450 students at a large public university in China for 150 teams, preregistered as 30 teams per
condition, which provides 80\% power to detect a standardized difference of $d\approx0.74$ in a pairwise contrast at
$\alpha=0.05$. Two teams did not submit study data because of technical failures and were excluded before rating,
leaving 148 teams and 444 participants (29 or 30 teams per condition). Participants were 21.6 years old on average, and they came from a range of majors and from the undergraduate to the doctoral level, and they used
generative AI frequently (median self-reported frequency of 4 on a five-point scale). Sessions took place in February
and March 2026.

Teams were assigned to condition and task by a pre-generated allocation sequence containing ten slots for each of the 15
condition-by-task cells in random order; enrolled participants were randomly formed into teams that filled the next
available slot (realized cell sizes 9--11; Online Appendix Table~\ref{tab:ec-alloc}). Participants worked at
partitioned workstations, offline communication was prohibited, and all team communication passed through the
platform's chat. Participants received a base payment of 45 RMB, and members of the ten highest-scoring teams received
a bonus of up to 50 RMB each. Teams were told that experts would score their submissions blind, primarily on creativity
and feasibility and, to a lesser extent, on quantity. The design, hypotheses, and analysis plan were preregistered
before data collection. The study was approved by the authors' institutional review
board, and all participants gave informed consent.

Randomization produced well-balanced conditions (Online Appendix Table~\ref{tab:balance}). None of the ten
pretreatment covariates differs across conditions ($p\geq0.114$), and an omnibus permutation test across all five
conditions yields $p=0.66$. Joint orthogonality tests of each treatment against the control and of C1 against C2 do not
reject in four of five comparisons ($p=0.35$--$0.87$). For full-process access against the control, the joint test
rejects ($p=0.015$), mainly because full-process teams scored lower on emotional stability; adjusting for all ten
covariates does not reduce the full-process effect (0.523 points, $p=0.001$; Table~\ref{tab:main}, column 3).

\subsection{Outcomes}

Four raters with five to ten years of industry experience in innovation management or industrial design scored every proposal while blind to condition and team. Each proposal was rated on
originality quality, appropriateness, elaboration, and text quality (1--7 scales), together with a capped count of
distinct valid features and a naming bonus. The creativity composite weights originality (0.90 quality, 0.10 capped
feature count, plus the naming bonus) by 0.70, appropriateness by 0.15, and elaboration by 0.15; text quality is
analyzed separately.

The primary outcome is the team's mean composite across the proposals in its submission, averaged over raters. It is
the expected rating of a proposal drawn from what the team chose to submit. This is the quantity raters were instructed to judge and the natural measure of performance when what a team commits to matters. Because every rater scored every
submission, additive differences in rater severity cancel in team means. Agreement among raters was modest
(ICC(2,$k$)$\,=0.44$ at the proposal level), as is common for expert creativity ratings. Rating noise is independent of
the randomly assigned condition, so it widens confidence intervals without biasing contrasts, and the timing contrast
holds for every rater subset (see Section~\ref{sec:robust}).

\subsection{Process Measures}

The platform logged 13,324 chat messages, 2,124 assistant exchanges, and 4,776 personal drafts, from which we
construct five classes of process measures:

\emph{Portfolio measures} are the rater-coded number of proposals per
submission, rated elaboration, submission length, and characters per proposal.

 \emph{Search measures} are the volume
of team chat during S1 and the semantic dispersion of the submission, defined as the mean pairwise TF--IDF cosine
distance among its paragraph-level idea units.

\emph{Request types} classify each assistant query into one of five categories: generating new options, developing a
specified option, evaluating or selecting among options, writing or polishing text for submission, and other
requests. A large language model coded the queries by applying a written codebook (Online
Appendix~\ref{ec:coding}) after the queries had been shuffled and stripped of team, condition, stage, and time stamps,
so the coding was blind to treatment. A second, independent blind coding of 200 random queries agreed with the codes in
87.0\% of cases (Cohen's $\kappa=0.82$; $\kappa=0.89$ for generation versus all other requests).

\emph{Funnel measures} reconstruct the pool of ideas each team held at the gate and follow it into the submission. The
pool is the set of distinct ideas in the members' last personal drafts of the search stage, split at list markers and
deduplicated within team; an idea is AI-derived if it matches the team's search-stage assistant output. We record the
share of pool ideas that reappear in the submission and the share of submission sentences that match any pre-gate
material of the team (drafts, chat, and assistant output).

\emph{Tracing measures} follow the assistant's output. We split each response into idea units at list items and
headings (8,603 units) and trace each unit into the final submission and into the team chat posted after it arrived. A
unit \emph{entered the submission} if one of its sentences has a TF--IDF cosine similarity of at least 0.35 with a
sentence of the team's submission, computed on character bigrams and trigrams, which require no word segmentation of
Chinese text. It entered \emph{near-verbatim} if the similarity is at least 0.70, and it was \emph{shared} if it
reached the first threshold with a later chat message. The threshold is conservative: applied to other teams'
submissions for the same task, it matches 0.4\% of unit--submission pairs. Online Appendices~\ref{ec:funnel}
and~\ref{ec:fate} report alternative thresholds and sentence-embedding versions of the funnel and tracing measures.

\subsection{Estimation and Inference}

We estimate
\begin{equation}\label{eq:main}
Y_i=\alpha+\sum_{c\in\{1,2,3,5\}}\pi_c\,\mathbf{1}[\text{C}c_i]+\lambda_{\text{task}(i)}+X_i'\psi+\varepsilon_i,
\end{equation}
where $Y_i$ is the outcome for team $i$, the coefficients $\pi_c$ are differences from the no-AI control,
$\lambda_{\text{task}(i)}$ are task fixed effects, and $X_i$ are optional team-mean pretreatment covariates. The timing
effect is $\pi_1-\pi_2$. We report the heteroskedasticity-robust standard errors for team-level models and
team-clustered standard errors for rater-level, proposal-level, and idea-unit-level models. For each condition
contrast, we also report a randomization-inference $p$ value from 10,000 re-randomizations of condition labels between
the two compared arms within task strata \citep{Young2019}. For the family of eight preregistered and exploratory
condition contrasts, we report Benjamini--Hochberg adjusted $p$ values \citep{BH1995}, the preregistered multiplicity
procedure, and, as conservative references, Holm and Westfall--Young step-down adjusted $p$ values
\citep{WestfallYoung1993,ListShaikhXu2019}.

\section{Results}\label{sec:results}

We first report how teams used the assistant and how the timing of access affected rated creativity
(Sections~\ref{sec:use}--\ref{sec:dims}). We then trace the mechanism through the idea funnel, the use of the assistant
in the selection window, and the fate of its output (Sections~\ref{sec:funnel}--\ref{sec:fate}), and close with the
warning (Section~\ref{sec:warning}).

\subsection{Assistant Use}\label{sec:use}

Teams used the assistant only within their assigned windows, and the two focal conditions used it with the same
intensity (Table~\ref{tab:design}): 448 queries (14.9 per team) under early-search access and 413 (13.8 per team)
under later-selection access. Teams with continuous access issued twice as many (30.0 per team) and, without any
instruction to do so, concentrated them in the search stage (438 in S1, 319 in S2, and 144 in S3).

\subsection{The Effect of Timing on Rated Creativity}\label{sec:main}

\begin{figure}[!htbp]
		\centerline{\includegraphics[width=0.85\textwidth]{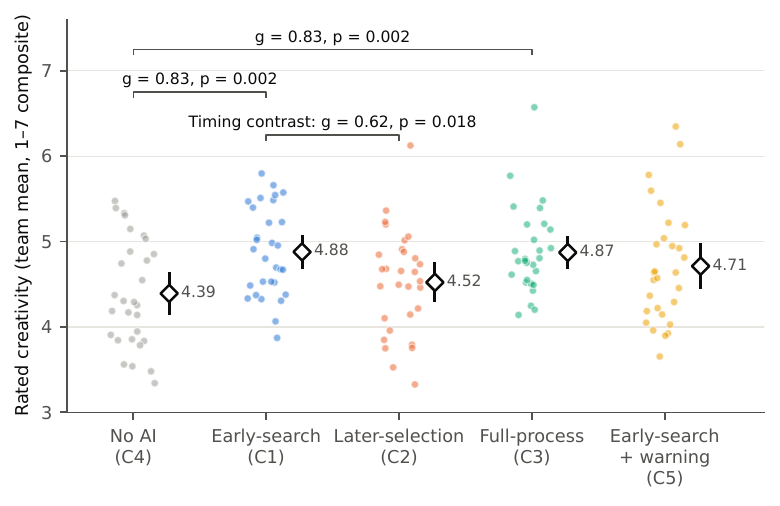}}
		\noindent\caption{{\bf Rated creativity by AI-access schedule.} Points are teams ($n=148$); diamonds are condition means with 95\% confidence intervals. Brackets report Hedges $g$ and two-sided Welch $p$ values.}
		\label{fig:main}
	\end{figure}

Access to the assistant raised rated creativity when it opened during search but not when it opened during selection
(Figure~\ref{fig:main} and Table~\ref{tab:main}). Submissions from teams with early-search access received a mean
rating of 4.88, compared with 4.52 for teams with later-selection access and 4.39 for teams without AI. The timing
contrast is 0.356 points (95\% CI $[0.064,0.648]$; $g=0.62$; Welch $p=0.018$), which supports
Hypothesis~\ref{hyp:timing}. Relative to the control, early-search access raised rated creativity by 0.486 points
($g=0.83$, $p=0.002$), which supports Hypothesis~\ref{hyp:access}, and continuous access raised it by 0.479 points
($g=0.83$, $p=0.002$). Later-selection access, with the same model for the same 15 minutes, produced a gain of 0.130
points that is statistically indistinguishable from zero ($g=0.21$, $p=0.43$). Moving the window from search to
selection thus forgoes about three-quarters of the gain from search-stage access.

\insertfloat{Figure~\ref{fig:main}}
\insertfloat{Table~\ref{tab:main}}

The timing contrast is stable across specifications: 0.361 (SE 0.153) with task fixed effects, 0.369 (SE 0.168) with
the ten pretreatment covariates, 0.361 (SE 0.148) at the rater-by-team level with rater fixed effects, and 0.481 (SE
0.171) at the proposal level, which weights teams by the number of proposals they submitted. Randomization inference
yields $p=0.021$, and the contrast survives the preregistered Benjamini--Hochberg correction across all eight condition
contrasts (adjusted $p=0.048$).

Continuous access performed as well as early-search access, and no better (difference $-0.007$; 95\% CI
$[-0.27,0.26]$). This null result is informative. If the benefit of early access arose because teams converged better
without an assistant during selection, continuous access should have resembled late access; it did not. Adding
selection-stage access to search-stage access had no detectable effect, whereas replacing search-stage access with
selection-stage access eliminated the gain. This is the pattern Proposition~\ref{prop:latesearch} predicts and Section~\ref{sec:mech} documents.

\subsection{Rated Dimensions}\label{sec:dims}

The timing gap appears on every rated dimension (Table~\ref{tab:dims}). Relative to later-selection access,
early-search access was rated higher by 0.37 points on elaboration, 0.36 on originality and on originality quality,
0.34 on appropriateness, and 0.32 on text quality. The contrasts in originality, appropriateness, and originality
quality are significant at the 5\% level, and those in elaboration ($p=0.061$) and text quality ($p=0.056$) at the 10\%
level; after a permutation-based Westfall--Young adjustment across the five dimensions, the contrasts in
appropriateness ($p=0.008$) and originality quality ($p=0.031$) remain significant. Relative to the control,
early-search access raised every dimension, most on elaboration (0.695 points) and appropriateness (0.530) and least on
originality (0.431). The timing gap is therefore as large in the originality and fit of the ideas as in their
elaboration. In terms of Proposition~\ref{prop:timing}, the data point to the screening premium and the dilution of
quality, rather than the dilution of development alone, as its source.

\insertfloat{Table~\ref{tab:dims}}

\subsection{The Idea Funnel}\label{sec:funnel}

The model locates the effect of timing at the screen: early access should enlarge the pool that the team screens, and
the team should prune it, whereas late access should leave the pool unchanged and bring options in after the screen.
Figure~\ref{fig:funnel} and Table~\ref{tab:funnel} show each team's funnel, reconstructed from the platform logs. The
pool at the gate is the set of distinct ideas in the members' last personal drafts of the search stage, where the
instructions asked teams to note their ideas. We measure the origin of the submission by matching each submitted
sentence with the team's search-stage drafts, chat, and assistant output; matched instead against other teams'
search-stage material for the same task, the rule matches fewer than 1\% of sentences.

\insertfloat{Figure~\ref{fig:funnel}}
\insertfloat{Table~\ref{tab:funnel}}

\paragraph{Early access widened the pool, and teams pruned it.} Early-search teams held 19.8 distinct ideas at the
gate, compared with 3.9 for teams without AI (difference 15.9, $p<0.001$, $g=1.23$). Most of these ideas were
AI-derived (18.6), and the teams' own drafted ideas fell from 3.9 to 1.3 ($p<0.001$): the assistant replaced part of
the team's own ideation in the written pool while multiplying the pool's size. Teams then pruned. About a third of the
pooled ideas (32\%) reappeared in the submission, the same share as without AI, and submissions contained 2.9 proposals. Warned and full-process teams show the same pattern with smaller pools (13.6 and 12.5 ideas). Under the
model's assumptions (ideas drawn from a common normal distribution whose spread equals the reliability-adjusted spread
of proposal ratings, and a screen that retains the two best ideas), a pool of 20 rather than 4 ideas implies a
screening premium of 0.46 points, close to the observed 0.49 (Online Appendix Table~\ref{tab:ec-funnel-cal}). Because
the calibration assumes perfect screening and AI ideas as good as the team's own, it checks orders of magnitude rather
than estimating the premium.

\paragraph{Late access replaced the screened pool.} Later-selection teams entered the gate with the same pool as teams
without AI (3.8 ideas compared with 3.9, $p=0.87$), as randomization implies, because neither had the assistant during
search. What happened next differed. Only 11\% of their pre-gate ideas reappeared in the submission, compared with 32\%
without AI ($p=0.005$), and only 5\% of their submitted text matched anything that existed before the gate, compared
with 22\% without AI ($p<0.001$) and 71\% under early-search access. The comparison with full-process teams holds the
window, the prompt, and the model fixed: full-process teams kept 25\% of their pre-gate ideas ($p=0.031$ for the
difference), and 32\% of their submitted text came from before the gate ($p<0.001$). Sentence-embedding similarity,
which recognizes paraphrase, yields the same pattern (7\% compared with 23\% without AI and 36\% under full-process
access), so the low share does not reflect the assistant rewording pre-gate ideas (Online Appendix~\ref{ec:funnel}).
Late access did not add options to a screened set; it replaced the set, as in the displacement case of
Remark~\ref{rem:displacement}.

\paragraph{What the replacement cost.} Remark~\ref{rem:displacement} implies that replacing the screened set forfeits
the team's own screening premium. We find no direct evidence of that loss. Proposals that originated before the gate
were rated no higher than proposals that originated after it (Online Appendix Table~\ref{tab:ec-funnel-prop}), and
late-access submissions were not rated below those of teams without AI (Section~\ref{sec:main}). Either teams screened
their own small pools imprecisely, or selection-stage assistance offset the loss. The quality cost of late access
therefore appears as the premium it forgoes, namely a pool five times larger to choose from, rather than as a measurable penalty for setting the team's own ideas aside. Two further cautions apply. Within conditions, neither pool size nor
the number of late generation requests predicts how much of the pool a team kept, so the funnel differences are
differences between randomly assigned schedules, not a causal chain traced within teams. And the pool is measured from
what teams wrote down, not from everything they discussed.

\subsection{Access History and the Use of the Assistant}\label{sec:mech}

Proposition~\ref{prop:latesearch} predicts that teams entering selection without AI-generated options use
selection-stage access to keep searching, whereas teams entering selection with such options use it to develop them.
Figure~\ref{fig:mech}(a) and Panel B of Table~\ref{tab:mech} test this prediction.

\insertfloat{Figure~\ref{fig:mech}}
\insertfloat{Table~\ref{tab:mech}}

During search, teams used the assistant in the same way regardless of condition: generation requests accounted for 55\%
of queries under early-search access, 52\% under the warning, and 58\% under full-process access. The conditions
diverge in the selection window. Teams with later-selection access devoted 30\% of their selection-stage queries to
generating new options and 40\% to developing a specified option. Teams with full-process access, using the same
assistant in the same window under the same system prompt, devoted 11\% to generation and 52\% to development. At the
team level, the generation share was 29.1\% compared with 11.0\% (difference 18.1 percentage points, $p<0.001$), and
teams issued 4.07 compared with 1.20 generation requests ($p<0.001$; $g=1.15$). Because the selection window was the
later-selection teams' first contact with the assistant, part of the difference could reflect orientation, such as
pasting the task prompt; setting aside each team's first three queries in the window leaves generation shares of
21.1\% and 8.9\% ($p=0.011$).

The two conditions differ only in whether the team also had search-stage (and review-stage) access, which was randomly
assigned, so the difference is a causal effect of access history on the use of an identical tool in an identical
window. The pattern is the one Proposition~\ref{prop:latesearch} predicts. Teams that had searched with the assistant
entered selection with larger pools and used the tool mostly to develop what they had; teams that had not devoted a
much larger share of the same window to further search, even though the instructions for this stage asked them to
select among their first-stage ideas and the assistant's prompt directed it toward refinement. The magnitudes are of
the same order as in the model's numerical example (4.8 compared with 2.1 late requests; Online
Appendix~\ref{ec:model}). Development nonetheless remained the most frequent request in both conditions.

\subsection{The Fate of Late Options}\label{sec:fate}

Lemma~\ref{lem:retention} implies that options obtained after the screen are kept without comparison and enter the
deliverable in the form in which they arrived, and Proposition~\ref{prop:composition} implies that they add breadth. We
test the first implication by tracing the assistant's output (Table~\ref{tab:fate}) and the second by counting
proposals. Because assistant responses were visible only to the member who asked, the team chat was the only place
where the team as a whole could compare an assistant option with its own; output that entered the submission without
ever appearing in the chat did not pass the team's screen.

\insertfloat{Table~\ref{tab:fate}}

\paragraph{Late output entered as delivered.} Output from the selection window entered submissions more often than
output from the search window: 23.7\% of later-selection teams' idea units, compared with 14.6\% of early-search teams'
units (unit-level difference 8.3 percentage points, $p=0.005$). It entered near-verbatim more often (15.8\% compared
with 8.7\%; 6.1 points, $p=0.014$) and more often without ever having been shared with teammates (9.6\% compared with
3.9\%; 5.1 points, $p=0.006$). The share of output shared in the chat did not differ (26\% compared with 28\%): late
output was not discussed less, but more of it went into the deliverable without discussion. Part of these differences
reflects what teams asked for, because selection-window requests more often asked for development and for text
intended for the submission. Holding the request type fixed halves the differences in adoption and near-verbatim use,
which are then no longer significant, whereas the difference in unshared adoption remains (3.3 points, $p=0.046$).
Full-process teams show the same pattern within team: 11.3\% of their selection-window output entered the submission
unshared, compared with 2.5\% of their search-window output ($p<0.001$).

The cleanest comparison holds the window and the prompt fixed. In the selection window, later-selection teams used the
assistant's output near-verbatim about twice as often as full-process teams (15.8\% compared with 8.2\%; 6.6 points,
$p=0.024$; 8.0 points, $p=0.009$, holding the request type fixed). Because access history was randomly assigned, this
difference indicates that teams that met the assistant only after the screen carried its output into the deliverable as
it arrived, whereas teams that had searched with it used it to rework what they already had. These differences hold at
other thresholds and with sentence-embedding similarity (Online Appendix Table~\ref{tab:ec-fate-rob}). One finer
implication of Lemma~\ref{lem:retention} is not clearly supported: the rate at which output entered the submission
unshared rose over the selection window only slightly faster than over the search window (difference in slopes 0.4
percentage points per minute, $p=0.09$; Online Appendix Table~\ref{tab:ec-fate-time}).

\begin{figure}[!htbp]
		\centerline{\includegraphics[width=\textwidth]{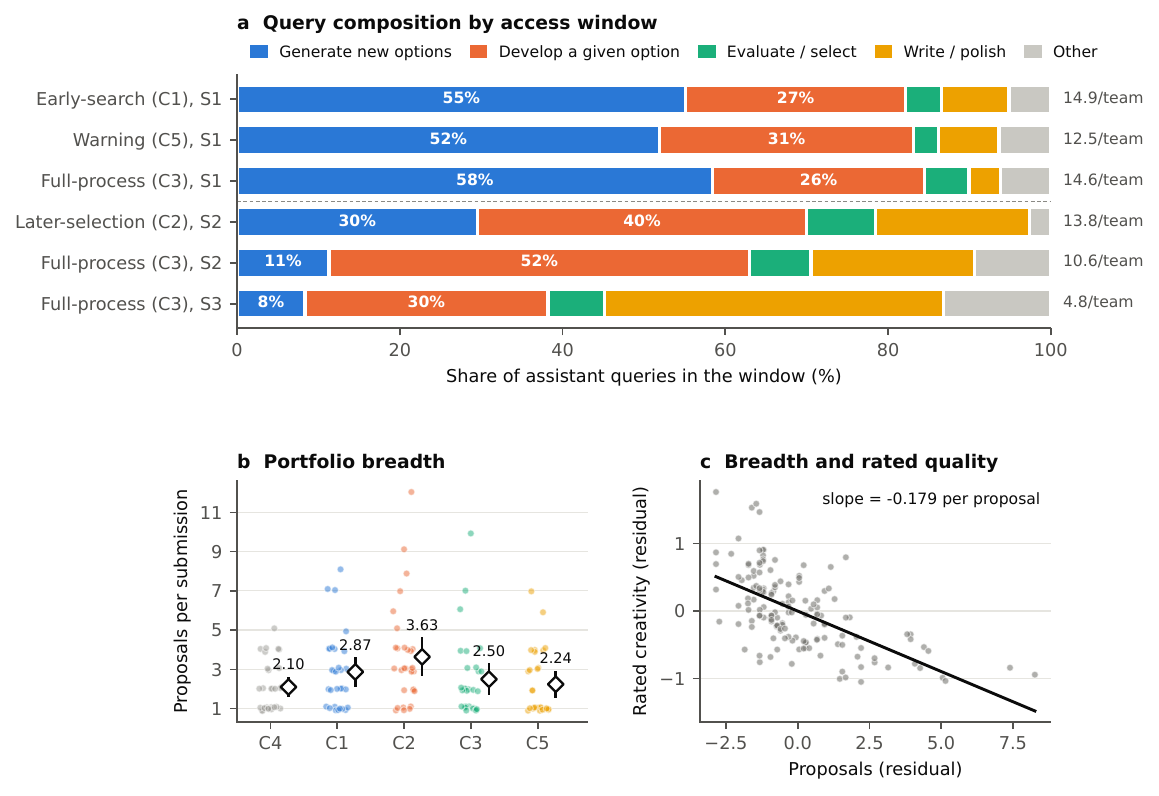}}
		\noindent\caption{{\bf How teams used access and what they submitted.} (a) Composition of the 2,124 assistant queries within each access window, coded blind to condition, stage, and team; numbers at right are queries per team in the window. (b) Rater-coded proposals per submission; points are teams, diamonds are means with 95\% confidence intervals. (c) Proposal count and rated creativity, both residualized on condition and task fixed effects; the line is the OLS fit.}
		\label{fig:mech}
	\end{figure}

\paragraph{More proposals, not better ones.} The submissions show the breadth that Proposition~\ref{prop:composition}
and Hypothesis~\ref{hyp:composition} predict (Figure~\ref{fig:mech}(b) and Table~\ref{tab:mech}, Panel A).
Later-selection access raised the number of proposals per submission by 1.53 relative to the control (SE 0.54,
$p=0.005$). This is an increase of 73\% over the control mean of 2.10 and twice the increase under early-search access (0.74,
$p=0.081$). The difference between the two windows, 0.78 proposals, has the predicted sign but is imprecisely estimated
($p=0.19$), and full-process teams, whose selection-stage access went mainly to development, submitted 2.50 proposals,
1.13 fewer than later-selection teams ($p=0.069$). The depth half of Hypothesis~\ref{hyp:composition} is supported only
against early access: early-search proposals were rated 0.37 points higher on elaboration ($p=0.061$), whereas
later-selection proposals were not less elaborated than control proposals (they were rated 0.33 points higher, $p=0.067$) and were about as long (15 more characters per proposal, $p=0.63$). Selection-stage assistance thus
appears to have offset the spreading of the development budget, as the model allows when $\beta$ approaches $rB/k$.
Across teams, each additional proposal was associated with a 0.179-point lower rating, conditional on condition and task
(Figure~\ref{fig:mech}(c)). Among teams with selection-stage access, however, the number of late generation requests
did not predict the number of proposals (Online Appendix~\ref{ec:coding}), so we interpret late search and late breadth
as parallel consequences of the schedule rather than as links in a measured chain. Nor did breadth reflect wider
search: the semantic dispersion of submissions rose by the same amount under both single-window conditions (0.033 and
0.032 relative to the control; difference $p=0.78$).

\paragraph{Breadth, not quality.} The cost of late access lies in what teams retained rather than in the assistant's
text as such. Within conditions, the share of a submission derived from assistant output, shared or not, is not
associated with lower ratings by either panel, whereas each additional proposal is associated with a human score 0.29
standard deviations lower (Online Appendix Table~\ref{tab:ec-unscreened}); and late-access submissions were no more
similar than early-access submissions to other teams' assistant output or submissions. How the portfolio is valued
matters accordingly, as Online Appendix Proposition~\ref{prop:aggregation} predicts (Online Appendix
Table~\ref{tab:portfolio}). The early-search advantage over later-selection access is 0.361 points in the mean
proposal, 0.221 points in the best proposal ($g=0.50$, $p=0.055$), and reverses in the total: later-selection
submissions summed to 2.06 points more ($g=-0.23$, $p=0.35$). Relative to no AI, later-selection access raised the best
proposal by 0.426 points and the total by 6.31 points (both $p=0.001$). Late access converts the assistant's options
into breadth rather than quality.

\subsection{Warnings}\label{sec:warning}

The standing warning reduced reliance on the assistant without changing rated quality, as Hypothesis~\ref{hyp:warning}
states. Warned teams issued 12.5 queries per team, against 14.9 without the warning ($p=0.09$). The share of submission
sentences that closely match the team's own assistant output, a measure defined for every assigned team and therefore a randomized comparison, fell from 65\% to 47\% (difference $-0.18$; 95\% CI $[-0.34,-0.03]$; $p=0.020$; Online
Appendix Table~\ref{tab:ec-unscreened}). Rated creativity was statistically indistinguishable from that of unwarned
teams ($-0.167$; 95\% CI $[-0.49,0.15]$; $p=0.30$), and so were the request profile (52\% compared with 55\%
generation) and the fate of the output (Table~\ref{tab:fate}). When options arrive during search, the team's own screen
already disciplines them, so a warning that reduces reliance has little left to improve (Online
Appendix~\ref{ec:process} reports the details).

\section{Robustness and Alternative Evaluators}\label{sec:robust}

\subsection{Robustness}\label{sec:robustness}

The timing contrast is positive, and its 95\% confidence interval excludes zero, in every specification in
Table~\ref{tab:robust}. It holds for the two industry experts alone (0.435) and for the two doctoral raters alone
(0.277), and it ranges from 0.298 to 0.420 when any one rater is dropped. It is essentially unchanged when we exclude
the two submissions with ambiguous proposal boundaries (0.360), a third submission on which raters disagreed about the
number of proposals (0.388), or the eight teams whose stages ran off schedule (0.380). It is also robust to a rank-based
test (Hodges--Lehmann shift 0.357, $p=0.029$), a team bootstrap (95\% CI $[0.070,0.639]$), randomization inference
($p=0.021$), and a crossed random-effects model of proposal-level ratings (0.390; 95\% CI $[0.109,0.672]$). The effects
of early-search and continuous access relative to the control are similarly robust, except that both are smaller and
not statistically significant when only the two doctoral raters are used.

\insertfloat{Table~\ref{tab:robust}}

Online Appendix Table~\ref{tab:mht} reports unadjusted, randomization-inference, false-discovery-rate, and family-wise
adjusted $p$ values for all eight condition contrasts. The effects of early-search and continuous access relative to the
control survive every adjustment, including Westfall--Young ($p=0.012$ for both). The timing contrast survives the
preregistered Benjamini--Hochberg procedure ($p=0.048$); under the most conservative family-wise procedures, which treat
all eight preregistered and exploratory contrasts as a single family, its adjusted $p$ values are 0.086
(Westfall--Young) and 0.107 (Holm). Task-specific estimates are positive in all three tasks (Online Appendix
Table~\ref{tab:ec-task}).

\subsection{Evaluators That Credit Breadth}\label{sec:breadth}

Corollary~\ref{cor:evaluator} predicts that crediting breadth narrows the timing gap. Adding a credit of $\gamma$ points
per proposal to each team's human score (Online Appendix Table~\ref{tab:ec-breadth}) reduces the early-search advantage
over later-selection access from 0.356 points at $\gamma=0$ to 0.279 at $\gamma=0.1$ ($p=0.019$) and 0.203 at
$\gamma=0.2$ ($p=0.076$), and the advantage vanishes at $\gamma^*=0.46$, the gap divided by the 0.78 additional proposals
submitted by later-selection teams. Breadth credit also changes the verdict on late access relative to no access: with
a credit of only 0.1 points per proposal, later-selection access outscores the control (0.283 points, $p=0.043$). A
conventional scoring rule that blends the number of proposals into originality and weights the three dimensions
equally yields the same pattern: the timing contrast is unchanged in standardized terms ($g=0.62$, $p=0.019$), but
later-selection access lies far above the control ($g=1.02$, $p<0.001$). An organization that rewards the number of
ideas will regard late access as beneficial; one that rewards the quality of what it commits to will not.

\subsection{Human Versus Language-Model Evaluators}\label{sec:llm}

We also scored every proposal with ten frontier language models from five model families, using the same rubric
(Table~\ref{tab:evaluators}). The model panel was internally consistent, its team scores correlate at 0.61 with those
of the human panel, and it agreed with the human panel on the gross effect of AI access, placing early-search,
later-selection, continuous, and warned access 0.58 to 1.41 standard deviations above the control. It did not,
however, distinguish early from late access. The timing contrast is 0.59 standard deviations of the team score under
human ratings ($p=0.019$) and $-0.08$ under the model panel ($p=0.77$). Excluding DeepSeek V3, the model closest to the
DeepSeek V3.2 assistant, leaves the contrast at $-0.10$, and excluding both DeepSeek models leaves it at $-0.08$. The
divergence appears on every dimension: the panel saw no difference between early- and late-access submissions in
originality, appropriateness, elaboration, or text quality, whereas the experts saw differences of 0.5 to 0.7 standard
deviations (Online Appendix Table~\ref{tab:ec-evaldecomp}).

\insertfloat{Table~\ref{tab:evaluators}}

Corollary~\ref{cor:evaluator} accounts for part of this divergence. On the original rating scales, human raters lowered
their score by 0.179 points for each additional proposal in a submission, whereas the model panel lowered it by 0.063
points, a difference of 0.116 (95\% CI $[0.082,0.150]$) that is essentially unchanged when one or both DeepSeek models
are excluded. Adding the proposal count to a model of the difference between the two panels' standardized scores
reduces the divergence on the timing contrast from 0.66 to 0.55 standard deviations; adding the share of AI-derived
sentences, a candidate channel for model self-preference \citep{Panickssery2024}, and rated text quality reduces it
further, to 0.46 (Online Appendix Table~\ref{tab:ec-evaldecomp}). Most of the divergence therefore reflects a difference
in how the two panels judge the same proposals: the model panel did not register the lower originality and fit that the
experts attributed to late-access proposals. Human ratings remain the appropriate primary outcome. They represent the expert judgment the task targeted, and the human penalty for breadth is estimated within conditions, where it cannot reflect differences between schedules. The comparison nonetheless shows what is at stake: an organization that evaluates
AI-assisted work with AI judges will see that assistance helps but not that its timing matters.

\section{Discussion}\label{sec:discussion}

 Generative AI changes what teams have to choose from, and the value of this change depends on when options arrive
 relative to the team's screen. The same model, available for the same time and used at similar rates, raised the
 quality of team output by more than half a standard deviation when teams had it during search and produced no
 detectable gain when they had it during selection.
 
 For managers, the access schedule is a design parameter, and an assistant left on by default sets it without anyone
 deciding. Teams should have the assistant while they build their option set and screen deliberately afterward.
 Continuous access did no harm in our experiment, because teams that had searched with the assistant later used it
 mainly for development. Where the assistant's first contribution comes late, teams can limit late use to development or
 require late output to be compared with what they have already chosen (Corollary~\ref{cor:gate}). Organizations should
 also decide whether they value depth or breadth before judging a deployment, since a small credit per proposal
 reverses the verdict on late access, and should validate AI evaluators against human judgment, since our
 language-model panel missed the cost of late timing. A warning against overreliance cut the assistant's share of
 submitted text by more than a quarter without changing quality, which suggests that process design is a stronger lever
 than exhortation. For research, the timing of AI contributions is a distinct treatment. Studies with continuous access estimate a mixture
 of early and late effects weighted by how participants schedule their own use, and studies that compare phases should
 report where the screen falls.
 
 Our design trades breadth of setting for clean identification. Participants were students working on a short,
 structured task, and the assistant was a single model. The model treats the team as a single decision maker that
 divides development effort equally, and our funnel and tracing measures see only what teams wrote down. Agreement among
 expert raters was modest, which is why we report every rater subset and an alternative panel of evaluators. The experiment also fixed when the screen fell, at minute 15 in every condition. It therefore identifies the effect of
 access before versus after a screen whose timing is set in advance, as it is in a stage-gate process. In this setting,
 the practical recommendation does not depend on why timing matters: teams should have the assistant during search.
 Which reference point governs the effect matters for theory and for settings in which the screen comes much earlier or
 later than it did here, and the design does not by itself separate the screen from elapsed time. One comparison
 already speaks against a purely clock-based account. Teams with late and with continuous access had the assistant
 during the same minutes and under the same prompt, yet teams with continuous access used it mainly for development and
 matched the creativity of early-access teams, whereas teams with late access used it to search again. How a team used
 the same minutes of access depended on what it held at the screen.
 
 The model points to tests that build directly on this study. Its two mechanisms rest on different reference points:
 the screening premium depends only on whether options arrive before the screen, whereas the retention of late options
 depends on how much time remains after it. Crossing the position of the access window
 relative to the screen with its position on the clock therefore separates the two accounts. The screen-relative
 account predicts a main effect of position relative to the screen and little effect of clock time; a clock-based
 account predicts the reverse. For the future study, such design can 
 enforce a screen on late output and remove the credit for quantity. Field settings with formal stage gates, longer projects with repeated screens, and
 other models would show how far the timing effect extends.

	\clearpage
	{\begin{singlespace}
			\small
			\bibliographystyle{apa}
			\bibliography{refs}
		\end{singlespace}
	}

	\clearpage

	\makeatletter
	\setlength{\@fptop}{0pt}
	\setlength{\@fpbot}{0pt plus 1fil}
	\makeatother
	\clearpage
	\clearpage
	\clearpage
	\begin{figure}[!htbp]
		\centerline{\includegraphics[width=\textwidth]{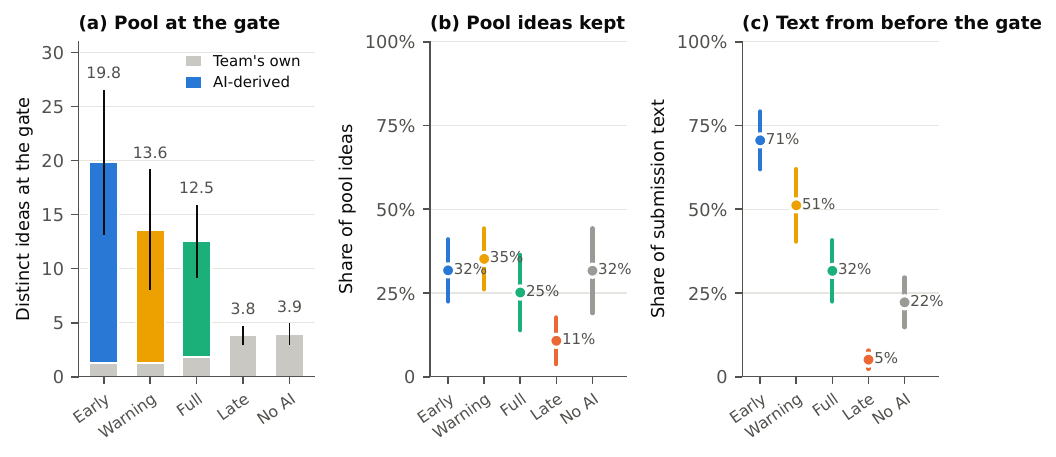}}
		\noindent\caption{{\bf The idea funnel.} (a) Distinct ideas in the members' last personal drafts of the search stage (the pool at the gate); gray segments are the team's own ideas and colored segments ideas derived from the assistant's search-stage output; bars are means and lines 95\% confidence intervals of the total. (b) Share of pool ideas that reappear in the final submission. (c) Share of submission sentences that match the team's search-stage drafts, chat, or assistant output. Early: early-search access (C1); Warning: early-search access with warning (C5); Full: full-process access (C3); Late: later-selection access (C2). Points are condition means with 95\% confidence intervals.}
		\label{fig:funnel}
	\end{figure}
	\clearpage

	\clearpage
	\begin{landscape}
\begin{table}[!htbp]
\small
\noindent\caption[Experimental conditions and observed assistant use]{{\bf Experimental conditions and observed assistant use.} The 40-minute task comprised a 15-minute divergent-search stage (S1), a 15-minute convergent-selection stage (S2), and a 10-minute final-review stage (S3). Assistant queries are logged requests to the integrated assistant (DeepSeek V3.2, temperature 0.1), after correction of one group-name typo in the log. No query was logged outside an assigned window. In C5 every assistant response was followed by a persistent statement that AI output is only an initial reference and should be discussed with teammates. One C5 team issued no query.\label{tab:design}}
\centering
\small \setlength{\tabcolsep}{3pt}
\begin{tabular}{llccccc}
\toprule
 & AI access & & & Assistant queries & Queries per & Teams with \\
Condition & window & Teams & Participants & (S1 / S2 / S3) & team (s.d.) & $\geq 1$ query \\
\midrule
Early-search access (C1) & Search (S1) & 30 & 90 & 448 / 0 / 0 & 14.9 (5.3) & 30 \\
Later-selection access (C2) & Selection (S2) & 30 & 90 & 0 / 413 / 0 & 13.8 (4.4) & 30 \\
Full-process access (C3) & S1--S3 & 30 & 90 & 438 / 319 / 144 & 30.0 (9.8) & 30 \\
Early-search + warning (C5) & Search (S1) & 29 & 87 & 362 / 0 / 0 & 12.5 (5.5) & 28 \\
No-AI control (C4) & None & 29 & 87 & 0 / 0 / 0 & -- & 0 \\
\midrule
Total & & 148 & 444 & 1248 / 732 / 144 & & 118 \\
\bottomrule
\end{tabular}
\end{table}
\end{landscape}
	\clearpage
	\begin{table}
\small
\noindent\caption[Effects of the AI-access schedule on rated creativity]{{\bf Effects of the AI-access schedule on rated creativity.} The dependent variable is the calibrated creativity composite from four blind expert raters, averaged over the proposals in a team's submission (columns 1--3), the rater's mean over a team's proposals (column 4), or the rating of each proposal (column 5). Coefficients are differences from the no-AI control (C4). Columns 1--3 report HC3 heteroskedasticity-robust standard errors; columns 4--5 report standard errors clustered by team (148 clusters). Column 5 weights teams by their number of proposals and therefore estimates a proposal-weighted rather than a team-weighted contrast. Team covariates (column 3) are the ten pretreatment measures in Table~\ref{tab:balance}, mean-centered. Randomization inference re-randomizes condition labels between the two compared arms within task strata (10,000 draws); the Westfall--Young step-down procedure re-randomizes all five labels within task strata (10,000 draws) and controls the family-wise error rate across all eight condition contrasts. All eight contrasts with Welch, RI, BH, Holm and WY $p$ values are reported in Online Appendix Table~\ref{tab:mht}. $^{*}p<0.10$, $^{**}p<0.05$, $^{***}p<0.01$ (two-sided).\label{tab:main}}
\centering
\small
\begin{tabular}{lccccc}
\toprule
 & (1) & (2) & (3) & (4) & (5) \\
\midrule
Early-search access (C1) & $0.486^{***}$ & $0.486^{***}$ & $0.441^{***}$ & $0.486^{***}$ & $0.488^{***}$ \\
 & $(0.154)$ & $(0.147)$ & $(0.168)$ & $(0.141)$ & $(0.163)$ \\[2pt]
Later-selection access (C2) & $0.130$ & $0.124$ & $0.073$ & $0.124$ & $0.007$ \\
 & $(0.165)$ & $(0.165)$ & $(0.195)$ & $(0.159)$ & $(0.182)$ \\[2pt]
Full-process access (C3) & $0.479^{***}$ & $0.473^{***}$ & $0.523^{***}$ & $0.473^{***}$ & $0.449^{***}$ \\
 & $(0.152)$ & $(0.148)$ & $(0.162)$ & $(0.142)$ & $(0.155)$ \\[2pt]
Early-search + warning (C5) & $0.319^{*}$ & $0.313^{*}$ & $0.280$ & $0.313^{*}$ & $0.204$ \\
 & $(0.177)$ & $(0.166)$ & $(0.178)$ & $(0.160)$ & $(0.167)$ \\[2pt]
\midrule
C1 $-$ C2 (timing contrast) & $0.356^{**}$ & $0.361^{**}$ & $0.369^{**}$ & $0.361^{**}$ & $0.481^{***}$ \\
 & $(0.148)$ & $(0.153)$ & $(0.168)$ & $(0.148)$ & $(0.171)$ \\[2pt]
C3 $-$ C1 & $-0.007$ & $-0.012$ & $0.081$ & $-0.012$ & $-0.039$ \\
 & $(0.134)$ & $(0.135)$ & $(0.146)$ & $(0.130)$ & $(0.144)$ \\[2pt]
C5 $-$ C1 & $-0.167$ & $-0.172$ & $-0.161$ & $-0.172$ & $-0.284^{*}$ \\
 & $(0.161)$ & $(0.155)$ & $(0.161)$ & $(0.149)$ & $(0.149)$ \\[2pt]
\midrule
\multicolumn{6}{l}{\textit{Inference for C1 $-$ C2 across the family of eight condition contrasts (column 1 sample)}} \\
\quad Randomization-inference $p$ & \multicolumn{5}{l}{$0.021$} \\
\quad Benjamini--Hochberg adjusted $p$ & \multicolumn{5}{l}{$0.048$} \\
\quad Westfall--Young adjusted $p$ & \multicolumn{5}{l}{$0.086$} \\
\midrule
Task fixed effects & No & Yes & Yes & Yes & Yes \\
Team covariates & No & No & Yes & No & No \\
Rater fixed effects & -- & -- & -- & Yes & Yes \\
Unit of observation & Team & Team & Team & Rater $\times$ team & Rater $\times$ proposal \\
Observations & 148 & 148 & 148 & 592 & 1{,}573 \\
Adjusted $R^2$ & 0.073 & 0.110 & 0.120 & 0.556 & 0.488 \\
Control (C4) mean & \multicolumn{5}{c}{4.392} \\
\bottomrule
\end{tabular}
\end{table}
	\clearpage
	\begin{table}
\small
\noindent\caption[Effects by rated dimension]{{\bf Effects by rated dimension.} OLS with task fixed effects and HC3 standard errors; each dependent variable is the team mean of a rated dimension (1--7 scale; originality also includes the capped feature-count score and naming bonus). Originality, appropriateness and elaboration enter the composite with weights 0.70, 0.15 and 0.15; originality quality is the main component of originality; text quality is rated separately and does not enter the composite. WY $p$ values adjust the C1--C2 contrast for the five dimensions by step-down max-$|t|$ over 10,000 stratified re-randomizations, using the same task-fixed-effects HC3 $t$ statistic as the table. $^{*}p<0.10$, $^{**}p<0.05$, $^{***}p<0.01$ (two-sided).\label{tab:dims}}
\centering
\small
\begin{tabular}{lccccc}
\toprule
 & (1) & (2) & (3) & (4) & (5) \\
 & Originality & Appropriateness & Elaboration & \shortstack{Originality\\quality} & \shortstack{Text\\quality} \\
\midrule
Early-search access (C1) & $0.431^{**}$ & $0.530^{***}$ & $0.695^{***}$ & $0.483^{***}$ & $0.502^{***}$ \\
 & $(0.172)$ & $(0.121)$ & $(0.157)$ & $(0.147)$ & $(0.130)$ \\[2pt]
Later-selection access (C2) & $0.068$ & $0.188$ & $0.326^{*}$ & $0.123$ & $0.183$ \\
 & $(0.192)$ & $(0.121)$ & $(0.178)$ & $(0.153)$ & $(0.135)$ \\[2pt]
Full-process access (C3) & $0.461^{***}$ & $0.428^{***}$ & $0.579^{***}$ & $0.442^{***}$ & $0.408^{***}$ \\
 & $(0.170)$ & $(0.126)$ & $(0.163)$ & $(0.143)$ & $(0.140)$ \\[2pt]
Early-search + warning (C5) & $0.332^{*}$ & $0.203$ & $0.338^{**}$ & $0.347^{**}$ & $0.456^{***}$ \\
 & $(0.188)$ & $(0.138)$ & $(0.166)$ & $(0.158)$ & $(0.120)$ \\[2pt]
\midrule
C1 $-$ C2 & $0.364^{**}$ & $0.342^{***}$ & $0.369^{*}$ & $0.360^{**}$ & $0.319^{*}$ \\
 & $(0.173)$ & $(0.111)$ & $(0.197)$ & $(0.142)$ & $(0.167)$ \\
\quad WY-adjusted $p$ (5 dimensions) & $0.076$ & $0.008$ & $0.079$ & $0.031$ & $0.079$ \\
\midrule
Control (C4) mean & 4.466 & 4.570 & 3.867 & 4.075 & 3.978 \\
Adjusted $R^2$ & 0.108 & 0.122 & 0.078 & 0.147 & 0.072 \\
Observations & 148 & 148 & 148 & 148 & 148 \\
\bottomrule
\end{tabular}
\end{table}
	\clearpage
	\begin{landscape}
\begin{table}[!htbp]
\small
\noindent\caption[The idea funnel: The pool at the gate, what survived, and where the submission came from]{{\bf The idea funnel: The pool at the gate, what survived, and where the submission came from.} The pool at the gate is the set of distinct ideas in the members' last personal drafts of the search stage, split at list markers and de-duplicated within team (TF--IDF cosine $\geq$ 0.5); an idea is AI-derived if it matches the team's search-stage assistant output. Pool ideas in submission are pool ideas that reappear in the final submission. Submission text from before the gate is the share of submission sentences that match the team's search-stage drafts, chat or assistant output (TF--IDF cosine $\geq$ 0.35); matched against other teams' search-stage material in the same task, the same rule matches 0.8\% of sentences. Online Appendix~\ref{ec:funnel} reports alternative de-duplication rules, ideas that appeared in any search-stage draft, and paraphrase-robust sentence-embedding versions of both measures. Proposals submitted are rater-coded.\label{tab:funnel}}
\centering
\small \setlength{\tabcolsep}{3pt}
\begin{adjustbox}{max width=\textwidth}
\begin{tabular}{lcccccc}
\toprule
 & \multicolumn{3}{c}{Pool at the gate (distinct ideas)} & Pool ideas in & Proposals & Submission text \\
\cmidrule(lr){2-4}
 & All & Team's own & AI-derived & submission & submitted & from before the gate \\
\midrule
\multicolumn{7}{l}{\textit{Panel A. Condition means}} \\[2pt]
Early-search (C1) & 19.8 & 1.3 & 18.6 & 5.1 & 2.87 & 71\% \\
Early-search + warning (C5) & 13.6 & 1.3 & 12.3 & 3.7 & 2.24 & 51\% \\
Full-process (C3) & 12.5 & 1.8 & 10.7 & 2.7 & 2.50 & 32\% \\
Later-selection (C2) & 3.8 & 3.8 & 0.0 & 0.4 & 3.63 & 5\% \\
No-AI control (C4) & 3.9 & 3.9 & 0.0 & 1.2 & 2.10 & 22\% \\
\midrule
\multicolumn{7}{l}{\textit{Panel B. Differences (Welch tests; $p$ in parentheses)}} \\[2pt]
Early-search $-$ no AI & $+15.9$ ($<$0.001) & $-2.7$ ($<$0.001) & $+18.6$ ($<$0.001) & $+3.9$ ($<$0.001) & $+0.76$ (0.077) & $+48$ pp ($<$0.001) \\
Early-search $-$ later-selection & $+16.0$ ($<$0.001) & $-2.6$ ($<$0.001) & $+18.6$ ($<$0.001) & $+4.7$ ($<$0.001) & $-0.77$ (0.198) & $+65$ pp ($<$0.001) \\
Later-selection $-$ no AI & $-0.1$ (0.870) & $-0.1$ (0.870) & -- & $-0.8$ (0.007) & $+1.53$ (0.006) & $-17$ pp ($<$0.001) \\
\bottomrule
\end{tabular}
\end{adjustbox}
\end{table}
\end{landscape}
	\clearpage
	\begin{landscape}
\begin{table}[!htbp]
\footnotesize
\noindent\caption[Submission composition, team process, and the use of assistant access]{{\bf Submission composition, team process, and the use of assistant access.} Panel A: OLS with task fixed effects and HC3 standard errors ($n=148$ teams); all outcomes are post-treatment. Proposals per submission is the rater-coded number of distinct proposals; characters per proposal is computed within team and then averaged; semantic dispersion is the mean pairwise TF--IDF cosine distance among paragraph-level idea units of the submission. Panel B: each of the 2,124 assistant queries was coded by a large language model applying a written codebook, blind to condition, stage and team, into one of five request types (Generate new options; Develop a specified option; Evaluate or select; Write or polish for submission; Other). Shares pool queries within the cell. A second, independent blind coding of 200 random queries agreed with the codes in 87.0\% of cases (Cohen's $\kappa=0.82$; $\kappa=0.89$ for generate versus other). In the selection window, later-selection teams devoted a team-average 29.1\% of queries to generating new options against 11.0\% for full-process teams (difference 18.1 percentage points, $p<0.001$), and issued 4.07 versus 1.20 generation queries per team ($p<0.001$; Welch $t$-tests, 30 teams each). Because the selection window was later-selection teams' first contact with the assistant, we also drop each team's first three queries in the window: the shares become 21.1\% and 8.9\% ($p=0.011$). $^{*}p<0.10$, $^{**}p<0.05$, $^{***}p<0.01$ (two-sided).\label{tab:mech}}
\centering
\footnotesize \setlength{\tabcolsep}{2.5pt}
\begin{tabular}{lcccccc}
\toprule
 & (1) & (2) & (3) & (4) & (5) & (6) \\
 & \shortstack{Proposals per\\submission} & \shortstack{Rated\\elaboration} & \shortstack{Submission\\length (chars)} & \shortstack{Characters per\\proposal} & \shortstack{Stage-1 chat\\(1,000 chars)} & \shortstack{Semantic\\dispersion} \\
\midrule
\multicolumn{7}{l}{\textit{Panel A. Submission and process measures}} \\[2pt]
Early-search access (C1) & $0.74^{*}$ & $0.70^{***}$ & $220.5^{***}$ & $28.6$ & $1.02^{***}$ & $0.033^{***}$ \\
 & $(0.43)$ & $(0.16)$ & $(72.7)$ & $(27.9)$ & $(0.23)$ & $(0.008)$ \\[2pt]
Later-selection access (C2) & $1.53^{***}$ & $0.33^{*}$ & $310.9^{***}$ & $15.2$ & $-0.02$ & $0.032^{***}$ \\
 & $(0.54)$ & $(0.18)$ & $(77.4)$ & $(31.7)$ & $(0.08)$ & $(0.009)$ \\[2pt]
Full-process access (C3) & $0.39$ & $0.58^{***}$ & $262.9^{**}$ & $120.2$ & $1.04^{***}$ & $0.037^{***}$ \\
 & $(0.46)$ & $(0.16)$ & $(107.8)$ & $(77.6)$ & $(0.19)$ & $(0.008)$ \\[2pt]
Early-search + warning (C5) & $0.12$ & $0.34^{**}$ & $129.6^{*}$ & $97.5^{*}$ & $1.14^{***}$ & $0.030^{***}$ \\
 & $(0.40)$ & $(0.17)$ & $(71.0)$ & $(58.4)$ & $(0.25)$ & $(0.009)$ \\[2pt]
C1 $-$ C2 & $-0.78$ & $0.37^{*}$ & $-90.5$ & $13.4$ & $1.04^{***}$ & $0.002$ \\
 & $(0.60)$ & $(0.20)$ & $(93.3)$ & $(36.0)$ & $(0.23)$ & $(0.006)$ \\[2pt]
C2 $-$ C3 & $1.13^{*}$ & $-0.25$ & $48.0$ & $-105.0$ & $-1.06^{***}$ & $-0.005$ \\
 & $(0.62)$ & $(0.20)$ & $(122.9)$ & $(81.7)$ & $(0.19)$ & $(0.006)$ \\[2pt]
Control (C4) mean & 2.10 & 3.87 & 442.2 & 243.9 & 0.60 & 0.916 \\
\midrule
\multicolumn{7}{l}{\textit{Panel B. Composition of assistant queries within the access window}} \\[2pt]
Condition, stage & Queries/team & Generate & Develop & Evaluate & Write & Other \\
\cmidrule(lr){1-7}
C1, S1 & 14.9 & 55.1\% & 27.0\% & 4.5\% & 8.3\% & 5.1\% \\
C5, S1 & 12.5 & 51.9\% & 31.2\% & 3.0\% & 7.5\% & 6.4\% \\
C3, S1 & 14.6 & 58.4\% & 26.0\% & 5.5\% & 3.9\% & 6.2\% \\
C2, S2 & 13.8 & 29.5\% & 40.4\% & 8.5\% & 18.9\% & 2.7\% \\
C3, S2 & 10.6 & 11.3\% & 51.7\% & 7.5\% & 20.1\% & 9.4\% \\
C3, S3 & 4.8 & 8.3\% & 29.9\% & 6.9\% & 41.7\% & 13.2\% \\
\bottomrule
\end{tabular}
\end{table}
\end{landscape}

	\clearpage
	\begin{landscape}
\begin{table}[!htbp]
\small
\noindent\caption[The fate of assistant output: Adoption, verbatim use, and sharing with teammates]{{\bf The fate of assistant output: Adoption, verbatim use, and sharing with teammates.} Each of the 2,124 assistant responses is split into idea units at list items and headings (8,603 units). A unit \emph{entered the submission} if one of its sentences has a TF--IDF (character 2--3-gram) cosine similarity of at least 0.35 with a sentence of the team's final submission, \emph{near-verbatim} if the similarity is at least 0.70, and \emph{shared} if it reaches the same threshold with a team-chat message posted after the response arrived. Assistant responses were visible only to the member who asked, so a unit that entered the submission without being shared never reached the team's common channel. Applied to other teams' submissions in the same task, the threshold matches 0.4\% of unit--submission pairs (Online Appendix Table~\ref{tab:ec-fate-cal}). Panel B regresses unit indicators on the window contrast with task fixed effects (and, in the second row of each pair, fixed effects for the five request types); standard errors are clustered by team. Online Appendix~\ref{ec:fate} reports alternative thresholds, sentence-embedding similarity, generation requests only, and timing within the window. $^{*}p<0.10$, $^{**}p<0.05$, $^{***}p<0.01$ (two-sided).\label{tab:fate}}
\centering
\small \setlength{\tabcolsep}{3pt}
\begin{tabular}{lccccc}
\toprule
 & & Entered the & Entered near- & Shared in & Entered without \\
 & Idea units & submission & verbatim & team chat & being shared \\
\midrule
\multicolumn{6}{l}{\textit{Panel A. Share of the assistant's idea units, by access window (team means)}} \\[2pt]
Early-search (C1), search window & 1,850 & 14.6\% & 8.7\% & 27.7\% & 3.9\% \\
Warning (C5), search window & 1,492 & 13.6\% & 8.1\% & 27.6\% & 4.1\% \\
Full-process (C3), search window & 1,798 & 6.1\% & 3.2\% & 22.2\% & 2.5\% \\
Later-selection (C2), selection window & 1,617 & 23.7\% & 15.8\% & 26.3\% & 9.6\% \\
Full-process (C3), selection window & 1,216 & 19.7\% & 8.2\% & 21.5\% & 11.3\% \\
Full-process (C3), review window & 630 & 50.1\% & 32.6\% & 23.9\% & 29.3\% \\
\midrule
\multicolumn{6}{l}{\textit{Panel B. Differences in percentage points (unit level, task fixed effects, team-clustered s.e.)}} \\[2pt]
Late vs.\ early options: C2 (S2) $-$ C1 (S1) &  & $+8.3$*** & $+6.1$** & $-0.8$ & $+5.1$*** \\
 &  & $(2.9)$ & $(2.4)$ & $(3.5)$ & $(1.8)$ \\[1pt]
\quad holding request type fixed &  & $+4.2$ & $+3.4$ & $-3.2$ & $+3.3$** \\
 &  & $(2.7)$ & $(2.3)$ & $(3.7)$ & $(1.6)$ \\[1pt]
Same window: C2 (S2) $-$ C3 (S2) &  & $+2.7$ & $+6.6$** & $+4.2$ & $-2.0$ \\
 &  & $(3.6)$ & $(2.8)$ & $(3.4)$ & $(2.5)$ \\[1pt]
\quad holding request type fixed &  & $+4.2$ & $+8.0$*** & $+4.0$ & $-1.2$ \\
 &  & $(3.6)$ & $(2.9)$ & $(3.4)$ & $(2.6)$ \\[1pt]
\bottomrule
\end{tabular}
\end{table}
\end{landscape}
	\clearpage
	\begin{landscape}
\begin{table}[!htbp]
\small
\noindent\caption[Robustness of the three primary contrasts]{{\bf Robustness of the three primary contrasts.} Entries are differences in creativity score points with 95\% confidence intervals (team-level unless stated). $^{\dagger}$ The interval includes zero. Raters are labeled A--D in order of increasing raw mean (severity), as in Online Appendix Table~\ref{tab:ec-raters}; A and D are the two industry experts and B and C the two doctoral raters. Leave-one-out and subset scores average the remaining raters. Severity calibration does not change team scores, because every rater scored every submission and additive severity differences cancel in team means. Randomization inference re-randomizes condition labels between the two compared arms within tasks (10,000 draws). The crossed random-effects row is a REML linear mixed model of proposal-level ratings with task fixed effects and random intercepts for teams, proposals and raters (Online Appendix~\ref{ec:robust}). The proposal-level row weights each team by its number of proposals. The off-schedule row drops teams whose logged search or selection stage deviated from 15 minutes by more than 3 minutes (most often because a member joined late).\label{tab:robust}}
\centering
\small \setlength{\tabcolsep}{3pt}
\begin{tabular}{lccc}
\toprule
 & Early-search vs. & Early-search vs. & Full-process vs. \\
Specification & later-selection (C1--C2) & no AI (C1--C4) & no AI (C3--C4) \\
\midrule
\multicolumn{4}{l}{\textit{Baseline}} \\
\quad Welch $t$-test, calibrated score & 0.356 [0.064, 0.648] & 0.486 [0.182, 0.790] & 0.479 [0.179, 0.779] \\
\multicolumn{4}{l}{\textit{Specification}} \\
\quad OLS, task FE & 0.361 [0.061, 0.662] & 0.486 [0.198, 0.773] & 0.473 [0.184, 0.763] \\
\quad OLS, task FE + team covariates & 0.369 [0.040, 0.698] & 0.441 [0.112, 0.771] & 0.523 [0.206, 0.839] \\
\multicolumn{4}{l}{\textit{Measurement}} \\
\quad Industry-expert raters only (A, D) & 0.435 [0.056, 0.814] & 0.778 [0.342, 1.213] & 0.700 [0.264, 1.136] \\
\quad Doctoral raters only (B, C) & 0.277 [0.005, 0.550] & 0.195 [$-$0.066, 0.456]$^{\dagger}$ & 0.258 [$-$0.012, 0.528]$^{\dagger}$ \\
\quad Leave out rater A & 0.298 [0.016, 0.580] & 0.358 [0.073, 0.643] & 0.459 [0.190, 0.727] \\
\quad Leave out rater B & 0.420 [0.105, 0.736] & 0.623 [0.277, 0.970] & 0.572 [0.238, 0.907] \\
\quad Leave out rater C & 0.344 [0.020, 0.669] & 0.544 [0.200, 0.887] & 0.533 [0.184, 0.882] \\
\quad Leave out rater D & 0.362 [0.066, 0.657] & 0.420 [0.129, 0.711] & 0.352 [0.053, 0.651] \\
\multicolumn{4}{l}{\textit{Level of analysis}} \\
\quad Rater $\times$ team, rater + task FE & 0.361 [0.069, 0.653] & 0.486 [0.207, 0.765] & 0.473 [0.193, 0.754] \\
\quad Proposal level (proposal-weighted) & 0.481 [0.144, 0.819] & 0.488 [0.165, 0.811] & 0.449 [0.143, 0.755] \\
\quad Crossed random effects & 0.390 [0.109, 0.672] & 0.486 [0.190, 0.781] & 0.472 [0.173, 0.772] \\
\multicolumn{4}{l}{\textit{Sample}} \\
\quad Excl.\ 2 ambiguously unitized teams & 0.360 [0.062, 0.659] & 0.486 [0.182, 0.790] & 0.479 [0.179, 0.779] \\
\quad Excl.\ 3 inconsistently unitized teams & 0.388 [0.091, 0.685] & 0.514 [0.212, 0.817] & 0.479 [0.179, 0.779] \\
\quad Excl.\ 8 teams with off-schedule stages & 0.380 [0.075, 0.685] & 0.473 [0.155, 0.792] & 0.453 [0.145, 0.761] \\
\multicolumn{4}{l}{\textit{Inference}} \\
\quad Mann--Whitney (Hodges--Lehmann shift) & 0.357 ($p=0.029$) & 0.497 ($p=0.003$) & 0.489 ($p=0.007$) \\
\quad Team bootstrap, percentile CI & 0.356 [0.070, 0.639] & 0.486 [0.201, 0.777] & 0.479 [0.197, 0.766] \\
\quad Randomization inference & $p=0.021$ & $p=0.002$ & $p=0.002$ \\
\bottomrule
\end{tabular}
\end{table}
\end{landscape}
	\clearpage
	\begin{landscape}
\begin{table}[!htbp]
\small
\noindent\caption[Human versus language-model evaluators]{{\bf Human versus language-model evaluators.} Panel A: OLS with task fixed effects and HC3 standard errors ($n=148$). Each evaluator's team score is standardized across teams (mean 0, s.d.\ 1); AI scores are standardized within model, averaged over models, and then re-standardized. Column 4 is the difference between the AI and human $z$-scores, so it measures relative rather than absolute leniency. The proposal-count row is from a separate model that adds the rater-coded proposal count to the condition and task indicators. Panel B stacks the raw human and pooled AI team scores (296 observations) and regresses them on proposal count, an AI indicator and their interaction, with condition and task fixed effects that are allowed to differ between the two scales, and team-clustered standard errors; brackets are 95\% confidence intervals. The ten models are Claude Sonnet 4.6, Claude Opus 4.6 and 4.7, GPT-5.5 and GPT-5.5 Pro, Gemini 3.1 Flash and Pro, DeepSeek V3 and V4 Pro, and Kimi K2.6. The experimental assistant ran on DeepSeek V3.2; column 3 drops DeepSeek V3, and the last row of Panel B drops both DeepSeek models. $^{*}p<0.10$, $^{**}p<0.05$, $^{***}p<0.01$ (two-sided).\label{tab:evaluators}}
\centering
\small \setlength{\tabcolsep}{2.5pt}
\begin{tabular}{lcccc}
\toprule
 & (1) & (2) & (3) & (4) \\
 & \shortstack{Human panel\\(4 raters)} & \shortstack{AI panel\\(10 models)} & \shortstack{AI panel\\excl. DeepSeek V3} & \shortstack{AI minus\\human} \\
\midrule
\multicolumn{5}{l}{\textit{Panel A. Condition effects by evaluator (standardized team scores)}} \\[2pt]
Early-search access (C1) & $0.788^{***}$ & $0.703^{***}$ & $0.715^{***}$ & $-0.085$ \\
 & $(0.238)$ & $(0.230)$ & $(0.228)$ & $(0.230)$ \\[2pt]
Later-selection access (C2) & $0.202$ & $0.779^{***}$ & $0.814^{***}$ & $0.577^{**}$ \\
 & $(0.267)$ & $(0.255)$ & $(0.248)$ & $(0.225)$ \\[2pt]
Full-process access (C3) & $0.768^{***}$ & $1.411^{***}$ & $1.429^{***}$ & $0.642^{***}$ \\
 & $(0.240)$ & $(0.224)$ & $(0.221)$ & $(0.218)$ \\[2pt]
Early-search + warning (C5) & $0.509^{*}$ & $0.577^{**}$ & $0.586^{**}$ & $0.068$ \\
 & $(0.269)$ & $(0.235)$ & $(0.231)$ & $(0.232)$ \\[2pt]
C1 $-$ C2 & $0.586^{**}$ & $-0.075$ & $-0.100$ & $-0.662^{***}$ \\
 & $(0.249)$ & $(0.256)$ & $(0.254)$ & $(0.222)$ \\[2pt]
Proposal count (added to model) & $-0.290^{***}$ & $-0.144^{***}$ & $-0.136^{***}$ & $0.146^{***}$ \\
 & $(0.034)$ & $(0.041)$ & $(0.040)$ & $(0.035)$ \\
\midrule
\multicolumn{5}{l}{\textit{Panel B. Penalty per additional proposal on the original rating scale}} \\[2pt]
 & Human panel & AI panel & Difference (AI $-$ human) & \\
All 10 models & $-$0.179 [$-$0.217, $-$0.141] & $-$0.063 [$-$0.097, $-$0.030] & 0.116 [0.082, 0.150]$^{***}$ & \\
Excluding DeepSeek V3 & $-$0.179 [$-$0.217, $-$0.141] & $-$0.063 [$-$0.097, $-$0.028] & 0.116 [0.082, 0.151]$^{***}$ & \\
Excluding both DeepSeek models & $-$0.179 [$-$0.217, $-$0.141] & $-$0.059 [$-$0.094, $-$0.024] & 0.120 [0.085, 0.155]$^{***}$ & \\
\bottomrule
\end{tabular}
\end{table}
\end{landscape}
	\clearpage

	\setlist{noitemsep}  %

	\pagenumbering{arabic}		
	\renewcommand*{\thepage}{Appendix-\arabic{page}}

	\counterwithin*{table}{section}
	\counterwithin*{figure}{section}
	\setcounter{section}{0}
	\setcounter{table}{0}
	\setcounter{figure}{0}
	\renewcommand\thesection{\Alph{section}}
	\renewcommand{\thetable}{\Alph{section}.\arabic{table}}
	\renewcommand{\thefigure}{\Alph{section}.\arabic{figure}}

	\title{
		{\bf \Large Online Appendix for \vspace{0.25cm}\\``Too Late to Prune: Generative AI, the Screen, and Team Creative Decisions'' }
	}

	\author{Difang Huang \and Shumin Yu \and Fadong Chen \and Weihua Zhou}

\date{October 10, 2026}

	\maketitle
	\singlespacing
	\thispagestyle{plain}
	\noindent This Online Appendix follows the order of the main text. Appendix~\ref{ec:model} contains the proofs and
	additional results for the model; Appendices~\ref{ec:materials}--\ref{ec:measurement} document the experiment and
	its measures; Appendix~\ref{ec:inference} supports the main results; Appendices~\ref{ec:funnel}--\ref{ec:process}
	support the mechanism analyses; Appendix~\ref{ec:evaluators} covers the alternative evaluators; and
	Appendix~\ref{ec:mods} reports an exploratory moderator screen. Condition codes: C1, early-search access; C2,
	later-selection access; C3, full-process access; C4, no-AI control; C5, early-search access with warning.

	\vspace{0.3cm}

\clearpage
\singlespacing
\section{Model}\label{ec:model}

\subsection{How the Portfolio Is Valued: Mean, Best, and Total}\label{ia:aggregation}

The mean is not the only way to value a portfolio. An organization that runs an innovation tournament may care about
its best idea \citep{Girotra2010,TerwieschUlrich2009}, and one that pays per idea may care about the total. Because late
access trades depth for breadth, the three aggregations rank the schedules differently.

\begin{proposition}[Aggregation]\label{prop:aggregation}
Let $\rho=1$, so that all late ideas are retained ($r=a$), and let $\beta\leq aB/k$.
(i) The advantage of early access in the \emph{best} proposal equals the development term alone,
$h(B/k)-h\bigl((B+\beta)/(k+a)\bigr)$; it is nonnegative and strictly smaller than the advantage in the mean.
(ii) The \emph{total} delivered quality is higher under late access whenever $a\mu\geq k\,[M_k(n+a)-M_k(n)]$.
\end{proposition}

Proposition~\ref{prop:aggregation} predicts a specific ordering: the timing advantage should be largest in the mean,
smaller but positive in the best proposal, and negative in the sum. Part (ii) adds levels and therefore depends on where
the quality scale starts; on a 1--7 rating scale, the mean is large relative to the screening premium, and the condition
holds. The ordering also holds when only some late ideas are retained ($\rho<1$), provided that development help is
small relative to the number of late ideas retained, $\beta<\rho aB/k$ (Appendix~\ref{ia:numerical}).
Table~\ref{tab:portfolio} in Appendix~\ref{ec:inference} reports the corresponding evidence. The ordering is a
prediction of the model rather than a choice of measure: the preregistered outcome is the mean, which corresponds to the
per-proposal judgment raters were asked to make and to what a team commits to the next stage, whereas organizations
that screen ideas downstream may value the best idea or the breadth of the funnel.

\subsection{Warnings}\label{ia:warningmodel}

\begin{remark}[Warnings]\label{rem:warning}
A warning against overreliance lowers the share $\alpha<1$ of assistant ideas that teams take seriously during search,
so mean quality becomes $M_k(n+\alpha a)+h(B/k)$. Because these ideas must still pass the team's screen, the warning
affects quality only through $M_k(n+a)-M_k(n+\alpha a)$. When the gains from a larger pool diminish, this loss is less
than proportional to the reduction in supply, whereas the assistant's share of the ideas the team works with falls in
proportion to it. A warning should therefore lower reliance on assistant output proportionally more than it lowers
rated quality.
\end{remark}

Remark~\ref{rem:warning} underlies Hypothesis~\ref{hyp:warning}; Appendix~\ref{ia:warning} reports the evidence.

\subsection{Proofs}\label{app:proofs}
Throughout, $\theta_{(1:N)}\geq\theta_{(2:N)}\geq\dots$ denote order statistics of $N$ i.i.d.\ draws from $F$, and
$M_k(N)=\E\bigl[\frac1k\sum_{i=1}^k\theta_{(i:N)}\bigr]$.

\paragraph{Preliminaries.} (a) $M_k(N+1)\geq M_k(N)$ with strict inequality for continuous $F$: adding a draw weakly
raises each of the top-$k$ order statistics and strictly raises their sum with positive probability. (b) For $N>k$,
$M_k(N)>\mu$, since the top $k$ of $N$ draws have a higher expected mean than $k$ draws chosen at random, whose mean
is $\mu$. (c) For $h$ strictly concave with $h(0)=0$, the function $\phi(K)=K\,h(B/K)$ is strictly increasing in $K>0$,
because $\phi'(K)=h(B/K)-(B/K)h'(B/K)>0$ by strict concavity and $h(0)=0$.

\paragraph{Proof of Lemma~\ref{lem:retention}.} Keeping without comparison yields $\E[\Delta(\theta,R)]=\Delta(\mu,R)$,
because $\Delta$ is linear in $\theta$, and discarding yields 0; (i) follows from $\Delta(\mu,R)\geq0$. Comparing yields
$\E[\max\{\Delta(\theta,R),0\}]-c(R)$. Since $-\Delta(\theta,R)=(\hat\theta(R)-\theta)/(K+1)$, comparing is preferred to
keeping without comparison if and only if $\E[\max\{-\Delta,0\}]=b(R)\geq c(R)$. The cost satisfies
$c'(R)=[h'(R/K)-h'((R-\tau)/K)]/K<0$ by strict concavity. For the benefit, write
$\hat\theta(R)=\bar Q-(K+1)[\,g(R)+\gamma_T]$ with $g(R)=h(R/(K+1))-h(R/K)$. With $y=R/(K+1)<x=R/K$,
$g'(R)=[\,y\,h'(y)-x\,h'(x)]/R\leq0$ because $e\,h'(e)$ is nondecreasing, so $\hat\theta$ is nondecreasing in $R$ and so is
$b(R)$, since $\E[(z-\theta)^+]$ is increasing in $z$. Hence $b-c$ is strictly increasing in $R$ on $R>\tau$; letting
$R^*=\inf\{R>\tau: b(R)\geq c(R)\}$ (with $R^*=\infty$ if the set is empty) gives (ii). For $R<R^*$ the option is kept
for sure; for $R\geq R^*$ it is compared and kept if and only if $\theta\geq\hat\theta(R)$, which gives (iii), with
$F(\hat\theta)>0$ when $F$ has full support. $\square$

\paragraph{Proof of Proposition~\ref{prop:timing}.} (i) $\bar Q_E-\bar Q_N=M_k(n+a)-M_k(n)>0$ by (a). (ii) Writing the
selection part of $\bar Q_L$ as $M_k(n)-\frac{r}{k+r}(M_k(n)-\mu)$ gives the stated decomposition. The first term is
positive by (a), the second by (b). The third is nonnegative if and only if $B/k\geq(B+\beta)/(k+r)$, that is,
$\beta\leq rB/k$, because $h$ is increasing. (iii) $\bar Q_L-\bar Q_N=-\frac{r}{k+r}(M_k(n)-\mu)
+h\bigl(\frac{B+\beta}{k+r}\bigr)-h(B/k)$. The first term is negative by (b). The second is nonpositive if
$\beta\leq rB/k$, in which case $\bar Q_L<\bar Q_N$; it is positive only if $\beta>rB/k$. $\square$

\paragraph{Proof of Proposition~\ref{prop:composition}.} $K_L=k+r>k$ for $r\geq1$. $e_L<e_E$ if and only if
$(B+\beta)/(k+r)<B/k$, i.e., $\beta<rB/k$. $\square$

\paragraph{Proof of Proposition~\ref{prop:aggregation}.} (i) With $\rho=1$, all $a$ late ideas are retained, so the best
idea in the late portfolio is $\max\{\theta_{(1:n)},\text{late draws}\}$, which has the distribution of $\theta_{(1:n+a)}$,
the best idea in the early portfolio. Each proposal in a portfolio receives the same development, so the best delivered
quality is $\theta_{(1:n+a)}+h(e)$ in both cases and the expected difference is $h(B/k)-h((B+\beta)/(k+a))\geq0$. The
mean difference equals this term plus the two strictly positive selection terms of Proposition~\ref{prop:timing}(ii), so
it is strictly larger. (ii) Total quality is $T_E=k\,M_k(n+a)+k\,h(B/k)$ and $T_L=k\,M_k(n)+a\mu+(k+a)\,h\bigl(\frac{B+\beta}{k+a}\bigr)$.
Then $T_L-T_E=a\mu-k[M_k(n+a)-M_k(n)]+[(k+a)h(\frac{B+\beta}{k+a})-k\,h(\frac{B}{k})]$. The last bracket is at least
$\phi(k+a)-\phi(k)>0$ by (c) and monotonicity of $h$. Hence $T_L>T_E$ whenever $a\mu\geq k[M_k(n+a)-M_k(n)]$. $\square$

\paragraph{Numerical analysis for $\rho<1$.} We simulated the model with $\theta\sim\mathcal N(4.3,0.6^2)$, $n=a=6$, $k=2$,
$B=1$, $h(e)=0.8\sqrt e$, $\beta\in\{0,0.3,0.6\}$ and $\rho\in\{0.6,1\}$ (100,000 replications per cell; Table~\ref{tab:ec-grid}). In every cell the early-minus-late gap is largest in the mean (0.74--0.97), positive and smaller in the
best proposal (0.20--0.29), and negative in the total. A wider grid of 432 parameter combinations finds the same
ordering in all 297 cells with $\beta<\rho aB/k$ (Appendix~\ref{ia:numerical}).

\paragraph{Remark~\ref{rem:displacement}.} With full displacement the late portfolio holds $r$ unscreened draws, whose
expected mean is $\mu$, and the budget $B+\beta$ is shared among $r$ proposals, so $\bar Q_L=\mu+h((B+\beta)/r)$.
Subtracting from $\bar Q_E=M_k(n+a)+h(B/k)$ gives the stated gap; writing $M_k(n+a)-\mu=[M_k(n+a)-M_k(n)]+[M_k(n)-\mu]$
shows that it includes the team's own screening premium. $\square$

\paragraph{Proof of Corollary~\ref{cor:gate}.} Under an enforced comparison, a late option is kept if and only if
$\Delta(\theta,R)\geq0$, that is, $\theta\geq\hat\theta(R)$ (Lemma~\ref{lem:retention}). The retained late options are
therefore draws from $F$ truncated below at $\hat\theta(R)$, with mean $\E[\theta\mid\theta\geq\hat\theta(R)]>\mu$ for
$F$ with full support, and their number $r'$ is a thinning of $r$. Replacing $\mu$ by this conditional mean in the
decomposition of Proposition~\ref{prop:timing}(ii) gives the stated term, which is nonpositive when
$\E[\theta\mid\theta\geq\hat\theta(R)]\geq M_k(n)$. Each comparison costs $c(R)$ in development, as in the lemma. $\square$

\paragraph{Proof of Proposition~\ref{prop:latesearch}.} (i) Since late options are kept without comparison, requesting and keeping one more option of
quality $\theta$ changes $W$ by $(\theta-\bar Q)/(K+1)+h((R-\delta)/(K+1))-h(R/K)+\gamma_T$ at a cost $\eta$. Because this
change is linear in $\theta$, its expectation net of $\eta$ is $G(\bar Q,K,R)$, whose derivative with respect to $\bar Q$ is
$-1/(K+1)<0$. (ii) Fix a sequence of late draws $\theta_1,\theta_2,\dots$. After $j$ requests the team
holds $K_j=K+j$ proposals, has $R_j=R-j\delta$ units of time, and has mean quality
$\bar Q_j=(K\bar Q_0+\sum_{i\leq j}\theta_i)/(K+j)$. Neither $K_j$ nor $R_j$ depends on $\bar Q_0$, and $\bar Q_j$ is
increasing in $\bar Q_0$. By (i), $G_j(\bar Q_0)\equiv G(\bar Q_j,K_j,R_j)$ is therefore decreasing in $\bar Q_0$ for every
$j$. The team stops at $m(\bar Q_0)=\min\{j\geq0:G_j(\bar Q_0)\leq0\text{ or }R_j\leq\delta\}$. If $\bar Q_0'>\bar Q_0$,
then $G_j(\bar Q_0')\leq G_j(\bar Q_0)$ for every $j$, so every index at which the team stops from $\bar Q_0$ is also an
index at which it would stop from $\bar Q_0'$, and $m(\bar Q_0')\leq m(\bar Q_0)$. (iii) Couple the schedules by giving
the team the same $n$ own ideas and adding the assistant's $a$ ideas under early access. Adding draws to a pool weakly
raises each of its $k$ largest order statistics, so the mean of the $k$ best of $n+a$ draws is at least the mean of the
$k$ best of the $n$ own draws, realization by realization. By (ii), the number of late requests under early access is
then at most the number without it, for every realization of the late draws. Retained late options equal the requests
(Lemma~\ref{lem:retention}, $R<R^*$) and inherit the ordering; with a fixed number of selection-stage queries, the
remaining queries go to development. $\square$

\paragraph{Proof of Corollary~\ref{cor:evaluator}.} $S_{\gamma,E}-S_{\gamma,L}=\bar Q_E-\bar Q_L+\gamma(K_E-K_L)=\bar Q_E-\bar Q_L-\gamma r$,
which is linear and strictly decreasing in $\gamma$ with root $\gamma^*$. $\square$

\paragraph{Remark~\ref{rem:warning}.} For $k=1$, $M_1(N+1)-M_1(N)=\int F(x)^N(1-F(x))\,dx$, which is decreasing in $N$:
the marginal value of an additional idea in the pool diminishes. With diminishing increments, $M_k(n+a)-M_k(n+\alpha a)
<(1-\alpha)\,[M_k(n+a)-M_k(n)]$, so reducing the effective supply from $a$ to $\alpha a$ lowers the screening premium by
less than the proportion $1-\alpha$. For $k>1$ we verified diminishing increments numerically for the normal
distribution used in the numerical analysis ($k=1,2,3$, $N\leq15$).

\subsection{Numerical Analysis}\label{ia:numerical}

Proposition~\ref{prop:aggregation} is stated for $\rho=1$. Table~\ref{tab:ec-grid} shows that the ordering it predicts
holds across retention probabilities and levels of selection-stage development help: early access dominates late access
in mean quality, the early-access advantage in the best proposal is positive but smaller, and late access yields a
larger total. A wider grid ($n\in\{4,6,8\}$, $a\in\{2,4,6\}$, $k\in\{1,2,3\}$, $\beta\in\{0,0.3,0.6,1\}$,
$\rho\in\{0.1,0.3,0.6,1\}$; 40,000 replications per cell) finds the
ordering in all 297 cells with $\beta<\rho aB/k$; it can fail when development help is large relative to the late ideas
retained. In every cell of Table~\ref{tab:ec-grid}, late access also falls below no access, as
Proposition~\ref{prop:timing}(iii) implies when $\beta\leq rB/k$. In the data, later-selection access is not
distinguishable from the control (0.13 points, 95\% CI $[-0.19,0.46]$); the model accommodates a positive difference
only if selection-stage development help is large.

Table~\ref{tab:ec-retention} illustrates Lemma~\ref{lem:retention}, which derives the retention of late options from
the trade-off between comparing a new option and developing the retained ones. With the parameters of
Table~\ref{tab:ec-grid} and a comparison time of one-tenth of the selection-stage budget, a team compares a new option
when at least about half of the stage remains ($R^*=0.48$) and keeps it without comparison otherwise. The benefit of
comparing rises and its cost falls with the time remaining, so the decision switches only once. The threshold rises
with the time a comparison takes.

The same numerical example illustrates Proposition~\ref{prop:latesearch}. A team enters selection holding the two best
of its ideas and 0.4 units of development time, so late options are kept without comparison. Each request costs
$\eta=0.05$ in attention and $\delta=0.05$ units of development time, and the perceived credit per proposal is
$\gamma_T=0.35$. Teams without early access, whose two best ideas come from six draws, make 4.8 late requests on
average (60\% make at least one); teams with early access, whose two best ideas come from twelve draws, make 2.1
(27\%). For common draws, the number of requests under early access never exceeds the number without it (100,000
replications). The observed numbers of selection-window generation requests, 4.07 for
later-selection teams and 1.20 for full-process teams, are of the same order.

\clearpage
\begin{table}[!htbp]
\noindent\caption[Numerical analysis of the search--selection model]{{\bf Numerical analysis of the search--selection model.} Parameters: $\theta\sim\mathcal N(4.3,0.6^2)$, $n=a=6$, $k=2$, $B=1$, $h(e)=0.8\sqrt e$; 100,000 replications per cell. In every cell early access dominates late access in mean quality, the gap in the best proposal is positive and smaller, and the gap in total quality is negative, as Proposition~\ref{prop:aggregation} states for $\rho=1$.\label{tab:ec-grid}}
\centering
\begin{tabular}{ccccccccccc}
\toprule
 & & \multicolumn{3}{c}{Mean quality} & \multicolumn{3}{c}{Early $-$ late gap} & & \multicolumn{2}{c}{Effort per proposal} \\
\cmidrule(lr){3-5}\cmidrule(lr){6-8}\cmidrule(lr){10-11}
$\beta$ & $\rho$ & None & Early & Late & Mean & Best & Total & $K_L$ & Early & Late \\
\midrule
0.0 & 1.0 & 5.439 & 5.691 & 4.726 & 0.965 & 0.284 & $-$26.43 & 8.00 & 0.500 & 0.125 \\
0.0 & 0.6 & 5.437 & 5.690 & 4.860 & 0.831 & 0.291 & $-$15.71 & 5.60 & 0.500 & 0.189 \\
0.3 & 1.0 & 5.438 & 5.689 & 4.767 & 0.922 & 0.242 & $-$26.76 & 8.00 & 0.500 & 0.163 \\
0.3 & 0.6 & 5.437 & 5.688 & 4.911 & 0.778 & 0.240 & $-$15.98 & 5.60 & 0.500 & 0.245 \\
0.6 & 1.0 & 5.438 & 5.689 & 4.801 & 0.888 & 0.208 & $-$27.03 & 8.00 & 0.500 & 0.200 \\
0.6 & 0.6 & 5.438 & 5.689 & 4.952 & 0.738 & 0.201 & $-$16.24 & 5.60 & 0.500 & 0.302 \\
\bottomrule
\end{tabular}
\end{table}

\clearpage
\begin{landscape}
\begin{table}[!htbp]
\noindent\caption[Optimal retention of a late option (Lemma~\ref{lem:retention})]{{\bf Optimal retention of a late option (Lemma~\ref{lem:retention}).} Parameters as in Table~\ref{tab:ec-grid}: $\theta\sim\mathcal N(4.3,0.6^2)$, $h(e)=0.8\sqrt e$, a portfolio of $K=2$ proposals holding the two best of six own ideas, perceived quantity credit $\gamma_T=0.35$, and comparison time $\tau=0.10$ (in units of the selection-stage development budget, $B=1$). The benefit of comparing is $\mathbb E[(\hat\theta-\theta)^+]/(K+1)$; the cost is $h(R/K)-h((R-\tau)/K)$. $\Delta(\mu,R)\ge0$ means that an average option is worth keeping if it is not compared. Retention is 1 when the option is kept without comparison and $1-F(\hat\theta)$ when it is compared.\label{tab:ec-retention}}
\centering
\begin{tabular}{ccccccc}
\toprule
Remaining time $R$ & Cutoff $\hat\theta$ & $\Delta(\mu,R)$ & Benefit of comparing & Cost of comparing & Decision & Retention $\rho(R)$ \\
\midrule
0.15 & 3.94 & 0.119 & 0.0340 & 0.0926 & Keep without comparing & 1.00 \\
0.20 & 3.96 & 0.113 & 0.0357 & 0.0741 & Keep without comparing & 1.00 \\
0.30 & 3.99 & 0.103 & 0.0388 & 0.0569 & Keep without comparing & 1.00 \\
0.40 & 4.02 & 0.094 & 0.0415 & 0.0479 & Keep without comparing & 1.00 \\
0.50 & 4.04 & 0.086 & 0.0441 & 0.0422 & Compare & 0.67 \\
0.60 & 4.06 & 0.079 & 0.0464 & 0.0382 & Compare & 0.65 \\
0.80 & 4.10 & 0.067 & 0.0509 & 0.0327 & Compare & 0.63 \\
1.00 & 4.13 & 0.056 & 0.0551 & 0.0290 & Compare & 0.61 \\
\midrule
\multicolumn{7}{l}{Threshold $R^*$ for comparison time $\tau$: $\tau=0.05$: 0.19; $\tau=0.10$: 0.48; $\tau=0.15$: 0.78; $\tau=0.20$: 1.09.} \\
\bottomrule
\end{tabular}
\end{table}
\end{landscape}

\clearpage
\section{Experimental Materials}\label{ec:materials}

\subsection{Tasks}
Each team received one of three object-combination tasks. The instruction read (translated from Chinese): ``Combine the
concepts of [object A] and [object B] to design one or more products or services that are both creative and
feasible.'' The pairs were (i) a backpack and a pet accessory, (ii) a refrigerator and a calendar, and (iii) shoes and a
map.

\subsection{Platform and Stages}
After logging in with their student ID, participants read condition-specific instructions, completed an interface
tutorial, and began the session after a three-second countdown that started only when all three members had finished
the tutorial. The interface had three panels: a team chat (left), the assistant panel in conditions with access
(center), and a personal draft editor that saved automatically every minute (right). The instructions (translated)
asked teams to work through three stages. \emph{Stage 1, divergence (15 minutes):} ``Brainstorm, drawing on the
features of the two objects, and generate as many preliminary product concepts as possible; you may note them briefly
in the document panel. Do not make final decisions at this stage; focus first on quantity.'' \emph{Stage 2, focusing
(15 minutes):} ``Select the best ideas from the first stage (in any number) and develop them. Settle the details of the
final proposals, name them, describe their functions (within 250 characters), and record them in the document panel.''
\emph{Stage 3, final review (10 minutes):} ``Review your proposals as a professional product designer would: check
feasibility and proofread, make sure the final proposals are fully recorded in the document panel, and choose one member
to submit.'' Each proposal was to be named and described in no more than 250 Chinese characters.

A tool notice told teams with access in which stage or stages they could use the assistant and that they could decide
for themselves how to use it; its wording was identical across conditions except for the window it named. Teams in the
no-AI condition were asked to work without online resources or external AI tools. The laboratory disabled access to
other applications and the internet; workstations were separated by partitions, and offline communication was
prohibited.

\subsection{Assistant Configuration}
The assistant was DeepSeek V3.2, called through the provider's API with the model identifier \texttt{deepseek-chat} at
temperature 0.1. Each team had its own conversation instance. Queries and responses were private to the member who
issued them; teammates saw assistant content only if it was shared in the chat. At each query, the model received the
team's 100 most recent chat messages and the following system prompt (translated):

\begin{quote}\small
You are a Product Design AI Assistant assisting a three-person team with a product innovation design task.
\emph{Task background.} Current task: \{task name\}. Task requirements: \{task description\}.
\emph{Your responsibilities.} Assist the team in creative ideation, solution optimization and problem analysis. Answer
product-design questions.
\emph{Current stage and behavioral guidelines.} Current stage: \{stage\}. If it is Stage 1, the divergence stage:
respond quickly to team requests and provide as many possible ideas as possible. If it is Stage 2, the convergence stage:
assist the team in refining product features, optimizing descriptions, evaluating market potential and screening for the
best solutions. If it is Stage 3, the final review stage: check for logical flaws, propose improvements and polish final
descriptions; conduct a final check to ensure the proposal is sound and error-free.
\emph{Response requirements.} Keep responses concise and professional, within 250 characters. Use a clear structure; lists or
bullet points are recommended.
\end{quote}

Outside a team's assigned window, the assistant panel was grayed out with an explanatory message; the no-AI condition
had no assistant panel. In the warning condition, every assistant response was followed by the statement, in bold
italics: ``AI serves only as an initial reference point and its outputs may not be accurate or comprehensive. We highly
encourage you to discuss these ideas further with your team members.''

\subsection{Incentives and Scoring Instructions}
Participants received 45 RMB for participation. Teams were told that product-design experts would score submissions
blind, primarily on creativity and feasibility and, to a lesser extent, on quantity. Members of the ten highest-scoring
teams received a bonus of up to 50 RMB each, and the strongest proposals could be forwarded to industry partners as
conceptual references. Bonus rankings used the same blinded scores as the analysis.

\clearpage
\section{Sample, Randomization, and Balance}\label{ec:sample}

The preregistered sample was 150 teams (30 per condition, ten per condition-by-task cell). Two teams did not submit
study data because of technical failures and were excluded before rating, leaving 148 teams and 444 participants. No
retained team was excluded. Table~\ref{tab:ec-alloc} shows the realized allocation.

Table~\ref{tab:balance} reports team-level balance on the ten pretreatment covariates, which Section~\ref{sec:design}
of the main text summarizes. Table~\ref{tab:ec-balance} reports participant-level balance. No continuous covariate
differs across conditions ($p\geq0.12$; largest standardized deviation 0.22). The categorical covariates are balanced on
gender ($p=0.33$) and study level ($p=0.09$), and discipline shows a borderline imbalance ($p=0.050$, Cram\'er's
$V=0.13$). Discipline does not enter any of our hypotheses and is not among the ten team covariates used in
Table~\ref{tab:main}, column 3, of the main text; we report the imbalance and do not adjust for it.

The platform advanced all three members of a team through the stages on a common clock that started when the last
member finished the tutorial. Measured from the first logged event of each stage, the search and selection stages
lasted 14.6 and 15.3 minutes for the median team. In eight teams, one of these stages deviated from 15 minutes by more
than three minutes, in most cases because a member joined late; excluding these teams leaves the results unchanged
(Table~\ref{tab:robust} of the main text).

\clearpage
\begin{table}[!htbp]
\noindent\caption[Realized allocation of teams to conditions and tasks]{{\bf Realized allocation of teams to conditions and tasks.} The allocation sequence contained ten slots for each of the 15 condition-by-task cells. Two recruited teams did not submit study data because of technical failures and were excluded before rating. Realized cell sizes range from 9 to 11; in C1 one team more than planned completed the backpack-and-pet task and one fewer the shoes-and-map task.\label{tab:ec-alloc}}
\centering
\begin{tabular}{lccccc}
\toprule
Condition & Backpack + pet & Refrigerator + calendar & Shoes + map & Teams & Participants \\
\midrule
C1 & 11 & 10 & 9 & 30 & 90 \\
C2 & 10 & 10 & 10 & 30 & 90 \\
C3 & 10 & 10 & 10 & 30 & 90 \\
C4 & 10 & 9 & 10 & 29 & 87 \\
C5 & 10 & 10 & 9 & 29 & 87 \\
\midrule
Total & 51 & 49 & 48 & 148 & 444 \\
\bottomrule
\end{tabular}
\end{table}
\clearpage
\begin{table}[!htbp]
\noindent\caption[Pretreatment balance across conditions (team means)]{{\bf Pretreatment balance across conditions (team means).} Entries are means of team-level averages of the registration survey ($n=148$ teams; 444 participants). Ranges in parentheses refer to the summed scale (a single item for AI-use frequency). The $p$ column reports a one-way ANOVA across the five conditions. Joint tests regress the treatment indicator on all ten covariates and task fixed effects within each pair of conditions (linear probability model, HC3 covariance) and test that all covariate coefficients are zero. The omnibus test sums the ten ANOVA $F$ statistics and compares the sum with 2,000 re-randomizations of condition labels within task strata.\label{tab:balance}}
\centering
\begin{tabular}{lcccccc}
\toprule
Covariate & C1 & C2 & C3 & C5 & C4 & $p$ \\
\midrule
Age (years) & 21.41 & 21.50 & 21.46 & 21.60 & 21.97 & 0.783 \\
Share female & 0.711 & 0.744 & 0.800 & 0.667 & 0.759 & 0.339 \\
Generative-AI use frequency (1--5) & 4.12 & 4.17 & 4.09 & 4.14 & 4.08 & 0.901 \\
AI acceptance (7--35) & 25.14 & 24.98 & 25.02 & 25.41 & 24.86 & 0.751 \\
Need for cognition (18--90) & 58.04 & 58.07 & 56.83 & 57.52 & 56.07 & 0.654 \\
Openness (2--10) & 6.88 & 6.94 & 6.61 & 6.91 & 6.57 & 0.349 \\
Conscientiousness (2--10) & 6.84 & 6.92 & 6.59 & 6.90 & 6.72 & 0.586 \\
Extraversion (2--10) & 5.87 & 5.99 & 5.47 & 5.70 & 5.86 & 0.318 \\
Agreeableness (2--10) & 7.66 & 7.53 & 7.80 & 7.89 & 7.80 & 0.402 \\
Emotional stability (2--10) & 6.29 & 5.99 & 5.73 & 6.25 & 6.15 & 0.114 \\
\midrule
\multicolumn{7}{l}{\textit{Joint orthogonality tests (all ten covariates)}} \\
\quad C1 vs. C4: $\chi^2(10)=11.15$ & \multicolumn{5}{l}{} & 0.346 \\
\quad C2 vs. C4: $\chi^2(10)=8.81$ & \multicolumn{5}{l}{} & 0.550 \\
\quad C3 vs. C4: $\chi^2(10)=22.02$ & \multicolumn{5}{l}{} & 0.015 \\
\quad C5 vs. C4: $\chi^2(10)=6.81$ & \multicolumn{5}{l}{} & 0.743 \\
\quad C1 vs. C2: $\chi^2(10)=5.27$ & \multicolumn{5}{l}{} & 0.872 \\
\quad Omnibus permutation test, all five conditions & \multicolumn{5}{l}{} & 0.657 \\
\bottomrule
\end{tabular}
\end{table}
\clearpage
\begin{table}[!htbp]
\noindent\caption[Participant-level pretreatment balance]{{\bf Participant-level pretreatment balance.} $n=444$ participants (age missing for two). Continuous variables: condition means, one-way ANOVA $p$, and the largest absolute standardized deviation of a condition mean from the sample mean. Categorical variables: shares, $\chi^2$ test $p$, and Cram\'er's $V$. The Big Five dimensions are two-item sums; need for cognition and AI acceptance are summed scales.\label{tab:ec-balance}}
\centering
\setlength{\tabcolsep}{4.5pt}
\begin{tabular}{lccccccc}
\toprule
Covariate & C1 & C2 & C3 & C5 & C4 & $p$ & Balance statistic \\
\midrule
Age (years) & 21.41 & 21.51 & 21.46 & 21.60 & 21.97 & 0.649 & 0.144 \\
Generative-AI use frequency (1--5) & 4.12 & 4.17 & 4.09 & 4.14 & 4.08 & 0.915 & 0.070 \\
AI acceptance (7--35) & 25.14 & 24.98 & 25.02 & 25.41 & 24.86 & 0.793 & 0.110 \\
Need for cognition (18--90) & 58.04 & 58.07 & 56.83 & 57.52 & 56.07 & 0.658 & 0.121 \\
Openness (2--10) & 6.88 & 6.94 & 6.61 & 6.91 & 6.57 & 0.306 & 0.139 \\
Conscientiousness (2--10) & 6.84 & 6.92 & 6.59 & 6.90 & 6.72 & 0.585 & 0.133 \\
Extraversion (2--10) & 5.87 & 5.99 & 5.47 & 5.70 & 5.86 & 0.320 & 0.176 \\
Agreeableness (2--10) & 7.66 & 7.53 & 7.80 & 7.89 & 7.80 & 0.336 & 0.162 \\
Emotional stability (2--10) & 6.29 & 5.99 & 5.73 & 6.25 & 6.15 & 0.115 & 0.220 \\
Trust in AI (1--5) & 3.38 & 3.20 & 3.38 & 3.38 & 3.28 & 0.543 & 0.140 \\
AI optimism (1--5) & 3.96 & 3.99 & 3.92 & 4.10 & 3.93 & 0.383 & 0.184 \\
Unease about AI (1--5) & 3.19 & 3.12 & 3.36 & 3.38 & 3.29 & 0.353 & 0.146 \\
Female (share) & 0.711 & 0.744 & 0.800 & 0.667 & 0.759 & 0.330 & $V=0.102$ \\
Study level & & & & & & 0.087 & $V=0.125$ \\
\quad Doctoral & 0.089 & 0.144 & 0.122 & 0.115 & 0.149 &  &  \\
\quad Master's & 0.211 & 0.189 & 0.289 & 0.333 & 0.368 &  &  \\
\quad Undergraduate & 0.700 & 0.667 & 0.589 & 0.552 & 0.483 &  &  \\
Discipline & & & & & & 0.050 & $V=0.133$ \\
\quad Agricultural sciences & 0.089 & 0.044 & 0.111 & 0.080 & 0.115 &  &  \\
\quad Engineering and technology & 0.233 & 0.322 & 0.144 & 0.207 & 0.184 &  &  \\
\quad Humanities and arts & 0.156 & 0.089 & 0.167 & 0.161 & 0.161 &  &  \\
\quad Medicine and health & 0.022 & 0.111 & 0.133 & 0.080 & 0.172 &  &  \\
\quad Natural sciences & 0.122 & 0.056 & 0.100 & 0.092 & 0.023 &  &  \\
\quad Social sciences & 0.378 & 0.378 & 0.344 & 0.379 & 0.345 &  &  \\
\bottomrule
\end{tabular}
\end{table}

\clearpage
\section{Measurement}\label{ec:measurement}

\subsection{Creativity: Rubric and Composite}
Submissions were grouped by task and numbered within task. Four raters scored every proposal in every submission
while blind to condition and team. Each rater completed one task before moving to the next and read all proposals in a
task before scoring any of them. The rubric was:
\begin{itemize}
\item \emph{Originality quality} (1--7): how novel and surprising the proposal is, beyond obvious combinations of the
two objects.
\item \emph{Originality quantity}: the number of distinct valid features, deduplicated within the submission, converted
to points (1 feature $=2$, 2 $=3$, 3 $=5$, 4 $=6$, $\geq5=7$).
\item \emph{Naming bonus} (0, 0.5, or 1) for a creative product name.
\item \emph{Appropriateness} (1--7): usefulness, feasibility, and fit with a plausible user need.
\item \emph{Elaboration} (1--7): how detailed and well developed the proposal is. Because of the length limit, longer
text was not rewarded; concise descriptions that make the functions understandable suffice.
\item \emph{Text quality} (1--7, supplementary): organization, readability, and coherence.
\end{itemize}
The composite is Originality $=0.90\times$quality $+\,0.10\times$quantity score $+$ naming bonus, and Creativity
$=0.70\times$Originality $+\,0.15\times$Appropriateness $+\,0.15\times$Elaboration. Team scores average each rater's
proposal scores within team and then average across raters.

\subsection{Unitization}
Rating templates segmented the 148 submissions into 394 proposal units before scoring. In two submissions that
numbered sections ``1'', ``2'' and ``3'' without labeling them as separate proposals, some raters treated the sections
as separate proposals and added rows, giving a union of 398 human-rated units; in a third submission, one rater
treated a multi-proposal document as a single proposal. Because the primary outcome averages within rater and team, we
retain each rater's own unitization. Excluding these submissions does not change the results (Table~\ref{tab:robust}
of the main text).

\subsection{Rater Severity and Agreement}
Raters differed in severity: mean composite scores ranged from 3.05 (rater A) to 5.36 (rater D). A least-squares model
with proposal and rater effects gives severity estimates of $-1.37$, $-0.24$, 0.67, and 0.94 for raters A--D. Because
every rater scored every submission, additive severity differences cancel in team means, and calibration leaves team
scores unchanged.

Table~\ref{tab:ec-rel} reports agreement. The ICC for the mean of the four raters is 0.44 for the composite at the
proposal level and 0.46 at the team level; the corresponding single-rater ICCs are 0.17 and 0.18. Agreement is
therefore modest. Krippendorff's $\alpha$, which counts differences in rater means as disagreement, is close to zero
because severity differs so much across raters.

The inter-rater correlations of team scores, 0.35--0.55 (Table~\ref{tab:ec-raters}), show that the raters ranked
submissions similarly but far from identically. Rating noise unrelated to the randomly assigned condition widens
confidence intervals but does not bias condition contrasts. Table~\ref{tab:ec-raters} also reports each rater's own
estimates: all four raters score early-search submissions above later-selection submissions ($g=0.24$--$0.72$), and
rater B's scores show little difference between AI and no-AI conditions.

\subsection{Coding of Assistant Requests}\label{ec:coding}
All 2,124 logged queries were coded into the five categories in Table~\ref{tab:ec-codebook}. Before coding, the
queries were assigned random identifiers, shuffled, and stripped of team, condition, stage, and time stamps. Queries
longer than 600 characters, which mostly contain pasted material, were shown as their first and last 300 characters,
because the request sits at the start or the end. A large language model (Claude) assigned the codes by applying the
written codebook in batches of about 270 queries. An earlier pass that showed only the first 140 characters of each
query misclassified long queries whose request followed pasted text; all 303 queries longer than 140 characters were
recoded on the full view, which changed 163 codes (mostly from Develop to Write). To assess reliability, a second,
independent coder of the same type coded a random sample of 200 queries blind to the existing codes. Agreement was
87.0\% (Cohen's $\kappa=0.82$; $\kappa=0.89$ for the binary distinction between generation and all other requests).
Both coders are language models; a human validation sample is a natural complement.

Table~\ref{tab:ec-qcounts} reports all counts. Generation dominated assistant use during search in every condition
(52--58\%). In the selection window, later-selection teams devoted 29.5\% of their queries to generation, against
11.3\% for full-process teams. At the team level, the difference in generation shares is 18.1 percentage points
($p<0.001$), and the difference in development shares runs the other way (41.0\% versus 52.0\%, $p=0.040$).
Development was nonetheless the most frequent request in both conditions. Later-selection teams issued 4.07 generation
queries per team in the selection window, against 8.23 per team during search under early-search access, so
selection-window generation was substantial but did not replicate search-stage use. For later-selection teams, the
selection window was the first contact with the assistant, and their early queries in the window include pasted task
prompts. Dropping each team's first three (five) queries in the window leaves generation shares of 21.1\% against
8.9\%, $p=0.011$ (18.6\% against 8.5\%, $p=0.024$).

Within the 60 teams with selection-stage access (C2 and C3), the number of late generation queries is not
significantly associated with the number of proposals submitted, conditional on condition and task ($b=0.097$ per
query, s.e.\ 0.125, $p=0.44$; within C2 alone, $b=0.111$, $p=0.53$). The composition result is therefore a
between-condition difference in how later access is used, consistent with Proposition~\ref{prop:latesearch}; the data
do not show that teams issuing more late generation requests submitted more proposals.

\clearpage
\begin{table}[!htbp]
\noindent\caption[Agreement among the four human raters]{{\bf Agreement among the four human raters.} ICC(2,1) and ICC(2,$k$) are two-way random-effects, absolute-agreement intraclass correlations for one rater and for the mean of $k=4$ raters, computed on units scored by all four raters. $\alpha$ is Krippendorff's interval alpha using all available ratings. Team-level units are the rater's mean over a team's proposals; proposal-level units are individual proposals. Fewer complete units are available for originality quality and text quality because some raters left these items blank for some proposals.\label{tab:ec-rel}}
\centering
\setlength{\tabcolsep}{5pt}
\begin{tabular}{lcccccccc}
\toprule
 & \multicolumn{4}{c}{Team level} & \multicolumn{4}{c}{Proposal level} \\
\cmidrule(lr){2-5}\cmidrule(lr){6-9}
Measure & ICC(2,1) & ICC(2,$k$) & $\alpha$ & $n$ & ICC(2,1) & ICC(2,$k$) & $\alpha$ & $n$ \\
\midrule
Creativity composite & 0.176 & 0.461 & 0.033 & 148 & 0.165 & 0.442 & 0.032 & 387 \\
Originality & 0.185 & 0.476 & 0.065 & 148 & 0.162 & 0.436 & 0.053 & 387 \\
Appropriateness & 0.070 & 0.232 & $-$0.105 & 148 & 0.081 & 0.260 & $-$0.077 & 387 \\
Elaboration & 0.183 & 0.473 & 0.044 & 148 & 0.216 & 0.525 & 0.092 & 387 \\
Originality quality & 0.095 & 0.296 & $-$0.070 & 148 & 0.091 & 0.285 & $-$0.055 & 387 \\
Text quality & 0.113 & 0.337 & $-$0.054 & 148 & 0.131 & 0.377 & $-$0.033 & 386 \\
\bottomrule
\end{tabular}
\end{table}
\clearpage
\begin{landscape}
\begin{table}[!htbp]
\noindent\caption[Rater characteristics, rater-specific effects, and inter-rater correlations]{{\bf Rater characteristics, rater-specific effects, and inter-rater correlations.} Raters are labeled in order of increasing mean (severity). Units are the proposals each rater scored (raters differed in how they segmented three submissions). Rater-specific effects are Welch contrasts on each rater's team means. Correlations are Pearson correlations of team means across raters. All four raters rank early-search above later-selection submissions; rater B's scores show little difference between AI and no-AI conditions.\label{tab:ec-raters}}
\centering
\begin{tabular}{llcccccccccc}
\toprule
 & & & Mean (s.d.) & \multicolumn{3}{c}{Hedges $g$ from rater's own scores} & \multicolumn{4}{c}{Correlation with rater} \\
\cmidrule(lr){5-7}\cmidrule(lr){8-11}
Rater & Type & Units & of composite & C1--C2 & C1--C4 & C3--C4 & A & B & C & D \\
\midrule
A & Industry expert & 396 & 3.05 (1.24) & 0.57 & 0.88 & 0.52 & -- & 0.46 & 0.46 & 0.53 \\
B & Doctoral researcher & 387 & 4.19 (0.92) & 0.24 & 0.12 & 0.30 & 0.46 & -- & 0.55 & 0.35 \\
C & Doctoral researcher & 394 & 5.10 (0.74) & 0.72 & 0.62 & 0.58 & 0.46 & 0.55 & -- & 0.44 \\
D & Industry expert & 396 & 5.36 (0.97) & 0.40 & 0.72 & 1.01 & 0.53 & 0.35 & 0.44 & -- \\
\bottomrule
\end{tabular}
\end{table}
\end{landscape}
\clearpage
\begin{table}[!htbp]
\noindent\caption[Codebook for assistant requests]{{\bf Codebook for assistant requests.} Decision rules: a query that proposes its own idea and also asks for more ideas is coded G; a query that pastes an existing idea and asks to refine it is coded D; naming a given product is D; a request dominated by a length or format instruction is W. Queries were shuffled and stripped of team, condition, stage and time stamps before coding.\label{tab:ec-codebook}}
\centering
\begin{tabular}{llll}
\toprule
Code & Category & Definition & Example queries (translated) \\
\midrule
G & \parbox[t]{3.2cm}{\raggedright Generate new options} & \parbox[t]{5.4cm}{\raggedright Requests new ideas, concepts, directions or alternative products, including the task prompt pasted with a request for designs, ``more'' or ``other'' ideas, and new angles.} & \parbox[t]{5.3cm}{\raggedright ``What innovative ideas come from combining a refrigerator and a calendar?''\\ ``Any other directions? This one is good. Give me some other directions.''} \\[4pt]
D & \parbox[t]{3.2cm}{\raggedright Develop a specified option} & \parbox[t]{5.4cm}{\raggedright Asks to elaborate, detail, extend, add features to, name, or plan the implementation of an idea already specified.} & \parbox[t]{5.3cm}{\raggedright ``Can the service be extended? How would it work?''\\ ``Help us define the core user group and the technical challenges.''} \\[4pt]
E & \parbox[t]{3.2cm}{\raggedright Evaluate or select} & \parbox[t]{5.4cm}{\raggedright Asks to compare options, judge feasibility, cost or market potential, choose the best, or identify flaws.} & \parbox[t]{5.3cm}{\raggedright ``Compare the pros and cons of a built-in versus an external calendar screen on the fridge.''\\ ``Compared with a real pet, what are its advantages and disadvantages?''} \\[4pt]
W & \parbox[t]{3.2cm}{\raggedright Write or polish} & \parbox[t]{5.4cm}{\raggedright Asks to rewrite into the submission format, shorten or expand to a length, polish wording, or merge options into one text.} & \parbox[t]{5.3cm}{\raggedright ``Expand it into a plan of about 250 characters.''\\ ``Write it as one paragraph, in the plainest language everyone can understand.''} \\[4pt]
I & \parbox[t]{3.2cm}{\raggedright Information or other} & \parbox[t]{5.4cm}{\raggedright Factual or background questions, clarifications, conversational or empty input.} & \parbox[t]{5.3cm}{\raggedright ``What are the uses and characteristics of pets?''\\ ``Why do you know we are having a team discussion?''} \\[4pt]
\bottomrule
\end{tabular}
\end{table}
\clearpage
\begin{landscape}
\begin{table}[!htbp]
\noindent\caption[Assistant requests by condition, stage, and type]{{\bf Assistant requests by condition, stage, and type.} Counts of coded queries with row percentages. No query occurred outside an assigned window; C4 had no assistant.\label{tab:ec-qcounts}}
\centering
\begin{tabular}{llcccccc}
\toprule
Condition & Stage & Generate & Develop & Evaluate & Write & Other & Total \\
\midrule
Early-search (C1) & S1 & 247 (55.1\%) & 121 (27.0\%) & 20 (4.5\%) & 37 (8.3\%) & 23 (5.1\%) & 448 \\
Later-selection (C2) & S2 & 122 (29.5\%) & 167 (40.4\%) & 35 (8.5\%) & 78 (18.9\%) & 11 (2.7\%) & 413 \\
Full-process (C3) & S1 & 256 (58.4\%) & 114 (26.0\%) & 24 (5.5\%) & 17 (3.9\%) & 27 (6.2\%) & 438 \\
Full-process (C3) & S2 & 36 (11.3\%) & 165 (51.7\%) & 24 (7.5\%) & 64 (20.1\%) & 30 (9.4\%) & 319 \\
Full-process (C3) & S3 & 12 (8.3\%) & 43 (29.9\%) & 10 (6.9\%) & 60 (41.7\%) & 19 (13.2\%) & 144 \\
Warning (C5) & S1 & 188 (51.9\%) & 113 (31.2\%) & 11 (3.0\%) & 27 (7.5\%) & 23 (6.4\%) & 362 \\
\midrule
All & & 861 & 723 & 124 & 283 & 133 & 2124 \\
\bottomrule
\end{tabular}
\end{table}
\end{landscape}

\clearpage
\section{Inference and Robustness for the Main Results}\label{ec:inference}

\subsection{The Family of Condition Contrasts}
Table~\ref{tab:mht} reports all eight condition contrasts on the primary outcome, with unadjusted Welch and
randomization-inference $p$ values and three multiplicity adjustments. The family combines the preregistered contrasts
(the four comparisons with the control, the timing contrast C1--C2, and the warning contrast C5--C1) with two
exploratory comparisons involving full-process access. Benjamini--Hochberg control of the false-discovery rate is the
preregistered procedure; Holm and Westfall--Young control the family-wise error rate and are reported as conservative
references. Table~\ref{tab:ec-pairwise} gives the underlying condition means and bootstrap intervals for Hedges $g$.
Table~\ref{tab:ec-dimfam} reports the 20 dimension-level contrasts with the control and their adjustment. The effects of
early-search and full-process access on appropriateness, elaboration, originality quality, and text quality, and the
effect of warned access on text quality, survive Holm adjustment across all 20 tests.

\subsection{Leave-One-Rater-Out Under Two Aggregation Rules}\label{ec:robust}
Table~\ref{tab:ec-loro} reports leave-one-rater-out estimates under two rules. The main text uses the mean of the
remaining raters' team scores, which applies the primary aggregation rule to three raters. The alternative rule
re-estimates rater severity from the remaining raters' proposal-level ratings in a linear mixed model before
aggregating. Both rules yield positive timing contrasts whose 95\% intervals exclude zero for every dropped rater
($g=0.54$--$0.69$).

\subsection{Crossed Random-Effects Model}
A linear mixed model of proposal-level ratings with condition indicators, task fixed effects, and crossed random
intercepts for teams, proposals within teams, and raters, estimated by restricted maximum likelihood (Wald inference),
yields a timing contrast of 0.390 points (95\% CI $[0.109, 0.672]$, $p=0.007$), an early-search effect of 0.486 points
(95\% CI $[0.190, 0.781]$, $p=0.001$), and a full-process effect of 0.472 points (95\% CI $[0.173, 0.772]$,
$p=0.002$). The estimated variance components are 0.208 (teams), 0.084 (proposals), 1.089 (raters), and 0.629
(residual).

\subsection{Task Heterogeneity}
Table~\ref{tab:ec-task} reports contrasts within each task. The timing contrast is positive in all three tasks, and a
test of task-by-condition interactions does not reject homogeneity ($p=0.29$).

\subsection{Mean, Best, and Total}
Table~\ref{tab:portfolio} reports the timing effect under the three aggregations of Proposition~\ref{prop:aggregation}.
The ordering follows the proposition, although only the difference in the mean is individually significant. In the mean
proposal score, the early-search advantage over later-selection access is 0.361 points ($g=0.62$). In the best proposal,
it is 0.221 points ($g=0.50$, $p=0.055$), positive and smaller. In the total score, the sign reverses: later-selection
submissions summed to 2.06 points more than early-search submissions ($g=-0.23$, $p=0.35$) because they contained more
proposals. Relative to no AI, later-selection access raised the best proposal by 0.426 points ($p=0.001$) and the total
by 6.31 points ($p=0.001$).

\clearpage
\begin{table}[!htbp]
\noindent\caption[Inference for the family of eight condition contrasts]{{\bf Inference for the family of eight condition contrasts.} The outcome is the calibrated team creativity score ($n=148$). RI is a randomization-inference $p$ value from 10,000 re-randomizations of condition labels between the two compared arms within task strata (studentized difference in means). BH controls the false-discovery rate; Holm and Westfall--Young (WY) step-down max-$|t|$ (10,000 re-randomizations of all five labels within task strata) control the family-wise error rate across all eight contrasts. The family combines confirmatory and exploratory contrasts, so the FWER adjustments are conservative for the focal C1--C2 test.\label{tab:mht}}
\centering
\setlength{\tabcolsep}{4pt}
\begin{tabular}{lccccccc}
\toprule
 & Difference & Hedges & \multicolumn{2}{c}{Unadjusted $p$} & \multicolumn{3}{c}{Adjusted $p$} \\
\cmidrule(lr){4-5}\cmidrule(lr){6-8}
Contrast & (points) & $g$ & Welch & RI & BH & Holm & WY \\
\midrule
Early-search vs.\ no AI (C1--C4) & 0.486 & 0.83 & $0.002$ & $0.002$ & $0.009$ & $0.018$ & $0.012$ \\
Later-selection vs.\ no AI (C2--C4) & 0.130 & 0.21 & $0.425$ & $0.433$ & $0.486$ & $1.000$ & $0.669$ \\
Full-process vs.\ no AI (C3--C4) & 0.479 & 0.83 & $0.002$ & $0.002$ & $0.009$ & $0.018$ & $0.012$ \\
Warning vs.\ no AI (C5--C4) & 0.319 & 0.48 & $0.071$ & $0.049$ & $0.142$ & $0.354$ & $0.251$ \\
Early-search vs.\ later-selection (C1--C2) & 0.356 & 0.62 & $0.018$ & $0.021$ & $0.048$ & $0.107$ & $0.086$ \\
Warning vs.\ early-search (C5--C1) & $-$0.167 & $-$0.27 & $0.297$ & $0.271$ & $0.417$ & $1.000$ & $0.669$ \\
Full-process vs.\ early-search (C3--C1) & $-$0.007 & $-$0.01 & $0.957$ & $0.956$ & $0.957$ & $1.000$ & $0.958$ \\
Full-process vs.\ warning (C3--C5) & 0.160 & 0.26 & $0.313$ & $0.301$ & $0.417$ & $1.000$ & $0.669$ \\
\bottomrule
\end{tabular}
\end{table}
\clearpage
\begin{landscape}
\begin{table}[!htbp]
\noindent\caption[Pairwise condition contrasts on rated creativity]{{\bf Pairwise condition contrasts on rated creativity.} Outcome: calibrated team creativity score. Differences and $p$ values from two-sided Welch $t$-tests; intervals for Hedges $g$ are percentile intervals from 5,000 team-level bootstrap resamples (seed 20261003).\label{tab:ec-pairwise}}
\centering
\setlength{\tabcolsep}{5pt}
\begin{tabular}{lcccccc}
\toprule
Contrast & $n$ (a/b) & Mean a (s.d.) & Mean b (s.d.) & Difference [95\% CI] & Hedges $g$ [95\% CI] & $p$ \\
\midrule
Early-search vs.\ no AI (C1--C4) & 30/29 & 4.88 (0.52) & 4.39 (0.64) & 0.486 [0.182, 0.790] & 0.83 [0.33, 1.41] & 0.002 \\
Later-selection vs.\ no AI (C2--C4) & 30/29 & 4.52 (0.61) & 4.39 (0.64) & 0.130 [$-$0.194, 0.455] & 0.21 [$-$0.29, 0.73] & 0.425 \\
Full-process vs.\ no AI (C3--C4) & 30/29 & 4.87 (0.50) & 4.39 (0.64) & 0.479 [0.179, 0.779] & 0.83 [0.35, 1.37] & 0.002 \\
Warning vs.\ no AI (C5--C4) & 29/29 & 4.71 (0.68) & 4.39 (0.64) & 0.319 [$-$0.028, 0.667] & 0.48 [$-$0.02, 0.99] & 0.071 \\
Early-search vs.\ later-selection (C1--C2) & 30/30 & 4.88 (0.52) & 4.52 (0.61) & 0.356 [0.064, 0.648] & 0.62 [0.13, 1.20] & 0.018 \\
Warning vs.\ early-search (C5--C1) & 29/30 & 4.71 (0.68) & 4.88 (0.52) & $-$0.167 [$-$0.485, 0.151] & $-$0.27 [$-$0.84, 0.22] & 0.297 \\
Full-process vs.\ early-search (C3--C1) & 30/30 & 4.87 (0.50) & 4.88 (0.52) & $-$0.007 [$-$0.272, 0.257] & $-$0.01 [$-$0.55, 0.48] & 0.957 \\
Full-process vs.\ warning (C3--C5) & 30/29 & 4.87 (0.50) & 4.71 (0.68) & 0.160 [$-$0.155, 0.475] & 0.26 [$-$0.23, 0.83] & 0.313 \\
\bottomrule
\end{tabular}
\end{table}
\end{landscape}
\clearpage
\begin{table}[!htbp]
\noindent\caption[Dimension-level contrasts with the no-AI control and multiplicity adjustment]{{\bf Dimension-level contrasts with the no-AI control and multiplicity adjustment.} Coefficients from OLS of each team-level dimension score on condition indicators and task fixed effects (HC3). The family comprises the 20 tests shown (four contrasts with the control $\times$ five dimensions); BH and Holm adjustments are applied across all 20.\label{tab:ec-dimfam}}
\centering
\begin{tabular}{llccccc}
\toprule
Dimension & Contrast & Coefficient [95\% CI] & Hedges $g$ & $p$ & BH $p$ & Holm $p$ \\
\midrule
Originality & C1--C4 & 0.431 [0.093, 0.769] & 0.62 & 0.012 & 0.023 & 0.124 \\
 & C2--C4 & 0.068 [$-$0.308, 0.443] & 0.10 & 0.724 & 0.724 & 0.839 \\
 & C3--C4 & 0.461 [0.128, 0.793] & 0.70 & 0.007 & 0.013 & 0.073 \\
 & C5--C4 & 0.332 [$-$0.036, 0.700] & 0.44 & 0.077 & 0.103 & 0.469 \\
\addlinespace
Appropriateness & C1--C4 & 0.530 [0.292, 0.768] & 1.16 & $<0.001$ & $<0.001$ & $<0.001$ \\
 & C2--C4 & 0.188 [$-$0.049, 0.425] & 0.42 & 0.120 & 0.150 & 0.602 \\
 & C3--C4 & 0.428 [0.182, 0.674] & 0.89 & $<0.001$ & 0.002 & 0.010 \\
 & C5--C4 & 0.203 [$-$0.067, 0.472] & 0.39 & 0.141 & 0.166 & 0.602 \\
\addlinespace
Elaboration & C1--C4 & 0.695 [0.388, 1.002] & 1.17 & $<0.001$ & $<0.001$ & $<0.001$ \\
 & C2--C4 & 0.326 [$-$0.023, 0.675] & 0.49 & 0.067 & 0.096 & 0.469 \\
 & C3--C4 & 0.579 [0.259, 0.898] & 0.92 & $<0.001$ & 0.002 & 0.006 \\
 & C5--C4 & 0.338 [0.012, 0.664] & 0.54 & 0.042 & 0.065 & 0.339 \\
\addlinespace
Originality quality & C1--C4 & 0.483 [0.195, 0.772] & 0.80 & 0.001 & 0.003 & 0.014 \\
 & C2--C4 & 0.123 [$-$0.176, 0.423] & 0.22 & 0.419 & 0.441 & 0.839 \\
 & C3--C4 & 0.442 [0.162, 0.723] & 0.82 & 0.002 & 0.005 & 0.026 \\
 & C5--C4 & 0.347 [0.036, 0.657] & 0.55 & 0.029 & 0.048 & 0.257 \\
\addlinespace
Text quality & C1--C4 & 0.502 [0.247, 0.758] & 1.02 & $<0.001$ & $<0.001$ & 0.002 \\
 & C2--C4 & 0.183 [$-$0.081, 0.447] & 0.36 & 0.175 & 0.195 & 0.602 \\
 & C3--C4 & 0.408 [0.133, 0.683] & 0.75 & 0.004 & 0.008 & 0.044 \\
 & C5--C4 & 0.456 [0.221, 0.692] & 1.01 & $<0.001$ & $<0.001$ & 0.003 \\
\bottomrule
\end{tabular}
\end{table}
\clearpage
\begin{landscape}
\begin{table}[!htbp]
\noindent\caption[Leave-one-rater-out estimates under two aggregation rules]{{\bf Leave-one-rater-out estimates under two aggregation rules.} Left panel: Hedges $g$ and Welch $p$ from contrasts on the mean of the three remaining raters' team scores (score-point estimates and intervals are in Table~\ref{tab:robust} of the main text). Right panel: Hedges $g$ with 95\% interval after re-estimating rater severity on the remaining three raters' proposal-level ratings (REML linear mixed model with crossed random intercepts for proposals, raters, teams and tasks; severities are the centered predicted rater effects), averaging the adjusted ratings to teams and bootstrapping teams 5,000 times. Under both rules, every interval for the C1--C2 contrast excludes zero.\label{tab:ec-loro}}
\centering
\begin{tabular}{lcccccc}
\toprule
 & \multicolumn{3}{c}{Mean of remaining raters (main text): $g$ ($p$)} & \multicolumn{3}{c}{Severity-adjusted mixed model: $g$ [95\% CI]} \\
\cmidrule(lr){2-4}\cmidrule(lr){5-7}
Dropped rater & C1--C2 & C1--C4 & C3--C4 & C1--C2 & C1--C4 & C3--C4 \\
\midrule
A & 0.54 (0.039) & 0.65 (0.015) & 0.88 (0.001) & 0.550 [0.045, 1.080] & 0.633 [0.139, 1.183] & 0.882 [0.410, 1.418] \\
B & 0.68 (0.010) & 0.93 ($<0.001$) & 0.89 (0.001) & 0.692 [0.218, 1.280] & 0.931 [0.419, 1.546] & 0.887 [0.391, 1.483] \\
C & 0.54 (0.038) & 0.82 (0.003) & 0.79 (0.003) & 0.544 [0.030, 1.115] & 0.804 [0.304, 1.397] & 0.790 [0.316, 1.369] \\
D & 0.62 (0.017) & 0.74 (0.006) & 0.61 (0.022) & 0.590 [0.117, 1.172] & 0.706 [0.216, 1.283] & 0.608 [0.134, 1.101] \\
\bottomrule
\end{tabular}
\end{table}
\end{landscape}
\clearpage
\begin{landscape}
\begin{table}[!htbp]
\noindent\caption[Condition contrasts within each task]{{\bf Condition contrasts within each task.} Entries are score-point differences, with Hedges $g$ and Welch $p$ in parentheses (9--11 teams per cell). The interaction test compares OLS models with and without task-by-condition interactions. The timing contrast is positive in all three tasks; cell-level estimates are imprecise by construction.\label{tab:ec-task}}
\centering
\begin{tabular}{lccccc}
\toprule
Task & C1--C2 & C1--C4 & C2--C4 & C3--C4 & C5--C4 \\
\midrule
Backpack + pet accessory & 0.17 (0.24; 0.594) & 0.58 (1.22; 0.009) & 0.41 (0.57; 0.207) & 0.65 (1.41; 0.004) & 0.31 (0.51; 0.250) \\
Refrigerator + calendar & 0.64 (1.20; 0.013) & 0.20 (0.31; 0.489) & $-$0.44 ($-$0.79; 0.101) & 0.07 (0.10; 0.815) & 0.27 (0.36; 0.417) \\
Shoes + map & 0.28 (0.57; 0.216) & 0.65 (1.09; 0.022) & 0.37 (0.63; 0.161) & 0.67 (1.21; 0.013) & 0.33 (0.57; 0.200) \\
\midrule
\multicolumn{6}{l}{Task $\times$ condition interaction: $F(8,133)=1.22$, $p=0.289$} \\
\bottomrule
\end{tabular}
\end{table}
\end{landscape}
\clearpage
\begin{table}[!htbp]
\noindent\caption[The value of the submitted portfolio under alternative aggregations]{{\bf The value of the submitted portfolio under alternative aggregations.} OLS with task fixed effects and HC3 standard errors. For each rater, the creativity composite of a team's proposals is averaged (column 1), maximized (column 2) or summed (column 3); the rater-level values are then averaged over the four raters. Column 1 reproduces the primary outcome. Column 4 is the rater-coded number of proposals. Columns 2--4 are implications of the search--selection framework (Section~2) rather than separate confirmatory outcomes. $^{*}p<0.10$, $^{**}p<0.05$, $^{***}p<0.01$ (two-sided).\label{tab:portfolio}}
\centering
\begin{tabular}{lcccc}
\toprule
 & (1) & (2) & (3) & (4) \\
 & \shortstack{Mean proposal\\score (primary)} & \shortstack{Best proposal\\score} & \shortstack{Total score\\(sum)} & \shortstack{Number of\\proposals} \\
\midrule
Early-search access (C1) & $0.486^{***}$ & $0.646^{***}$ & $4.25^{***}$ & $0.74^{*}$ \\
 & $(0.147)$ & $(0.129)$ & $(1.63)$ & $(0.43)$ \\[2pt]
Later-selection access (C2) & $0.124$ & $0.426^{***}$ & $6.31^{***}$ & $1.53^{***}$ \\
 & $(0.165)$ & $(0.129)$ & $(1.95)$ & $(0.54)$ \\[2pt]
Full-process access (C3) & $0.473^{***}$ & $0.541^{***}$ & $2.85$ & $0.39$ \\
 & $(0.148)$ & $(0.126)$ & $(1.89)$ & $(0.46)$ \\[2pt]
Early-search + warning (C5) & $0.313^{*}$ & $0.366^{***}$ & $1.00$ & $0.12$ \\
 & $(0.166)$ & $(0.138)$ & $(1.52)$ & $(0.40)$ \\[2pt]
\midrule
C1 $-$ C2 & $0.361^{**}$ & $0.221^{*}$ & $-2.06$ & $-0.78$ \\
 & $(0.153)$ & $(0.115)$ & $(2.20)$ & $(0.60)$ \\
\quad Hedges $g$, C1 vs.\ C2 & 0.62 & 0.50 & $-$0.23 & $-$0.33 \\[2pt]
C3 $-$ C1 & $-0.012$ & $-0.105$ & $-1.41$ & $-0.35$ \\
 & $(0.135)$ & $(0.111)$ & $(2.13)$ & $(0.52)$ \\
\midrule
Control (C4) mean & 4.392 & 4.699 & 8.81 & 2.10 \\
Adjusted $R^2$ & 0.110 & 0.195 & 0.062 & 0.049 \\
Observations & 148 & 148 & 148 & 148 \\
\bottomrule
\end{tabular}
\end{table}

\clearpage
\section{The Idea Funnel}\label{ec:funnel}

This appendix documents the funnel measures behind Table~\ref{tab:funnel} and Figure~\ref{fig:funnel} of the main
text.

\subsection{Construction}
The platform saved each member's personal draft every minute. We build the pool at the gate from each member's last
draft of the search stage, which the stage instructions asked teams to use for noting their concepts. Drafts are split
into idea items at numbered or bulleted list markers (items shorter than eight non-space characters are dropped).
Items are pooled within team and deduplicated: an item is kept only if its TF--IDF cosine similarity (character
2--3-grams) with every item already kept is below 0.5. An item is AI-derived if any of its sentences reaches a
similarity of 0.35 with a sentence of the team's search-stage assistant output, and the team's own otherwise. A pool
item survives if any of its sentences reaches 0.35 with a sentence of the final submission. The pre-gate text share is
the share of submission sentences that reach 0.35 with any sentence of the team's search-stage drafts, chat messages,
or assistant output. To calibrate it, we also match each submission with the pre-gate material of three other teams
that worked on the same task; at the main threshold, fewer than 1\% of sentences match in this placebo comparison.  

\subsection{Robustness}
Table~\ref{tab:ec-funnel-rob} reports the funnel contrasts under alternative rules. The early-access pool is three and a
half to five times the control pool under every rule: when ideas that appeared in any search-stage draft are counted,
under alternative deduplication thresholds, and when items are deduplicated with sentence embeddings, which treat
paraphrases as duplicates. The pre-gate text share of later-selection submissions is far below that of control
submissions at every TF--IDF threshold and with sentence-embedding similarity calibrated to placebo match rates of 5\%
and 1\%. Because embeddings recognize paraphrase, the low share does not reflect the assistant rewording pre-gate
ideas.

\subsection{Proposal-Level Origin and Within-Condition Associations}
Table~\ref{tab:ec-funnel-prop} links the funnel to the human ratings of individual proposals and reports
within-condition associations. Proposals of pre-gate origin are not rated differently from proposals of post-gate
origin, either with condition fixed effects or within teams that submitted both kinds. Within the later-selection and
full-process conditions, neither late generation requests nor pool size is associated with how much of the pool a team
kept. The funnel differences in the main text are therefore differences between randomly assigned schedules; the data
do not trace them to variation within conditions.

\subsection{Calibration of the Screening Premium}
Table~\ref{tab:ec-funnel-cal} asks whether the measured pools are large enough, under the model, to account for the
effect of early access. With ideas drawn from a normal distribution whose standard deviation equals the
reliability-adjusted standard deviation of proposal ratings, retaining the two best ideas of the early-search pool
rather than of the control pool yields a premium of the same size as the observed effect. The calibration assumes that
teams screen perfectly and that AI and own ideas are equally good; imperfect screening would lower the implied premium,
and better AI ideas would raise it.

\clearpage
\begin{landscape}
\begin{table}[!htbp]
\noindent\caption[Robustness of the funnel measures]{{\bf Robustness of the funnel measures.} Welch tests. Pool measures count distinct ideas in the members' search-stage drafts under alternative rules; pre-gate text shares are the share of submission sentences matching the team's search-stage drafts, chat or assistant output. Placebo match rates of the TF--IDF rule against other teams' pre-gate material in the same task: 2.3\% (0.25), 0.8\% (0.35, main) and 0.2\% (0.50). Embedding thresholds (0.836 and 0.885) are the cosine similarities exceeded by 5\% and 1\% of placebo matches; the embedding model is multilingual MiniLM, which recognizes paraphrase.\label{tab:ec-funnel-rob}}
\centering
\begin{tabular}{lcccccc}
\toprule
 & \multicolumn{3}{c}{Condition means} & \multicolumn{3}{c}{Differences ($p$)} \\
\cmidrule(lr){2-4}\cmidrule(lr){5-7}
Measure & C1 & C2 & C4 & C1 $-$ C4 & C1 $-$ C2 & C2 $-$ C4 \\
\midrule
Pool: ideas in any search-stage draft & 23.50 & 6.30 & 6.69 & $+16.81$ ($<0.001$) & $+17.20$ ($<0.001$) & $-0.39$ (0.659) \\
Pool: de-duplication at 0.40 & 19.57 & 3.77 & 3.76 & $+15.81$ ($<0.001$) & $+15.80$ ($<0.001$) & $+0.01$ (0.989) \\
Pool: de-duplication at 0.60 & 19.90 & 3.87 & 4.00 & $+15.90$ ($<0.001$) & $+16.03$ ($<0.001$) & $-0.13$ (0.820) \\
Pool: embedding de-duplication (0.85) & 18.53 & 3.37 & 3.69 & $+14.84$ ($<0.001$) & $+15.17$ ($<0.001$) & $-0.32$ (0.527) \\
Pre-gate text share: TF--IDF threshold 0.25 & 0.74 & 0.07 & 0.26 & $+0.48$ ($<0.001$) & $+0.67$ ($<0.001$) & $-0.19$ ($<0.001$) \\
Pre-gate text share: TF--IDF threshold 0.50 & 0.65 & 0.03 & 0.16 & $+0.49$ ($<0.001$) & $+0.62$ ($<0.001$) & $-0.12$ ($<0.001$) \\
Pre-gate text share: embeddings, 5\% placebo rate & 0.63 & 0.07 & 0.23 & $+0.40$ ($<0.001$) & $+0.55$ ($<0.001$) & $-0.16$ ($<0.001$) \\
Pre-gate text share: embeddings, 1\% placebo rate & 0.54 & 0.04 & 0.11 & $+0.42$ ($<0.001$) & $+0.50$ ($<0.001$) & $-0.08$ (0.004) \\
\bottomrule
\end{tabular}
\end{table}
\end{landscape}
\clearpage
\begin{landscape}
\begin{table}[!htbp]
\noindent\caption[Proposal-level origin and within-condition associations]{{\bf Proposal-level origin and within-condition associations.} Each submission is segmented into the number of units the raters scored (353 units in 142 teams; 86\% of teams segmented exactly by their own headers or as a single unit, the rest by length). Within multi-unit teams, unit length correlates 0.27 with the raters' feature count and 0.37 with rated elaboration, which supports the alignment. A unit is of pre-gate origin if at least half of its sentences match the team's pre-gate material. Panel A: OLS with team-clustered standard errors; the within-team model uses teams with units of both origins. All associations are descriptive.\label{tab:ec-funnel-prop}}
\centering
\begin{tabular}{llccc}
\toprule
\multicolumn{5}{l}{\textit{Panel A. Pre-gate origin and proposal rating (creativity composite, rater mean)}} \\
Sample & Model & Estimate (s.e.) & $p$ & Units (teams) \\
all units & condition + task FE & $+0.022$ (0.090) & 0.805 & 353 (142) \\
all units & within team & $+0.027$ (0.128) & 0.837 & 167 (42) \\
exact or single segmentation & condition + task FE & $+0.058$ (0.096) & 0.544 & 292 (127) \\
exact or single segmentation & within team & $+0.089$ (0.136) & 0.515 & 141 (37) \\
\midrule
\multicolumn{5}{l}{\textit{Panel B. Share of proposals of pre-gate origin, by condition (all units)}} \\
C1: 0.80 & C5: 0.57 & C3: 0.28 & C2: 0.07 & C4: 0.26 \\
\midrule
\multicolumn{5}{l}{\textit{Panel C. Within-condition correlations (Pearson $r$, $p$)}} \\
Condition & $x$ & $y$ & $r$ ($p$) & $n$ \\
C2 & late generation requests & share of pool kept & $+0.08$ (0.667) & 29 \\
C2 & late generation requests & pre-gate text share & $+0.18$ (0.331) & 30 \\
C2 & pool at the gate & share of pool kept & $-0.03$ (0.868) & 29 \\
C2 & pool at the gate & pre-gate text share & $+0.15$ (0.433) & 30 \\
C3 & late generation requests & share of pool kept & $+0.05$ (0.810) & 30 \\
C3 & late generation requests & pre-gate text share & $+0.07$ (0.733) & 30 \\
C3 & pool at the gate & share of pool kept & $-0.18$ (0.353) & 30 \\
C3 & pool at the gate & pre-gate text share & $+0.31$ (0.100) & 30 \\
\bottomrule
\end{tabular}
\end{table}
\end{landscape}
\clearpage
\begin{table}[!htbp]
\noindent\caption[Screening premium implied by the measured pools]{{\bf Screening premium implied by the measured pools.} The implied premium is $\sigma_\theta[M_k(N)-M_k(N_0)]$ for standard normal order statistics, where $N$ and $N_0$ are the mean pool sizes at the gate (rounded) in the condition and the control, $k=2$ ideas are retained, and $\sigma_\theta$ is the standard deviation of the rater-mean proposal composite multiplied by the square root of its reliability (ICC(2,$k$) $=0.44$). It assumes perfect screening and equal quality of AI and own ideas; it is a check on magnitudes, not an estimate.\label{tab:ec-funnel-cal}}
\centering
\begin{tabular}{cccccccc}
\toprule
Condition & Pool & Control pool & $k$ & $\sigma_\theta$ & \shortstack{Implied premium\\(perfect screen)} & \shortstack{Observed\\vs.\ control} & Ratio \\
\midrule
C1 & 20 & 4 & 2 & 0.473 & 0.461 & 0.486 & 1.06 \\
C3 & 13 & 4 & 2 & 0.473 & 0.356 & 0.479 & 1.35 \\
C5 & 14 & 4 & 2 & 0.473 & 0.375 & 0.319 & 0.85 \\
\bottomrule
\end{tabular}
\end{table}

\clearpage
\section{Tracing Assistant Output}\label{ec:fate}

This appendix documents the tracing measures behind Table~\ref{tab:fate} of the main text and reports their
robustness.

\subsection{Construction}
We split each of the 2,124 assistant responses into idea units at numbered list items, markdown headings, and bold
headings, merging units shorter than 15 non-space characters into the preceding text; this gives 8,603 units. Each unit
and each final submission are split into sentences at Chinese and Western sentence punctuation and line breaks
(sentences of at least eight characters). Similarity is the cosine similarity of TF--IDF vectors of character bigrams
and trigrams (sublinear term frequency, minimum document frequency 2), which requires no word segmentation of Chinese
text. A unit \emph{entered the submission} if any of its sentences reaches the threshold with any sentence of its own
team's final submission; it entered \emph{near-verbatim} if the maximum similarity is at least 0.70; and it was
\emph{shared} if any of its sentences reaches the threshold with a chat message that the team posted after the response
arrived. Because assistant responses were visible only to the member who asked, a unit that entered the submission
without being shared did not pass through the team's common channel. The window of a unit is the access window in which
the response was received; minutes into the window are measured from the first logged event of the stage. 

\subsection{Calibration}
Table~\ref{tab:ec-fate-cal} compares each unit with the submission of every other team that worked on the same task.
At the main threshold of 0.35, 0.4\% of such placebo pairs match, against 16.4\% of units matching their own team's
submission. Sentence-embedding thresholds (multilingual MiniLM) are chosen to reproduce the TF--IDF placebo match
rates.

\subsection{Robustness}
Table~\ref{tab:ec-fate-rob} repeats the comparisons of Table~\ref{tab:fate} of the main text at three TF--IDF thresholds
and three placebo-matched embedding thresholds, with and without fixed effects for the five request types. Three
patterns are stable. First, output from later-selection teams' selection window entered submissions, entered
near-verbatim, and entered unshared more often than output from early-search teams' search window under every measure.
With request-type fixed effects, these differences shrink by about half, and at the 5\% level they remain significant
only for unshared entry under the TF--IDF measures at thresholds of 0.25 and 0.35. Second, in the same selection
window, later-selection teams used the assistant's output near-verbatim more often than full-process teams under both
similarity measures, with or without request-type fixed effects. Third, in raw terms, full-process teams'
selection-window output entered their submissions, and entered unshared, more often than their search-window output.
Restricted to units from requests for new options, the differences between the windows have the same sign throughout;
at the lowest threshold, the differences in adoption and in unshared adoption are significant.

\subsection{Timing Within the Window}
Lemma~\ref{lem:retention} implies that, within the selection stage, options obtained closer to the deadline are less
likely to be compared. Table~\ref{tab:ec-fate-time} estimates how adoption changes with minutes elapsed in the window,
with team fixed effects. Output received later in either window was more likely to enter the submission, as expected,
because later output is closer to the final text. The slope for unshared entry is steeper in the selection window than
in the search window, but the difference is not significant ($p=0.09$ for all units). We therefore do not claim support
for this finer implication.

\subsection{AI-Derived Content, Its Pricing, and Typicality}
Table~\ref{tab:ec-unscreened} links the tracing measures to the submissions. Panel A matches each submission sentence to
its most similar assistant unit. Later-selection submissions contain more AI-derived text than early-search
submissions, mostly text that had been shared in the chat; full-process submissions contain the most unshared
AI-derived text, reflecting their review-stage use of the assistant for polishing. Warned teams' submissions contain
less AI-derived text than unwarned early-search teams' submissions (47\% against 65\%, $p=0.020$). Because the share is
defined for every assigned team, this is a randomized comparison of reliance, complementing the observational
proximity comparison in Appendix~\ref{ec:proximity}.

Panel B shows that, within conditions, AI-derived text, shared or not, is associated with higher scores from both
panels, whereas the number of proposals is associated with lower scores, more strongly for the human panel. These
associations are descriptive, because teams choose how much AI-derived text to use. They indicate that the cost of late
access comes not from the assistant's text as such but from what teams retained. Panel C measures typicality for every
team, including the control. Early- and late-access submissions are both more similar than control submissions to the
assistant output received by other teams in the same task, but the two single-window conditions do not differ from
each other, and neither differs from the control in similarity to other teams' submissions.

\clearpage
\begin{landscape}
\begin{table}[!htbp]
\noindent\caption[Calibration of the tracing rule]{{\bf Calibration of the tracing rule.} Each assistant idea unit is compared with its own team's final submission and with the submission of every other team that worked on the same task. The placebo match rate is the share of unit--other-team pairs in which some sentence pair reaches the threshold; it is the false-positive rate of the rule. Embedding thresholds (multilingual MiniLM, cosine similarity of normalized sentence embeddings) are set so that their placebo match rates equal those of the TF--IDF thresholds named in parentheses.\label{tab:ec-fate-cal}}
\centering
\begin{tabular}{lccc}
\toprule
Similarity measure & Threshold & \shortstack{Placebo match rate\\(other teams' submissions)} & \shortstack{Match rate\\(own submission)} \\
\midrule
TF--IDF (character 2--3 grams) & 0.250 & 1.1\% & 21.1\% \\
TF--IDF (character 2--3 grams) & 0.300 & 0.6\% & 18.3\% \\
TF--IDF (character 2--3 grams) \textbf{(main)} & 0.350 & 0.4\% & 16.4\% \\
TF--IDF (character 2--3 grams) & 0.400 & 0.2\% & 15.0\% \\
TF--IDF (character 2--3 grams) & 0.500 & 0.1\% & 12.8\% \\
TF--IDF (character 2--3 grams) & 0.700 & 0.0\% & 9.6\% \\
Sentence embeddings (matched to TF--IDF 0.25) & 0.864 & 1.1\% & 14.8\% \\
Sentence embeddings (matched to TF--IDF 0.35) & 0.896 & 0.4\% & 11.6\% \\
Sentence embeddings (matched to TF--IDF 0.50) & 0.924 & 0.1\% & 9.3\% \\
\bottomrule
\end{tabular}
\end{table}
\end{landscape}
\clearpage
\begin{table}[!htbp]
\noindent\caption[Robustness of the tracing results]{{\bf Robustness of the tracing results.} Entries are differences in percentage points of idea units, from unit-level linear probability models with task fixed effects and team-clustered standard errors; ``Type FE'' adds fixed effects for the five request types. Near-verbatim uses a TF--IDF similarity of 0.70 (for embeddings, the threshold with the same placebo match rate). Generation rows restrict to units from requests for new options, so request-type fixed effects do not apply. $^{*}p<0.10$, $^{**}p<0.05$, $^{***}p<0.01$.\label{tab:ec-fate-rob}}
\centering
\setlength{\tabcolsep}{3.5pt}
\begin{tabular}{>{\raggedright\arraybackslash}p{2.2in}cccccc}
\toprule
 & \multicolumn{2}{c}{Entered submission} & \multicolumn{2}{c}{Near-verbatim} & \multicolumn{2}{c}{Entered unshared} \\
\cmidrule(lr){2-3}\cmidrule(lr){4-5}\cmidrule(lr){6-7}
Contrast (window) & Raw & Type FE & Raw & Type FE & Raw & Type FE \\
\midrule
\multicolumn{7}{l}{\textit{TF--IDF, threshold 0.25}} \\
\quad C2 (S2) $-$ C1 (S1) & $+10.7$*** & $+6.3$* & $+6.1$** & $+3.4$ & $+4.5$*** & $+3.8$** \\
\quad C2 (S2) $-$ C1 (S1), generation requests & $+8.0$** & -- & $+2.2$ & -- & $+3.7$** & -- \\
\quad C2 (S2) $-$ C3 (S2) & $+0.8$ & $+1.9$ & $+6.6$** & $+8.0$*** & $-5.1$** & $-4.9$* \\
\quad C3 (S2) $-$ C3 (S1) & $+17.3$*** & $+10.0$*** & $+4.9$*** & $+1.6$ & $+10.1$*** & $+8.8$*** \\
\multicolumn{7}{l}{\textit{TF--IDF, threshold 0.35 (main)}} \\
\quad C2 (S2) $-$ C1 (S1) & $+8.3$*** & $+4.2$ & $+6.1$** & $+3.4$ & $+5.1$*** & $+3.3$** \\
\quad C2 (S2) $-$ C1 (S1), generation requests & $+4.1$ & -- & $+2.2$ & -- & $+2.3$ & -- \\
\quad C2 (S2) $-$ C3 (S2) & $+2.7$ & $+4.2$ & $+6.6$** & $+8.0$*** & $-2.0$ & $-1.2$ \\
\quad C3 (S2) $-$ C3 (S1) & $+13.0$*** & $+7.0$** & $+4.9$*** & $+1.6$ & $+8.5$*** & $+5.9$*** \\
\multicolumn{7}{l}{\textit{TF--IDF, threshold 0.50}} \\
\quad C2 (S2) $-$ C1 (S1) & $+7.1$*** & $+3.7$ & $+6.1$** & $+3.4$ & $+6.7$*** & $+3.8$* \\
\quad C2 (S2) $-$ C1 (S1), generation requests & $+2.6$ & -- & $+2.2$ & -- & $+2.0$ & -- \\
\quad C2 (S2) $-$ C3 (S2) & $+5.6$* & $+6.9$** & $+6.6$** & $+8.0$*** & $+3.5$ & $+4.8$ \\
\quad C3 (S2) $-$ C3 (S1) & $+7.8$*** & $+3.7$ & $+4.9$*** & $+1.6$ & $+6.4$*** & $+2.2$ \\
\multicolumn{7}{l}{\textit{Embeddings, placebo-matched to TF--IDF 0.25}} \\
\quad C2 (S2) $-$ C1 (S1) & $+7.6$** & $+4.4$ & $+6.2$** & $+3.8$* & $+6.5$** & $+3.5$ \\
\quad C2 (S2) $-$ C1 (S1), generation requests & $+5.1$* & -- & $+2.3$ & -- & $+3.6$ & -- \\
\quad C2 (S2) $-$ C3 (S2) & $+2.8$ & $+3.6$ & $+6.4$** & $+7.5$*** & $+1.5$ & $+2.6$ \\
\quad C3 (S2) $-$ C3 (S1) & $+10.3$*** & $+4.5$* & $+3.6$*** & $+1.5$ & $+9.1$*** & $+3.5$ \\
\multicolumn{7}{l}{\textit{Embeddings, placebo-matched to TF--IDF 0.35}} \\
\quad C2 (S2) $-$ C1 (S1) & $+7.4$** & $+4.5$* & $+6.2$** & $+3.8$* & $+6.0$** & $+3.0$ \\
\quad C2 (S2) $-$ C1 (S1), generation requests & $+4.1$* & -- & $+2.3$ & -- & $+2.7$ & -- \\
\quad C2 (S2) $-$ C3 (S2) & $+4.7$ & $+5.8$ & $+6.4$** & $+7.5$*** & $+3.7$ & $+4.9$ \\
\quad C3 (S2) $-$ C3 (S1) & $+7.6$*** & $+3.5$ & $+3.6$*** & $+1.5$ & $+6.7$*** & $+2.4$ \\
\multicolumn{7}{l}{\textit{Embeddings, placebo-matched to TF--IDF 0.50}} \\
\quad C2 (S2) $-$ C1 (S1) & $+6.8$** & $+4.4$* & $+6.2$** & $+3.8$* & $+6.2$** & $+3.6$ \\
\quad C2 (S2) $-$ C1 (S1), generation requests & $+3.7$* & -- & $+2.3$ & -- & $+2.8$ & -- \\
\quad C2 (S2) $-$ C3 (S2) & $+5.6$* & $+6.4$** & $+6.4$** & $+7.5$*** & $+5.1$* & $+6.0$** \\
\quad C3 (S2) $-$ C3 (S1) & $+5.8$*** & $+3.0$ & $+3.6$*** & $+1.5$ & $+5.2$*** & $+2.4$ \\
\bottomrule
\end{tabular}
\end{table}
\clearpage
\begin{table}[!htbp]
\noindent\caption[Timing within the access window]{{\bf Timing within the access window.} Unit-level linear probability models of the outcome on minutes elapsed in the window, a selection-window indicator and its interaction with minutes, with team fixed effects and team-clustered standard errors ($p$ in parentheses). Units from the search windows of C1, C3 and C5 and the selection windows of C2 and C3. Lemma~\ref{lem:retention} implies a steeper slope in the selection window for options kept without comparison; the differences are small, all but one positive, and none is significant at the 5\% level.\label{tab:ec-fate-time}}
\centering
\setlength{\tabcolsep}{3.5pt}
\begin{tabular}{llccc}
\toprule
Outcome & Units & \shortstack{Slope per minute,\\search window (pp)} & \shortstack{Difference in slope,\\selection window (pp)} & Units \\
\midrule
Entered submission & all units & $+1.04$ ($<0.001$) & $+0.39$ (0.259) & 7,973 \\
Entered submission & generation requests only & $+0.48$ (0.010) & $+0.55$ (0.510) & 3,551 \\
Near-verbatim & all units & $+0.70$ ($<0.001$) & $+0.24$ (0.413) & 7,973 \\
Near-verbatim & generation requests only & $+0.37$ (0.010) & $-0.12$ (0.794) & 3,551 \\
Entered unshared & all units & $+0.38$ (0.002) & $+0.38$ (0.088) & 7,973 \\
Entered unshared & generation requests only & $+0.07$ (0.361) & $+0.93$ (0.175) & 3,551 \\
\bottomrule
\end{tabular}
\end{table}
\clearpage
\begin{landscape}
\begin{table}[!htbp]
\noindent\caption[AI-derived content, its pricing by the two panels, and typicality]{{\bf AI-derived content, its pricing by the two panels, and typicality.} Panel A: each submission sentence is matched to its most similar assistant idea unit of the same team (TF--IDF, threshold 0.35); it is AI-derived if the match reaches the threshold and never shared if that unit did not reappear in the team chat; $p$ values in parentheses. Matched against the assistant output of five other teams in the same task, the same rule classifies 1.4\% of submission sentences as AI-derived (against 71.7\% for the team's own output). Panel B: OLS of standardized team scores on the three variables with condition and task fixed effects ($n=148$; HC3 standard errors in parentheses); because condition is held fixed, these are descriptive associations. Panel C: mean TF--IDF similarity of a submission to the other submissions in its task, and to the pooled assistant output received by other teams in its task; defined for every team, so the contrasts preserve randomization ($p$ in parentheses).\label{tab:ec-unscreened}}
\centering
\begin{tabular}{lcccc}
\toprule
\multicolumn{5}{l}{\textit{Panel A. Content shares (team means; Welch tests)}} \\
 & C2 $-$ C1 & C2 $-$ C3 & C1 $-$ C4 & C5 $-$ C1 \\
AI-derived share of submission sentences & $+0.167$ (0.008) & $-0.110$ (0.009) & $+0.650$ ($<0.001$) & $-0.182$ (0.020) \\
\quad of which never shared in chat & $+0.062$ (0.236) & $-0.201$ (0.005) & $+0.133$ (0.001) & $-0.049$ (0.260) \\
\quad of which shared in chat & $+0.104$ (0.088) & $+0.091$ (0.187) & $+0.517$ ($<0.001$) & $-0.133$ (0.058) \\
\midrule
\multicolumn{5}{l}{\textit{Panel B. Association with standardized team scores within conditions}} \\
 & Human panel & Language-model panel & AI $-$ human & \\
AI-derived, never shared & $+1.392$ (0.363) & $+1.785$ (0.367) & $+0.393$ (0.365) & \\
AI-derived, shared & $+1.115$ (0.311) & $+0.976$ (0.352) & $-0.140$ (0.323) & \\
Number of proposals & $-0.290$ (0.030) & $-0.159$ (0.038) & $+0.131$ (0.038) & \\
\midrule
\multicolumn{5}{l}{\textit{Panel C. Typicality (TF--IDF similarity; differences from the no-AI control, task fixed effects)}} \\
 & C1 & C2 & C2 $-$ C1 & \\
Similarity to other teams' submissions & $+0.0030$ (0.120) & $+0.0023$ (0.232) & $-0.0007$ (0.754) & \\
Similarity to other teams' assistant output & $+0.0297$ ($<0.001$) & $+0.0325$ ($<0.001$) & $+0.0028$ (0.634) & \\
\bottomrule
\end{tabular}
\end{table}
\end{landscape}

\clearpage
\section{Process Traces, Engagement, and the Warning}\label{ec:process}

\subsection{Process Traces}
The platform logged 13,324 chat messages, 2,124 assistant exchanges, 21,354 activity-log rows, and 4,776 personal
drafts for the 148 analyzed teams. Stage-1 chat volume is the character count of human chat during Stage~1. Semantic
dispersion is the mean pairwise TF--IDF cosine distance among paragraph-level idea units in the final submission
(Chinese text segmented with \texttt{jieba}; unigrams and bigrams; minimum document frequency 2, maximum 0.90).

Table~\ref{tab:ec-paths} reports product-of-coefficients decompositions of condition effects through process
variables. They are descriptive: the process variables are post-treatment, and the decompositions do not identify
mediation. The proposal-count row quantifies the breadth channel of Proposition~\ref{prop:composition}: the
selection-stage access indicator is associated with creativity through proposal count by $-0.273$ points, offsetting a
positive direct association of 0.398. Total interaction volume, measured by chat messages, chat characters, or
assistant queries over the whole session, carries no indirect association.

\subsection{Engagement with Assistant Outputs}\label{ec:proximity}
For the 118 teams that used the assistant, we measure how closely the final submission follows the assistant outputs
the team received. Each sentence of the submission is matched to its most similar source among the team's assistant
outputs, chat, and drafts, using sentence embeddings (SentenceTransformers 5.1.2) and, as a robustness check, TF--IDF
vectors. Semantic proximity is the mean maximum similarity to assistant outputs, and the AI-derived sentence share is
the share of sentences whose nearest source is an assistant output.

Placebo baselines qualify what the measure captures. A proposal is more similar to its own team's assistant outputs
than to the team's own chat (0.043, 95\% CI [0.024, 0.062], $p<0.001$) or drafts (0.051, 95\% CI [0.024, 0.077],
$p<0.001$), but only marginally more similar to them than to assistant outputs received by other teams (0.013, 95\% CI
[$-$0.0001, 0.027], $p=0.054$). Proximity therefore reflects reliance on assistant-style content in general as well as
on the team's own exchanges, and we interpret it as a measure of borrowing from the assistant rather than of
team-specific uptake.

Table~\ref{tab:ec-prox} reports the results. Later-selection and full-process submissions borrow more from assistant
output than early-search submissions. Among users, proximity is positively associated with rated creativity, also after
controlling for interaction volume. Because teams choose how closely to follow the assistant, this association is
descriptive.

\subsection{The Warning Condition}\label{ia:warning}
The standing warning reduced reliance on the assistant without changing rated quality, as Remark~\ref{rem:warning}
predicts. Warned teams used the assistant somewhat less (12.5 compared with 14.9 queries per team, about one-sixth
fewer; Welch $p=0.09$) and borrowed less from its output. The share of submission sentences that closely match the
team's own assistant output, a measure defined for every assigned team and therefore a randomized comparison, fell
from 65\% under unwarned early-search access to 47\% under the warning (difference $-0.18$; 95\% CI $[-0.34,-0.03]$;
$p=0.020$; Table~\ref{tab:ec-unscreened}). Among teams that used the assistant, the semantic proximity between their
submissions and the assistant outputs they received was 0.064 lower ($p=0.005$; Table~\ref{tab:ec-prox}); because this
comparison is restricted to teams that used the assistant, it is observational. Rated creativity was statistically
indistinguishable from that of unwarned teams ($-0.167$; 95\% CI $[-0.49,0.15]$; $p=0.30$), the request profile was
nearly identical (52\% compared with 55\% generation), and so was the fate of the output (Table~\ref{tab:fate} of the
main text). When options arrive during search, the team's own screen already disciplines them, so a warning that
reduces reliance has little left to improve: reducing borrowing and improving the work are different objectives.

\clearpage
\begin{landscape}
\begin{table}[!htbp]
\noindent\caption[Exploratory decomposition of condition effects through process variables]{{\bf Exploratory decomposition of condition effects through process variables.} The $a$ path is the task-adjusted effect of the access indicator on the process variable (in the variable's units); the $b$ path is the adjusted association between the process variable and creativity; total and direct effects are the indicator's coefficients without and with the process variable; indirect effects use percentile intervals from 5,000 team bootstraps. Stage-1 and Stage-2 indicators follow the saturated coding (C1, C3, C5 and C2, C3, respectively). All process variables are post-treatment, so these are descriptive decompositions rather than mediation estimates. Rows below the line show that total interaction volume carries no indirect association.\label{tab:ec-paths}}
\centering
\setlength{\tabcolsep}{5pt}
\begin{tabular}{llcccccccc}
\toprule
Process variable & Indicator & $n$ & $a$ & $p_a$ & $b$ & $p_b$ & Total & Direct & Indirect $a\times b$ [95\% CI] \\
\midrule
Semantic dispersion & Stage-1 AI & 148 & 0.033 & $<0.001$ & 4.191 & 0.020 & 0.486 & 0.346 & 0.140 [0.019, 0.268] \\
Stage-1 chat (1,000 characters) & Stage-1 AI & 148 & 1.022 & $<0.001$ & 0.115 & 0.031 & 0.486 & 0.368 & 0.117 [0.021, 0.247] \\
Rater-coded proposal count & Stage-2 AI & 148 & 1.526 & 0.005 & $-$0.179 & $<0.001$ & 0.124 & 0.398 & $-$0.273 [$-$0.462, $-$0.097] \\
\midrule
Chat messages, all stages & Stage-1 AI & 148 & $-$15.9 & 0.109 & 0.0009 & 0.468 & 0.486 & 0.501 & $-$0.015 [$-$0.072, 0.028] \\
Chat messages, all stages & Stage-2 AI & 148 & $-$13.3 & 0.242 & 0.0009 & 0.468 & 0.124 & 0.137 & $-$0.012 [$-$0.066, 0.030] \\
Chat characters, all stages (1,000) & Stage-1 AI & 148 & 2.037 & $<0.001$ & 0.030 & 0.360 & 0.486 & 0.425 & 0.060 [$-$0.041, 0.186] \\
Chat characters, all stages (1,000) & Stage-2 AI & 148 & 1.802 & $<0.001$ & 0.030 & 0.360 & 0.124 & 0.071 & 0.053 [$-$0.034, 0.182] \\
Assistant queries, all stages & Stage-1 AI & 148 & 15.0 & $<0.001$ & 0.012 & 0.143 & 0.486 & 0.311 & 0.175 [$-$0.061, 0.426] \\
Assistant queries, all stages & Stage-2 AI & 148 & 13.8 & $<0.001$ & 0.012 & 0.143 & 0.124 & $-$0.036 & 0.161 [$-$0.055, 0.399] \\
\bottomrule
\end{tabular}
\end{table}
\end{landscape}
\clearpage
\begin{landscape}
\begin{table}[!htbp]
\noindent\caption[Engagement with assistant outputs]{{\bf Engagement with assistant outputs.} Semantic proximity is the similarity between a team's final submission and the assistant outputs it received (sentence embeddings, cosine similarity; robustness with TF--IDF vectors). Embedding- and TF--IDF-based team measures correlate at $r=0.91$--$0.92$. All models use HC3 standard errors. Proximity is undefined for the no-AI control, so Panels B and C are observational comparisons among assistant users and do not preserve randomization; one assigned C5 team that never used the assistant is excluded.\label{tab:ec-prox}}
\centering
\begin{tabular}{lccc}
\toprule
\multicolumn{4}{l}{\textit{Panel A. Share of submission sentences whose nearest source is an assistant output}} \\
Condition & Embedding & TF--IDF & \\
\midrule
Early-search (C1) & 0.130 [0.073, 0.187] & 0.116 [0.053, 0.179] & \\
Later-selection (C2) & 0.258 [0.176, 0.340] & 0.260 [0.171, 0.350] & \\
Full-process (C3) & 0.391 [0.296, 0.487] & 0.388 [0.265, 0.511] & \\
Warning (C5) & 0.125 [0.069, 0.180] & 0.060 [0.027, 0.093] & \\
\midrule
\multicolumn{4}{l}{\textit{Panel B. Condition differences in semantic proximity among assistant users ($n=118$; reference C1)}} \\
Contrast & Estimate (s.e.) & 95\% CI & $p$ \\
C2 vs.\ C1 & 0.051 (0.017) & [0.017, 0.084] & 0.003 \\
C3 vs.\ C1 & 0.081 (0.016) & [0.050, 0.113] & $<0.001$ \\
C5 vs.\ C1 & $-$0.064 (0.023) & [$-$0.109, $-$0.020] & 0.005 \\
\midrule
\multicolumn{4}{l}{\textit{Panel C. Association of proximity with rated creativity (condition and task fixed effects)}} \\
Measure & $n$ & Estimate [95\% CI] & $p$ \\
Embedding proximity (standardized) & 118 & 0.349 [0.116, 0.583] & 0.003 \\
\quad adding three interaction-volume controls & 118 & 0.431 [0.187, 0.676] & $<0.001$ \\
TF--IDF proximity (score points per s.d.) & 118 & 0.198 [0.060, 0.336] & 0.005 \\
Maximum sentence proximity (score points per s.d.) & 118 & 0.182 [0.073, 0.292] & 0.001 \\
AI-derived sentence share, embedding & 148 & $-$0.059 [$-$0.168, 0.050] & 0.286 \\
\bottomrule
\end{tabular}
\end{table}
\end{landscape}

\clearpage
\section{Alternative Evaluators}\label{ec:evaluators}

\subsection{The Language-Model Evaluator Panel}\label{ec:aipanel}
Ten models from five families scored every proposal once at temperature 0.1: Claude Sonnet 4.6, Claude Opus 4.6, and
Claude Opus 4.7; GPT-5.5 and GPT-5.5 Pro; Gemini 3.1 Flash and Gemini 3.1 Pro; DeepSeek V3 and DeepSeek V4 Pro; and
Kimi K2.6. For each submission, the model received the full text, the expected number of proposals from the locked
template, the human rubric with its 1--7 scales, and a JSON response schema; the system instruction designated the
model as a strict structured evaluator. Responses were validated against the expected proposal count. The AI composite
uses the same dimensions, weights, and aggregation as the human composite; scores are standardized within model,
averaged within model and team, and then averaged across models.

Table~\ref{tab:ec-aipanel} reports scale use, agreement with the human panel, and reliability. Models differ
substantially in leniency (mean composites from 4.53 to 6.14), so standardizing within model is essential.
Proposal-level correlations with the human composite range from 0.40 to 0.57.

\subsection{Evaluators That Credit Breadth}
Table~\ref{tab:ec-breadth} reports the timing contrast and the contrast of later-selection access with the control when
the human score credits each proposal with $\gamma$ points (Corollary~\ref{cor:evaluator} of the main text), and under
an equal-weight composite whose originality score credits the number of proposals.

\subsection{Sources of the Divergence Between the Panels}
Table~\ref{tab:ec-evaldecomp} decomposes the divergence between the language-model and human panels. Panel A regresses
the difference between the two panels' standardized team scores on condition indicators and adds, in turn, the
proposal count, the share of AI-derived sentences, and rated text quality. The proposal count accounts for about a
sixth of the divergence on the timing contrast, and the three variables together for about a third. Panel B shows that
the divergence appears on every rated dimension: the model panel does not distinguish early- from late-access
submissions on any of them.

\clearpage
\begin{landscape}
\begin{table}[!htbp]
\noindent\caption[The language-model evaluator panel]{{\bf The language-model evaluator panel.} Each model scored the 394 proposal units of the locked template once at temperature 0.1, using the human rubric, the full submission text, the expected number of proposals, and a JSON response schema. Mean and s.d.\ refer to the 1--7 composite; share $\geq6$ is the fraction of proposals scored 6 or higher. Agreement statistics compare each model with the mean human composite for the same proposal; mean difference is AI minus human. Because ICC(2,$k$) rises with the number of raters, the panel's higher reliability is not evidence of validity.\label{tab:ec-aipanel}}
\centering
\begin{tabular}{llcccccc}
\toprule
Model & Family & Mean & s.d. & Share $\geq 6$ & Pearson $r$ & Spearman $\rho$ & Mean diff. \\
\midrule
Claude Sonnet 4.6 & Claude & 4.53 & 0.80 & 0.010 & 0.483 & 0.427 & 0.11 \\
Claude Opus 4.6 & Claude & 4.72 & 0.76 & 0.028 & 0.395 & 0.325 & 0.30 \\
Claude Opus 4.7 & Claude & 5.42 & 0.75 & 0.208 & 0.516 & 0.430 & 0.99 \\
GPT-5.5 & GPT & 5.17 & 0.78 & 0.117 & 0.565 & 0.514 & 0.74 \\
GPT-5.5 Pro & GPT & 5.16 & 0.81 & 0.137 & 0.548 & 0.499 & 0.74 \\
Gemini 3.1 Flash & Gemini & 4.86 & 0.81 & 0.048 & 0.562 & 0.503 & 0.44 \\
Gemini 3.1 Pro & Gemini & 5.08 & 0.84 & 0.102 & 0.548 & 0.484 & 0.65 \\
DeepSeek V3 & DeepSeek & 6.14 & 0.88 & 0.635 & 0.442 & 0.399 & 1.72 \\
DeepSeek V4 Pro & DeepSeek & 5.62 & 0.82 & 0.299 & 0.508 & 0.437 & 1.20 \\
Kimi K2.6 & Kimi & 5.01 & 0.91 & 0.135 & 0.402 & 0.352 & 0.58 \\
\midrule
\multicolumn{8}{l}{Mean cross-family correlation between models: $r=0.673$ (39 pairs; range 0.384--0.799); within family: $r=0.755$ (6 pairs).} \\
\multicolumn{8}{l}{Reliability: AI panel ICC(2,$k$)$=0.913$ (10 models), 0.939 excluding DeepSeek V3; human panel ICC(2,$k$)$=0.442$ (4 raters).} \\
\multicolumn{8}{l}{Measurement invariance across evaluator types (CFA): metric CFI $=0.991$, RMSEA $=0.103$; scalar CFI $=0.985$, RMSEA $=0.093$.} \\
\bottomrule
\end{tabular}
\end{table}
\end{landscape}
\clearpage
\begin{table}[!htbp]
\noindent\caption[Evaluators that credit breadth]{{\bf Evaluators that credit breadth.} Upper panel: the human team score plus $\gamma$ points per rater-coded proposal (Corollary~\ref{cor:evaluator}). The timing gap vanishes at $\gamma^*=0.46$, the C1--C2 difference in the human score divided by the C2--C1 difference in proposals. Lower panel: originality $=0.6\times$quality $+\,0.4\times$ a score for the number of proposals (1 $=2$, 2 $=4$, 3 $=6$, $\geq4=7$), and creativity is the unweighted mean of originality, appropriateness and elaboration, computed per rater and averaged. Welch tests.\label{tab:ec-breadth}}
\centering
\begin{tabular}{lccccc}
\toprule
Credit $\gamma$ per proposal & C1 $-$ C2 [95\% CI] & $p$ & Hedges $g$ & C2 $-$ C4 & $p$ \\
\midrule
0.0 & 0.356 [0.064, 0.648] & 0.018 & 0.62 & 0.130 & 0.425 \\
0.1 & 0.279 [0.047, 0.512] & 0.019 & 0.61 & 0.283 & 0.043 \\
0.2 & 0.203 [$-$0.022, 0.428] & 0.076 & 0.46 & 0.436 & 0.001 \\
0.3 & 0.126 [$-$0.147, 0.399] & 0.359 & 0.24 & 0.589 & $<0.001$ \\
0.4 & 0.049 [$-$0.306, 0.405] & 0.782 & 0.07 & 0.742 & $<0.001$ \\
0.5 & $-$0.027 [$-$0.481, 0.427] & 0.905 & $-$0.03 & 0.895 & $<0.001$ \\
\midrule
\multicolumn{6}{l}{\textit{Equal-weight composite whose originality score credits the number of proposals}} \\
Contrast & Difference [95\% CI] & $p$ & Hedges $g$ & & \\
C1--C2 & 0.231 [0.040, 0.423] & 0.019 & 0.62 & & \\
C1--C4 & 0.604 [0.433, 0.774] & $<0.001$ & 1.82 & & \\
C2--C4 & 0.372 [0.185, 0.560] & $<0.001$ & 1.02 & & \\
C3--C4 & 0.445 [0.265, 0.624] & $<0.001$ & 1.27 & & \\
\bottomrule
\end{tabular}
\end{table}
\clearpage
\begin{landscape}
\begin{table}[!htbp]
\noindent\caption[Sources of the divergence between the human and language-model panels]{{\bf Sources of the divergence between the human and language-model panels.} Panel A: OLS with task fixed effects and HC3 standard errors ($n=148$); the dependent variable is the difference between the standardized model-panel and human-panel team scores (Table~\ref{tab:evaluators}, column 4 of the main text). Proposal count is rater-coded; AI share is the share of submission sentences whose nearest source is an assistant output (sentence embeddings; zero without access); rated text quality is the human team mean, centered. Panel B: Welch contrasts of standardized team scores ($p$ in parentheses); model-panel dimension scores are standardized within model, averaged over models, and re-standardized; the last column is the correlation of the two panels' team scores across all 148 teams. $^{*}p<0.10$, $^{**}p<0.05$, $^{***}p<0.01$.\label{tab:ec-evaldecomp}}
\centering
\begin{tabular}{lcccp{2.6in}c}
\toprule
\multicolumn{6}{l}{\textit{Panel A. AI-minus-human gap (standardized) regressed on condition and submission features}} \\
Specification & C1 $-$ C2 & C2 & C3 & Added variables & Adj.\ $R^2$ \\
Conditions only & $-0.662$*** & $+0.577$** & $+0.642$*** & -- & 0.107 \\
+ proposal count & $-0.547$** & $+0.354$ & $+0.585$*** & proposals $+0.146$*** & 0.207 \\
+ AI-derived sentence share & $-0.591$*** & $+0.434$* & $+0.426$* & AI share $+0.553$ & 0.115 \\
+ both & $-0.527$** & $+0.312$ & $+0.514$** & proposals $+0.142$***; AI share $+0.187$ & 0.203 \\
+ both + rated text quality & $-0.455$** & $+0.431$* & $+0.629$** & proposals $+0.086$*; AI share $+0.283$; text quality $-0.322$** & 0.228 \\
\midrule
\multicolumn{6}{l}{\textit{Panel B. Timing contrast C1 $-$ C2 by dimension (standardized team scores)}} \\
Dimension & Human panel & Model panel & AI $-$ human & \multicolumn{2}{c}{Correlation of panels} \\
Originality quality & $+0.60$ (0.017) & $+0.02$ (0.947) & $-0.58$ (0.020) & \multicolumn{2}{c}{0.49} \\
Appropriateness & $+0.69$ (0.002) & $-0.16$ (0.472) & $-0.84$ (0.005) & \multicolumn{2}{c}{0.13} \\
Elaboration & $+0.52$ (0.053) & $+0.13$ (0.602) & $-0.38$ (0.060) & \multicolumn{2}{c}{0.64} \\
Text quality & $+0.54$ (0.048) & $+0.18$ (0.461) & $-0.36$ (0.260) & \multicolumn{2}{c}{0.12} \\
Composite & $+0.58$ (0.018) & $-0.08$ (0.741) & $-0.66$ (0.003) & \multicolumn{2}{c}{0.61} \\
\bottomrule
\end{tabular}
\end{table}
\end{landscape}

\clearpage
\section{Exploratory Moderator Screen}\label{ec:mods}

We screened team means, standard deviations, and ranges of all registration-survey scales as moderators of the
condition effects. Table~\ref{tab:ec-mods} lists the eight interactions with the smallest $p$ values. Given the number
of tests, none is distinguishable from chance, and we do not interpret them.

\clearpage
\begin{landscape}
\begin{table}[!htbp]
\noindent\caption[Exploratory moderator screen (eight smallest $p$ values)]{{\bf Exploratory moderator screen (eight smallest $p$ values).} Each row is a separate model that adds the moderator and its interactions with the Stage-1, Stage-2 and warning indicators to the saturated condition model with task fixed effects (HC3); the moderator is standardized, and cells give $b$ ($p$). The screen covered all team means, standard deviations and ranges of the registration-survey scales. No adjustment for multiplicity is applied, and none of these interactions would survive one; we report them for transparency and do not interpret them.\label{tab:ec-mods}}
\centering
\begin{tabular}{lccc}
\toprule
Team-level moderator & Stage-1 AI $\times$ moderator & Stage-2 AI $\times$ moderator & Warning $\times$ moderator \\
\midrule
AI-use frequency, mean & $-$0.021 (0.844) & 0.332 (0.002) & 0.281 (0.105) \\
Conscientiousness, s.d. & 0.031 (0.795) & $-$0.141 (0.218) & $-$0.306 (0.013) \\
Conscientiousness, range & 0.032 (0.784) & $-$0.141 (0.208) & $-$0.311 (0.016) \\
AI optimism, mean & $-$0.052 (0.576) & 0.205 (0.026) & 0.197 (0.246) \\
Prefer AI interaction, mean & $-$0.062 (0.599) & $-$0.241 (0.029) & $-$0.220 (0.290) \\
Emotional stability, mean & $-$0.250 (0.034) & 0.017 (0.883) & 0.144 (0.291) \\
Openness, range & $-$0.017 (0.877) & 0.221 (0.036) & $-$0.064 (0.608) \\
AI better at work, mean & $-$0.264 (0.036) & 0.158 (0.199) & 0.252 (0.228) \\
\bottomrule
\end{tabular}
\end{table}
\end{landscape}

\end{document}